\documentclass[11pt,a4paper]{article}

\usepackage{amsthm}
\usepackage{amsmath}
\usepackage{natbib}
\usepackage[colorlinks,citecolor=blue,urlcolor=blue,filecolor=blue,backref=page]{hyperref}
\usepackage{graphicx}
\usepackage{enumitem}
\usepackage{verbatim}
\usepackage{graphics}
\usepackage{color}
\usepackage{caption}
\usepackage{comment}
\usepackage{amssymb}
\usepackage{subcaption}
\usepackage{tikz}
\usepackage{algorithm}
\usepackage{multirow}
\usepackage{titling}

\title{Bayesian inference for diffusion-driven mixed-effects models}
\author{Gavin A.~Whitaker$^1$ \and\ Andrew Golightly$^1$\thanks{email: \texttt{andrew.golightly@ncl.ac.uk}} \and\ Richard J.~Boys$^1$ \and\ Chris Sherlock$^2$}
\date{\small $^{1}$School of Mathematics \& Statistics, Newcastle University,\\
  Newcastle-upon-Tyne, NE1 7RU, UK \\
$^{2}$Department of Mathematics and Statistics, Lancaster University, Lancaster, LA1 4YF}

\newcommand{\indep}{\overset{indep}{\sim }}
\newcommand{\half}{{1/2}}

\begin{document}
\maketitle
\begin{abstract}
Stochastic differential equations (SDEs) provide a natural framework
for modelling intrinsic stochasticity inherent in many continuous-time
physical processes. When such processes are observed in multiple
individuals or experimental units, SDE driven mixed-effects models
allow the quantification of both between and within individual
variation. Performing Bayesian inference for such models using
discrete-time data that may be incomplete and subject to measurement
error is a challenging problem and is the focus of this paper. We
extend a recently proposed MCMC scheme to include the SDE driven
mixed-effects framework.  Fundamental to our approach is the
development of a novel construct that allows for efficient sampling of
conditioned SDEs that may exhibit nonlinear dynamics between
observation times. We apply the resulting scheme to synthetic data
generated from a simple SDE model of orange tree growth, and real data
on aphid numbers recorded under a variety of different treatment
regimes. In addition, we provide a systematic comparison of our
approach with an inference scheme based on a tractable approximation
of the SDE, that is, the linear noise approximation.
\end{abstract}


\section{Introduction}\label{intro}

Diffusion processes satisfying It\^o stochastic differential equations (SDEs) 
are a class of continuous-time, continuous-valued Markov stochastic processes
that can be used to model a wide range of phenomena. Examples include (but are
not limited to) epidemics, financial time series, population dynamics (including
predator-prey systems) and intra-cellular processes. When repeated measurements 
on a system of interest are made, differences between individuals or experimental 
units can be incorporated through random effects. Quantification of both system 
(intrinsic) variation and variation between units leads to a stochastic differential 
mixed-effects model (SDMEM). 


Unfortunately, analytic 
intractability of SDEs governing most nonlinear multivariate diffusions can make likelihood-based 
inference methods problematic. Methods to overcome this difficulty include 
closed-form expansion of the transition density \citep{Ait-Sahalia_2002,Ait-Sahalia_2008,
Stramer_Bognar_Schneider2010}, exact simulation approaches 
\citep{Beskos_2006,Sermaidis_2013} and use of the Euler-Maruyama 
approximation coupled with data augmentation \citep{pederen_1995b,
Elerian_2001,Eraker_2001,Durham_2001,Golightly_Wilkinson_2008,Stramer_Bognar_2011,Kou_2012}. Difficulty 
in performing inference for SDEs has resulted in relatively little work on SDMEMs. 

\cite{Picchini_2010} propose a procedure for obtaining approximate maximum likelihood 
estimates for SDMEM parameters based on a two step approach; they use a closed-form Hermite expansion 
\citep{Ait-Sahalia_2002,Ait-Sahalia_2008} to approximate the transition density,  
before using Gaussian quadrature to numerically integrate the conditional likelihood 
with respect to the random parameters. As noted by \cite{Picchini_2011}, the approach is, in practice, 
limited to a scalar random effect parameter since Gaussian quadrature is computationally inefficient 
when the dimension of the random effect parameter grows. The methodology is extended in \cite{Picchini_2011} to deal with 
multiple random effects. A number of limitations remain however. In particular 
a reducible diffusion process is required, that is, one which can be 
transformed to give a unit diffusion coefficient. Another drawback is that the method 
cannot account for measurement error. A promising approach appears to be the 
use of the extended Kalman filter (EKF) to provide a tractable approximation to 
the SDMEM. This has been the focus of 
\cite{Overgaard_et_al_2005}, \cite{Tornoe_et_al_2005} and \cite{Berglund_et_al_2011}. 
The \verb+R+ package \verb+PSM+ \citep{Klim_et_al_2009} uses the EKF to estimate 
SDMEMs. Unfortunately, a quantification of the effect of using these approximate 
inferential models appears to be missing from the literature. \cite{Donnet_Foulley_Samson_2010} 
discuss inference for SDMEMs in a Bayesian framework, and implement a Gibbs sampler when the SDE 
(for each experimental unit) has an explicit solution. When no explicit solution exists 
they suggest that a solution might be found using the Euler-Maruyama discretisation.



\subsection{Contributions and organisation of the paper}\label{cont}

In this article we provide a method that permits (simulation-based)
Bayesian inference for a large class of multivariate SDMEMs using
discrete-time observations that may be incomplete (so that only a
subset of model components are observed) and subject to measurement
error. The method makes use of a novel scheme that allows for
observations made sparsely in time, as the process of interest may
exhibit nonlinear dynamics between measurement times.

As a starting point, we consider a data augmentation approach that 
adopts an Euler-Maruyama approximation of unavailable transition densities and augments low 
frequency data with additional time points over which the approximation is satisfactory. Although 
a discretisation bias is introduced, this can be made arbitrarily small (at greater computational expense). 
Moreover, the approach is flexible, and is not restricted to reducible diffusions. A Bayesian approach 
then aims to construct the joint posterior density for parameters and the components of 
the latent process. The intractability of the posterior density necessitates simulation 
techniques such as Markov chain Monte Carlo. As is well documented in \cite{Roberts_Stramer_2001}, 
care must be taken in the design of the MCMC sampler due to dependence between the parameters 
entering the diffusion coefficient and the latent process. We therefore adapt the 
reparameterisation technique (known as the modified innovation scheme) of 
\cite{Golightly_Wilkinson_2008} and \cite{Golightly_Wilkinson_2010} 
(see also \cite{Stramer_Bognar_2011,Fuchs_2013,Papaspiliopoulos_2013}) 
to the SDMEM framework. A key requirement of the scheme is the ability 
to sample the latent process between two fixed values. Previous approaches have typically 
focused on the modified diffusion bridge construct of \cite{Durham_2001}. For the SDMEM 
considered in Section~\ref{aphid} we find that this construct fails to capture the 
nonlinear dynamics exhibited between observation times. We therefore develop a novel 
bridge construct that is simple to implement and can capture nonlinear behaviour. 

Finally, we provide a systematic comparison of our approach with an
inference scheme based on a linear noise approximation (LNA) of the
SDE. The LNA approximates transition densities as Gaussian, and when
combined with Gaussian measurement error, allows the latent process to
be integrated out analytically. Essentially a forward (Kalman) filter
can be implemented to calculate the marginal likelihood of all
parameters of interest, allowing a marginal Metropolis-Hastings scheme
targeting their posterior distribution. It should be noted, however,
that evaluation of the Gaussian transition densities under the LNA
require the solution of an ordinary differential equation (ODE) system
whose order grows quadratically with the number of components (say
$d$) governed by the SDE. The computational efficiency of an LNA based
inference scheme will therefore depend on $d$, and on whether or not
the ODE system can be solved analytically.

We apply the methods to two examples. First, we consider a synthetic
dataset generated from an SDMEM driven by the simple univariate model
of orange tree growth presented in \cite{Picchini_2010} and
\cite{Picchini_2011}. The ODE system governing the LNA solution is tractable 
in this example. Second, we fit a model of aphid growth to both real 
and synthetic data. The real data are taken from
\cite{Matis_2008} and consist of Cotton aphid (\emph{Aphis gossypii})
counts in the Texas High Plains obtained for three different levels of
irrigation water, nitrogen fertiliser and block. This application is
particularly challenging, due to the nonlinear drift and diffusion
coefficients governing the SDMEM and the ability to only observe one of
the model components (with error). Moreover, the ODE system governing 
the LNA solution is intractable and a numerical solver must be used. 
 
The remainder of the article is organised as follows. The SDMEM
framework is introduced in Section~\ref{sdmem}.  Section~\ref{bayes}
provides MCMC methods for Bayesian inference, with a novel bridge
construct outlined in Section~\ref{dgode}. The linear noise
approximation and its application as an inferential model is discussed
in Section~\ref{lna}. The methods are applied in Section~\ref{apps}
before conclusions are drawn in Section~\ref{disc}.
  
\section{Stochastic Differential Mixed-effects models}\label{sdmem}

Consider the case where we have $N$ experimental units randomly chosen
from a theoretical population, and associated with each unit $i$ is a
continuous-time $d$-dimensional It\^o process $\{X_t^i, t\geq 0\}$
governed by the SDE
\begin{equation}
dX_t^i=\alpha(X_t^i,\theta,b^i)\,dt+\sqrt{\beta(X_t^i,\theta,b^i)}\,dW_t^i,
\quad X_0^i=x_0^i, \quad i=1,\ldots,N. \label{eqn:sdmem}
\end{equation}
Here, $\alpha$ is a $d$-vector of drift functions, the 
diffusion coefficient $\beta$ is a $d \times d$ positive 
definite matrix with a square root representation 
$\sqrt{\beta}$ such that $\sqrt{\beta}\sqrt{\beta}^T=\beta$ 
and $W_t^i$ is a $d$-vector of (uncorrelated) standard 
Brownian motion processes. The $p$-vector parameter 
$\theta=(\theta_{1},\ldots,\theta_{p})^T$ is common 
to all units whereas the $q$-vectors $b^i=(b_{1}^{i},\ldots,b_{q}^{i})^T$, 
$i=~1,\ldots,N$, are unit-specific effects, which may 
be fixed or random. In the most general random effects 
scenario we let $\pi(b^i|\psi)$ denote the joint distribution 
of $b^i$, parameterised by the $r$-vector $\psi=(\psi_{1},\ldots,\psi_{r})^T$. 
The model defined by (\ref{eqn:sdmem}) allows for differences between 
experimental units through different realisations of the Brownian motion paths $W_t^i$ 
and the random effects $b^i$, accounting for inherent stochasticity within 
a unit, and variation between experimental units respectively. 

We assume that each experimental unit $\{X_t^i,t\geq 0\}$ cannot be
observed exactly, but observations
$y^i=(y_{t_0}^i,y_{t_1}^i,\ldots,y_{t_n}^i)^T$ are available and these
are conditionally independent (given the latent process). We link the
observations to the latent process via
\begin{equation}\label{eqn:obs}
Y_t^i = F^T X_t^i + \epsilon_t, \qquad \epsilon_t|\Sigma\indep N(0,\Sigma),
\end{equation}
where $Y_t^i$ is a $d_o$-vector, $F$ is a constant $d\times d_o$
matrix and $\epsilon_t$ is a random $d_o$-vector. Note that this setup
allows for only observing a subset of components ($d_o<d$) and this
aspect is explored further in Section~\ref{aphid}.

Together \eqref{eqn:sdmem} and \eqref{eqn:obs} completely specify the
stochastic differential mixed-effects model. However, for most
problems of interest the form of the SDE associated with each unit
will not permit an analytic solution, precluding straightforward
inference for the unknown parameters. We therefore work with the
Euler-Maruyama approximation
\[
\Delta X_t^i\equiv X^i_{t+\Delta t}-X^i_t=\alpha(X_t^i,\theta,b^i)\,\Delta t+\sqrt{\beta(X_t^i,\theta,b^i)}\,\Delta W_t^i
\]
where $\Delta W_t^i\sim N(0,I_d\Delta t)$ and $\Delta t$ is the length
of time between observations, assumed equally spaced for notational
simplicity. It is, of course, unlikely that this approximation will be
sufficiently accurate over the intervals between observation times and
so we adopt a data augmentation scheme. Partitioning [$t_j,t_{j+1}$]
as
\[
t_j=\tau_{j,0}<\tau_{j,1}<\tau_{j,2}<\ldots<\tau_{j,m-1}<\tau_{j,m}=t_{j+1}
\]
introduces $m-1$ intermediate time points with interval widths of length 
\begin{equation}\label{eqn:part}
\Delta\tau\equiv\tau_{j,k+1}-\tau_{j,k}=\frac{t_{j+1}-t_j}{m}.
\end{equation}
The Euler-Maruyama approximation can then be applied over each interval 
of width $\Delta\tau$, and the associated discretisation bias can be made 
arbitrarily small at the expense of having to impute $\{X_t^i\}$ at the 
intermediate times. We adopt the shorthand notation 
\[
x_{[t_j,t_{j+1}]}^i\equiv x_{[j,j+1]}^i = ( x_{\tau_{j,0}}^i,x_{\tau_{j,1}}^i,\ldots,x_{\tau_{j,m}}^i )^T
\]
for the 
latent process over the time interval $[t_j,t_{j+1}]$ for unit $i$. 
Hence, the complete latent trajectory associated with unit $i$ is given by
\[
(x^i)^T=((x_{[0,1]}^i)^T,(x_{(1,2]}^i)^T \ldots,(x_{(n-1,n]}^i)^T)
\]
and we stack all unit-specific trajectories into a matrix
$x=(x^1,\ldots,x^N)$.  Likewise the matrix $y=(y^1,\ldots,y^N)$
denotes the entire set of observations. Next we focus on how to
perform Bayesian inference for the model quantities $x$, $\theta$,
$b=(b^1,\ldots,b^N)^T$, $\psi$ and $\Sigma$.

\section{Bayesian inference}\label{bayes}

The joint posterior for the common parameters $\theta$, fixed/random
effects $b$, hyperparameters $\psi$, measurement error variance
$\Sigma$ and latent values $x$ is given by
\begin{equation}
\pi(\theta,\psi,\Sigma,b,x|y)\propto
\pi(\theta)\pi(\psi)\pi(\Sigma)\pi(b|\psi)\pi(x|\theta,b)\pi(y|x,\Sigma) \label{eqn:posterior}
\end{equation}
where $\pi(\theta)\pi(\psi)\pi(\Sigma)$ is the joint prior density ascribed to $\theta$, 
$\psi$ and $\Sigma$. In addition we have that
\begin{equation}\label{eqn:xdens}
\pi(x|\theta,b)=\prod\limits_{i=1}^{N}\prod\limits_{j=0}^{n-1}\prod\limits_{k=1}^{m}\pi(x^i_{\tau_{j,k}}|x^i_{\tau_{j,k-1}},\theta,b^i)
\end{equation}
where
\[
\pi(x^i_{\tau_{j,k}}|x^i_{\tau_{j,k-1}},\theta,b^i)= N\left(x^i_{\tau_{j,k}}\,;\, 
x^i_{\tau_{j,k-1}}+\alpha(x^i_{\tau_{j,k-1}}\,\theta,b^i)\Delta\tau,~
\beta(x^i_{\tau_{j,k-1}},\theta,b^i)\Delta\tau\right)
\]
and $N(\cdot\,;\,m,V)$ denotes the multivariate Gaussian density with
mean $m$ and variance~$V$. Similarly
\[
\pi(y|x,\Sigma)= \prod\limits_{i=1}^{N}\prod\limits_{j=0}^{n}\pi(y_{t_{j}}^i|x_{t_{j}}^i,\Sigma)
\]
where $\pi(y_{t_j}^i|x_{t_j}^i,\Sigma)=N(y_{t_{j}}^i\,;\,x_{t_{j}}^i,\Sigma)$. 
Given the intractability of
the joint posterior distribution in (\ref{eqn:posterior}) we aim to
construct a Markov chain Monte Carlo (MCMC) scheme which generates
realisations from this posterior. The form of the SDMEM admits a Gibbs
sampling strategy with blocking that sequentially takes draws from the
full conditionals \\
\begin{minipage}[b]{0.5\linewidth}
\vspace{0.5em}
\begin{enumerate}
\item[1.] $\pi(x|\theta,\psi,\Sigma,b,y)=\pi(x|\theta,\Sigma,b,y)$,
\item[3.] $\pi(\theta|\psi,\Sigma,b,x,y)=\pi(\theta|b,x)$, 
\item[5.] $\pi(\psi|\theta,\Sigma,b,x,y)=\pi(\psi|b)$. 
\item[]  \vspace{-1em}
\end{enumerate} 
\end{minipage} 
\begin{minipage}[b]{0.5\linewidth}
\vspace{0.5em}
\begin{enumerate}
\item[2.] $\pi(\Sigma|\theta,\psi,b,x,y)=\pi(\Sigma|x,y)$,
\item[4.] $\pi(b|\theta,\psi,\Sigma,x,y)=\pi(b|\theta,\psi,x)$, 
\item[]  
\item[]  \vspace{-1em}
\end{enumerate} 
\end{minipage} 
Further blocking strategies that exploit the conditional dependencies
between the model parameters and latent trajectories can be used. For
example, in step 1 the latent trajectories can be updated separately
for each experimental unit.  Likewise, the unit-specific random
effects can be updated separately.  Where necessary, Metropolis-within-Gibbs 
updates can be used.  We note that as written, this scheme will
mix intolerably poorly as the degree of augmentation $m$ is increased
due to dependence between the latent values~$x$ and the parameters
entering the diffusion coefficient (namely $\theta$ and~$b$). We refer
the reader to \cite{Roberts_Stramer_2001} for a detailed discussion of
this problem. A simple mechanism for overcoming this issue is to
update the parameters and latent trajectories jointly (and this has
been considered for SDE models by \cite{Stramer_Bognar_2011} and
\cite{Golightly_Wilkinson_2011}).  For SDMEMs a joint update of
$\theta$, $b$ and $x$ is likely to result in a sampler with low
acceptance rates. We therefore wish to preserve the blocking structure
described above and instead adapt the reparameterisation of
\cite{Golightly_Wilkinson_2008} to our problem. In what follows, we
describe in detail each step of the Gibbs sampler.
 
\subsection{Path updates}\label{path}

The full conditional density of the latent paths for all experimental units 
is given by
\begin{align*}
\pi(x|\theta,\Sigma,b,y)&\propto \pi(x|\theta,b)\pi(y|x,\Sigma)
=\prod\limits_{i=1}^{N}\pi(x^i|\theta,b^i)\pi(y^i|x^i,\Sigma)
\end{align*}
which suggests a scheme where unit-specific paths are updated separately. We now
focus on an updating scheme for a single path, and drop $i$ from the notation, writing 
$x$ in place of $x^i$ and  $x_{[j,j+1]}$ in place of $x_{[j,j+1]}^i$. Since the parameters 
are fixed throughout this updating step, we also drop them from the notation.

\begin{figure}[t!]
\centering
\begin{tikzpicture}[scale=5,>=latex]
     \node[draw,circle,fill=black,inner sep=0.5mm, label=above:{\large $x_{t_j}$}]
         (xo0) at (0,0.5) {};
     \node[draw,circle,fill=black,inner sep=0.5mm, label=above:{\large $X_{t_{j+1}}$}]
         (xo1) at (1,0.5) {};
     \node[draw,circle,fill=black,inner sep=0.5mm, label=above:{\large $x_{t_{j+2}}$}]
         (xo2) at (2,0.5) {};
     \node[draw,circle,fill=black,inner sep=0.5mm, label=below:{\large $y_{t_j}$}]
         (xu0) at (0,0.15) {};
     \node[draw,circle,fill=black,inner sep=0.5mm, label=below:{\large $y_{t_{j+1}}$}]
         (xu1) at (1,0.15) {};
     \node[draw,circle,fill=black,inner sep=0.5mm, label=below:{\large $y_{t_{j+2}}$}]
         (xu2) at (2,0.15) {};

	\draw[thick,->]
         (0,0.25) -- (0,0.4);
	\draw[thick,->]
         (1,0.25) -- (1,0.4);
	\draw[thick,->]
         (2,0.25) -- (2,0.4);

     \draw[blue!80!,densely dotted]
	  (0.1,0.35) -- (0.1,0.65) -- (1.10,0.65) -- (1.10,0.45) -- (0.93,0.45) -- (0.93,0.35) -- cycle;

    \draw[blue!80!black,densely dotted]
         (1.12,0.35) rectangle (1.9,0.65);

     \draw[red!70!black,<->]
         (0.15,0.5) -- (0.85,0.5)
         node[midway, below, black] {\footnotesize$m-1$ latent values};

     \draw[red!70!black,<->]
         (1.15,0.5) -- (1.85,0.5)
         node[midway, below, black] {\footnotesize$m-1$ latent values};

     \node at (0.5, 0.3) {{\small Propose using $q_1$}};
     \node at (1.5, 0.3) {{\small Propose using $q_2$}};
\end{tikzpicture}
\caption{Path update illustration over a block of size $2m+1$.} \label{pic-2blocknoise}
\end{figure}
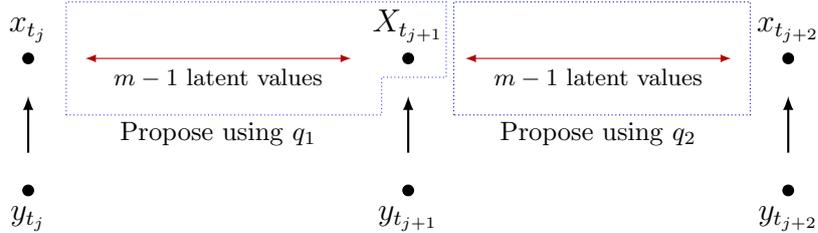

Following \cite{Golightly_Wilkinson_2008} we update $x$ in overlapping
blocks of size $2m+1$. Consider times $t_j$ and $t_{j+2}$ at which the
current values of the latent process are $x_{t_j}$ and
$x_{t_{j+2}}$. The full conditional density of the latent process over
the interval $(t_j,t_{j+2})$ is given by
\begin{equation}\label{eqn:FCDpath}
\pi(x_{(j,j+2)}|x_{t_j},y_{t_{j+1}},x_{t_{j+2}})
\propto \pi(y_{t_{j+1}}|x_{t_{j+1}})\prod\limits_{l=j}^{j+1}\prod\limits_{k=1}^{m}\pi(x_{\tau_{l,k}}|x_{\tau_{l,k-1}}).
\end{equation}
Under the nonlinear structure of the diffusion process, this full
conditional is intractable and so we use a Metropolis-Hastings step to
generate draws from (\ref{eqn:FCDpath}). We use an independence
sampler with proposal density of the form
\begin{equation}\label{pathprop}
q(x_{(j,j+2)}|x_{t_j},y_{t_{j+1}},x_{t_{j+2}})=q_1(x_{(j,j+1]}|x_{t_j},y_{t_{j+1}})\,
q_2(x_{(j+1,j+2)}|x_{t_{j+1}},x_{t_{j+2}}).
\end{equation}
Figure~\ref{pic-2blocknoise} gives an illustration of the updating
procedure which can be applied over intervals $(t_{j},t_{j+2})$,
$j=0,1,\ldots,n-2$, with two additional Metropolis-Hastings steps
(such as those described in \cite{Golightly_Wilkinson_2006}) that
allow for updating $x$ at times $t_0$ and $t_n$. Deriving appropriate
forms for $q_1$ and $q_2$ requires the ability to (approximately)
generate a discrete-time realisation of a diffusion process between
two time points at which the process is either observed exactly or
subject to Gaussian noise. The resulting trajectory is typically
referred to as a diffusion bridge.

Several strategies for constructing diffusion bridges 
have been proposed in the literature. For example,
\cite{pederen_1995b} used the Euler-Maruyama scheme to generate
bridges myopically of the end point. \cite{Durham_2001} 
use a linear Gaussian approximation of the 
distribution of the process conditional on the value at a
previous and future time point, giving a construct known as the
modified diffusion bridge. Extensions of this construct to the case 
of partial observation with additive Gaussian noise can be 
found in \cite{Golightly_Wilkinson_2008}. Whilst this construct 
can, in principle, be applied to arbitrary nonlinear multivariate 
diffusion processes, the effect of the Gaussian approximation is 
to guide the bridge towards the observation in a linear way, unless there is large 
uncertainty in the observation process. This effect is exacerbated in 
the case of no measurement error, in which case the resulting construct 
is independent of the drift of the target process. Consequently, 
use of the modified diffusion bridge as a proposal mechanism (in a 
Metropolis-Hastings independence sampler) is likely to result in 
low acceptance rates, unless the drift is of little importance 
in dictating the dynamics of the target process between observation 
times. Several attempts to overcome this issue have been 
proposed in the recent literature. 

A time-dependent combination of
the Pedersen and modified diffusion bridge approaches was proposed by
\cite{Lindstrom_2012}.  However, the resulting construct requires a 
model specific tuning parameter governing the relative weight of 
each contribution (either Pedersen or modified diffusion bridge). Moreover, the 
optimal value (in terms of maximising acceptance rate) may vary between 
observation intervals. \cite{Beskos_2013} use Hybrid Monte Carlo (HMC) 
on pathspace to generate SDE sample paths under various observation 
regimes. For the applications considered, the authors found reasonable 
gains in overall efficiency (as measured by minimum effective sample 
size per CPU time) over an independence sampler with a Brownian bridge 
proposal. However, we note that HMC also requires careful choice of tuning parameters 
(namely the number of steps (and their size) in the leapfrog integrator) 
to maximise efficiency. \cite{Schauer_2014} (see also \cite{Papaspiliopoulos_2012}) 
combine the ideas of \cite{Delyon_2006} and \cite{Clark_1990} 
to obtain a bridge based on the addition of a guiding term 
to the drift of the target SDE. The guiding term requires a tractable 
approximation of the unavailable transition densities governing the target process over
the length of the inter-observation interval. \cite{Schauer_2014} 
suggest using the transition densities associated with a class of linear processes, 
although we note that finding an approximation that is both accurate and computationally 
efficient may be difficult in practice. Moreover, such an approximation can suffer 
from computational efficiency due to the fact that it must be obtained 
at each intermediate time point.  

In the next section we describe a novel bridge construct
that requires no tuning parameters, is simple to implement (even when
only a subset of components are observed with Gaussian noise), computationally 
efficient and explicitly allows for the effect of the drift governing the target 
SDE.

\subsection{An improved bridge construct}\label{dgode}
Consider a typical interval $[t_j,t_{j+1}]$, partitioned into $m$
sub-intervals as in (\ref{eqn:part}), over which we wish to generate a
realisation of $\{X_t,t\in [t_{j},t_{j+1}]\}$ conditional on
$x_{t_{j}}\equiv x_{\tau_{j,0}}$ and the noisy measurement
$y_{t_{j+1}}\equiv y_{\tau_{j,m}}$. Our approach builds on the 
modified diffusion bridge of \cite{Durham_2001}, which we briefly review before 
describing our extension. 

\subsubsection{Modified diffusion bridge}

Key to constructing the modified diffusion bridge is an 
approximation of the joint distribution of $X_{\tau_{j,k}}$ and 
$Y_{t_{j+1}}$ (conditional on $x_{\tau_{j,k-1}}$). Under
the Euler-Maruyama approximation $X_{\tau_{j,k}}| X_{\tau_{j,k-1}}$
and $Y_{t_{j+1}}| X_{\tau_{j,k}}$ are Gaussian,
with the expressions for the mean and variance of the latter evaluated
at $X_{\tau_{j,k-1}}$ to give a linear Gaussian structure. This leads to the 
approximation
\begin{align}\label{eqn:mdbJ}
\begin{pmatrix}
X_{\tau_{j,k}} \\
Y_{t_{j+1}} \end{pmatrix}\bigg| x_{\tau_{j,k-1}}&\sim N\left\{\begin{pmatrix}
x_{\tau_{j,k-1}}+\alpha_{j,k-1}\Delta\tau  \\[0.2em]
F^T [x_{\tau_{j,k-1}}+\alpha_{j,k-1}\Delta^-] \end{pmatrix},\right. \nonumber\\
& \quad \qquad \qquad \qquad \qquad \left. \begin{pmatrix}
\beta_{j,k-1} \Delta\tau & \beta_{j,k-1} F\Delta\tau \\
F^T\beta_{j,k-1}\Delta\tau & F^T\beta_{j,k-1} F\Delta^- + \Sigma \end{pmatrix}\right\}
\end{align}
where $\Delta^- =t_{j+1}-\tau_{j,k-1}$ and we have used the shorthand
notation $\alpha(x_{\tau_{j,k-1}})=\alpha_{j,k-1}$ and
$\beta(x_{\tau_{j,k-1}})=\beta_{j,k-1}$. Conditioning further on $y_{t_{j+1}}$ gives 
a Gaussian approximation of \linebreak $\pi(x_{\tau_{j,k}}| x_{\tau_{j,k-1}},y_{t_{j+1}})$, 
denoted $\widehat{\pi}(x_{\tau_{j,k}}| x_{\tau_{j,k-1}},y_{t_{j+1}})$, 
which can be sampled recursively to give a bridge 
$x_{\tau_{j,0}},\ldots,x_{\tau_{j,m}}$. In the case of no measurement error and observation of all components 
(so that $y_{t_{j+1}}=x_{t_{j+1}}$ and $F=I_d$, the $d\times d$ identity 
matrix), we obtain
\[
\widehat{\pi}(x_{\tau_{j,k}}| x_{\tau_{j,k-1}},x_{t_{j+1}})=
N\left(x_{\tau_{j,k}}\,;\, x_{\tau_{j,k-1}}+\frac{x_{t_{j+1}}-x_{\tau_{j,k-1}}}{t_{j+1}-\tau_{j,k-1}}\Delta\tau
\,,\,\frac{t_{j+1}-\tau_{j,k}}{t_{j+1}-\tau_{j,k-1}}\beta_{j,k-1}\Delta\tau\right)
\]
which is the form of the modified diffusion bridge first described by
\cite{Durham_2001}. In this case, $\widehat{\pi}(x_{\tau_{j,k}}| x_{\tau_{j,k-1}},y_{t_{j+1}})$ 
can be seen as a linear approximation of the Brownian bridge SDE
\begin{equation}\label{eqn:SDEbridge0}
dX_t=\frac{X_{t_{j+1}}-X_t}{t_{j+1}-t}\,dt+\sqrt{\beta(X_t)}\,dW_t.
\end{equation}
Use of (\ref{eqn:SDEbridge0}) has been justified by \cite{Delyon_2006}, who 
show that the distribution of the target process (conditional on $x_{t_{j+1}}$) 
is absolutely continuous with respect to the distribution of the 
solution to (\ref{eqn:SDEbridge0}). We may therefore expect that 
a Metropolis-Hastings scheme that uses a proposal based on a discretisation of 
(\ref{eqn:SDEbridge0}) will yield a non-zero acceptance rate 
as $\Delta\tau\to 0$ (for a rigorous treatment of the
limiting forms, we refer the reader to \cite{Delyon_2006},
\cite{Stramer_Yan_2007} and to \cite{Papaspiliopoulos_2012} for a
recent discussion). However, it should also be noted that the 
linear drift function governing (\ref{eqn:SDEbridge0}) 
is independent of the the drift function $\alpha(\cdot)$ 
governing the target process. Consequently, in situations where 
realisations of the target SDE (with the same initial condition) 
exhibit strong and similar nonlinearity over the inter-observation 
time, the modified diffusion bridge is likely to be unsatisfactory.   

\subsubsection{Residual bridge}

\begin{figure}[t!]
\centering
\includegraphics[scale=0.3]{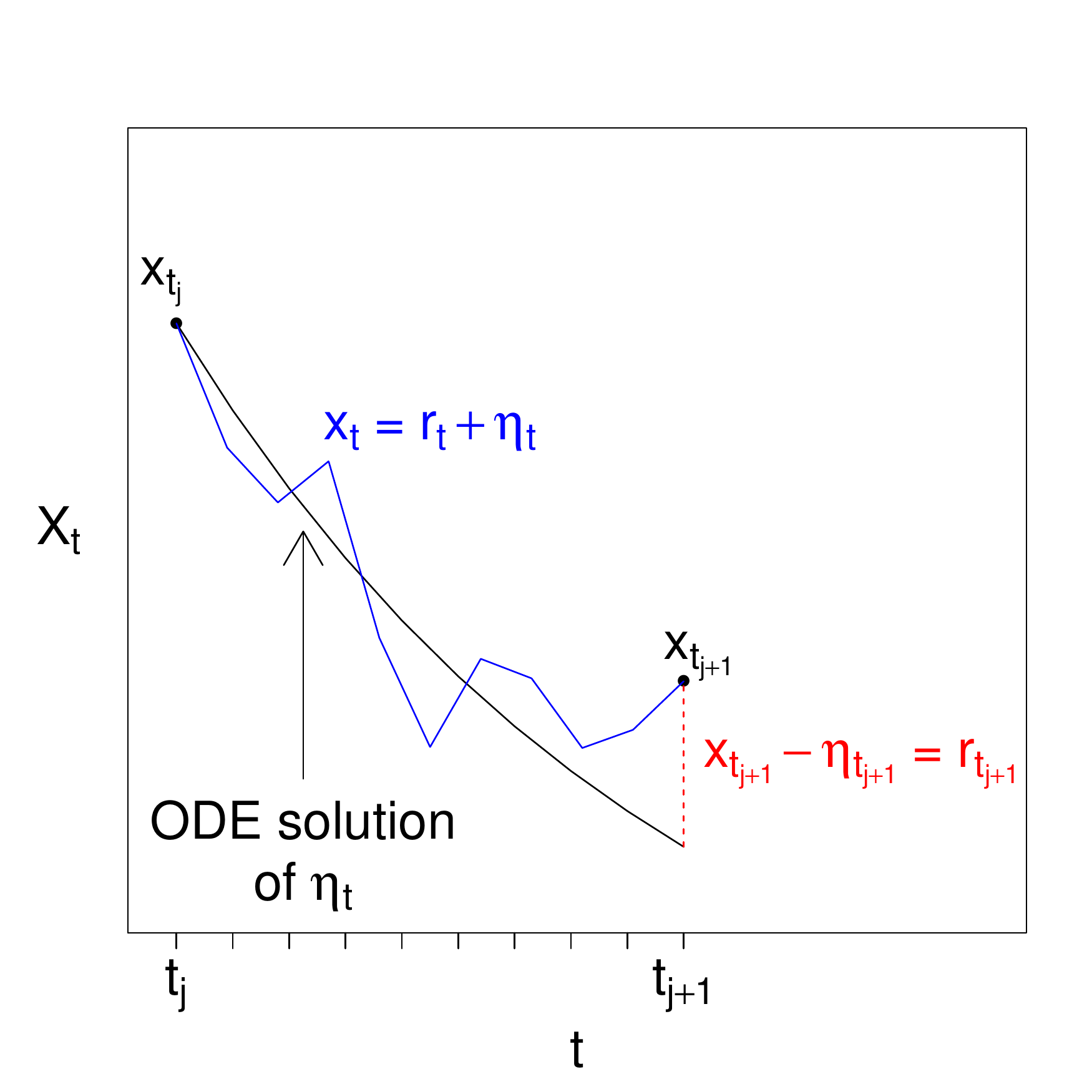}
\includegraphics[scale=0.3]{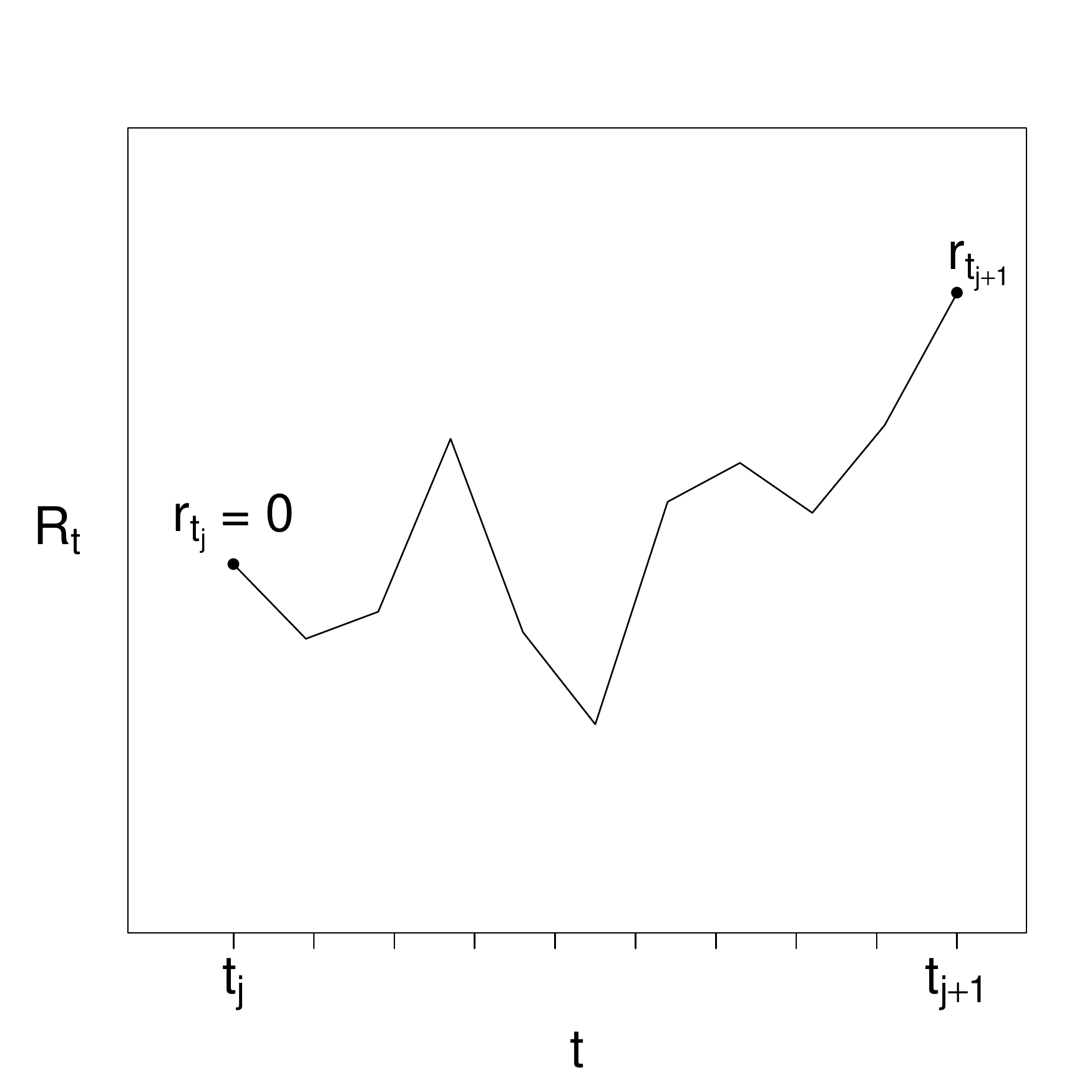}
	\caption{An illustration of the improved bridge construct. Left: The full bridge. Right: A sample path of $R_t$.}\label{fig:dgodeexplain}
\end{figure}

To allow explicitly 
for dynamics based on the drift, we partition 
$X_t$ into two parts, one that accounts for the drift in 
a deterministic way, and another as a residual stochastic process. 
The modified diffusion bridge is then applied to the \emph{residual 
stochastic process} rather than the target process itself. The 
partition we require is 
\begin{equation}\label{eqn:partx}
X_t=\eta_t+R_t,
\end{equation}
where $\{\eta_t,t\in [t_{j},t_{j+1}]\}$ is a deterministic 
process satisfying the ODE
\begin{equation}
\frac{d\eta_t}{dt} = \alpha(\eta_t),\quad \eta_{t_{j}}=x_{t_{j}}, \label{eqn:dgode_eta}
\end{equation}
and $\{R_t,t\in [t_{j},t_{j+1}]\}$ is a residual stochastic
process satisfying
\begin{align}
dR_t&\equiv dX_t-d\eta_t=\{\alpha(X_t)-\alpha(\eta_t)\}dt+\sqrt{\beta(X_t)}\,dW_t. \label{eqn:dgode_r}
\end{align}
We note that the partition in (\ref{eqn:partx}) is used by 
\cite{Fearnhead_2014} (see also Section~\ref{lna}) to derive 
a tractable approximation to the intractable transition densities governing 
$X_t$, whereas our primary motivation for (\ref{eqn:partx}) is the 
application of the modified diffusion bridge construct to the 
residual process, thus giving a proposal that is likely to perform 
well for arbitrarily fine discretisations and explicitly 
incorporates the drift of the target SDE. Therefore, we aim to 
derive an approximation $\widehat{\pi}(r_{\tau_{j,k}}| r_{\tau_{j,k-1}},y_{t_{j+1}})$, that
can be sampled recursively for $k=1,\ldots,m$ and combined with the
deterministic process (through numerical solution of
(\ref{eqn:dgode_eta})) via (\ref{eqn:partx}) to give a bridge
$x_{\tau_{j,0}},\ldots,x_{\tau_{j,m}}$. The scheme is illustrated in
Figure~\ref{fig:dgodeexplain}. 

The initial condition $\eta_{t_{j}}=x_{t_{j}}$ together with the Gaussian 
measurement error process imply that $r_{t_{j}}=0$ and 
\[
Y_{t_{j+1}}-F^T\eta_{t_{j+1}}=F^T R_{t_{j+1}} + \epsilon_{t_{j+1}},\qquad \epsilon_{t_{j+1}}|\Sigma\indep N(0,\Sigma).
\]
Hence, it should be clear that the joint distribution 
of $R_{\tau_{j,k}}$ and $Y_{t_{j+1}}-F^T\eta_{t_{j+1}}$
conditional on $R_{\tau_{j,k-1}}=r_{\tau_{j,k-1}}$ can be approximated as
\begin{align*}
\begin{pmatrix}
R_{\tau_{j,k}} \\
Y_{t_{j+1}}-F^T\eta_{t_{j+1}} \end{pmatrix}\bigg| r_{\tau_{j,k-1}}&\sim N\left\{\begin{pmatrix}
r_{\tau_{j,k-1}}+(\alpha_{j,k-1}-\alpha^{\eta}_{j,k-1})\Delta\tau  \\[0.2em]
F^T r_{\tau_{j,k-1}}+F^T (\alpha_{j,k-1}-\alpha^\eta_{j,k-1})\Delta^- \end{pmatrix},\right.\\
& \quad \qquad \qquad \qquad \left. \begin{pmatrix}
\beta_{j,k-1} \Delta\tau & \beta_{j,k-1} F\Delta\tau \\
F^T\beta_{j,k-1}\Delta\tau & F^T\beta_{j,k-1} F\Delta^- + \Sigma \end{pmatrix}\right\}
\end{align*}
where $\alpha(\eta_{\tau_{j,k-1}})=\alpha^{\eta}_{j,k-1}$. Conditioning further on
$y_{t_{j+1}}-F^T\eta_{t_{j+1}}$ gives
\begin{equation}
\widehat{\pi}(r_{\tau_{j,k}}| r_{\tau_{j,k-1}},y_{t_{j+1}}) = N(r_{\tau_{j,k}};\,\mu_{j,k}\,,\,\Psi_{j,k}), \label{eqn:dgodebridge1}
\end{equation}
where
\begin{align}
\begin{split}
\mu_{j,k}&=r_{\tau_{j,k-1}}+(\alpha^x_{j,k-1}-\alpha^\eta_{j,k-1})\Delta\tau+
\beta_{j,k-1} F\Delta\tau(F^T\beta_{j,k-1} F\Delta^- + \Sigma)^{-1} \\
&\qquad\qquad\qquad\times(y_{t_{j+1}}-F^T\eta_{t_{j+1}}-
\{F^T r_{\tau_{j,k-1}}+F^T (\alpha^x_{j,k-1}-\alpha^\eta_{j,k-1})\Delta^-\})
\end{split} 
\label{eqn:dgodebridge2}\\[0.5em]
\intertext{and}
\Psi_{j,k}&=\beta_{j,k-1}\Delta\tau-\beta_{j,k-1} F\Delta\tau(F^T\beta_{j,k-1} F\Delta^- + \Sigma)^{-1}F^T\beta_{j,k-1}\Delta\tau. 	\label{eqn:dgodebridge3}
\end{align}
Together \eqref{eqn:dgodebridge1}-\eqref{eqn:dgodebridge3} define our
bridge construct. These can be used to define the proposal mechanism
in \eqref{pathprop} for generating $\{X_t,t\in[t_j,t_{j+2}]\}$ by
taking
\begin{align*}
q_1(x_{(j,j+1]}|x_{t_j},y_{t_{j+1}})&=
\prod\limits_{k=1}^m\widehat{\pi}(x_{\tau_{j,k}}-\eta_{\tau_{j,k}}| r_{\tau_{j,k-1}},y_{t_{j+1}})
\intertext{and}
q_2(x_{(j+1,j+2)}| x_{t_{j+1}},x_{t_{j+2}})&=
\prod\limits_{k=1}^{m-1}\widehat{\pi}(x_{\tau_{j+1,k}}-\eta_{\tau_{j+1,k}}| r_{\tau_{j+1,k-1}},x_{t_{j+2}}), 
\end{align*}
where $\widehat{\pi}(x_{\tau_{j+1,k}}-\eta_{\tau_{j+1,k}}| r_{\tau_{j+1,k-1}},x_{t_{j+2}})$ can be sampled 
using (\ref{eqn:partx}) and (\ref{eqn:dgodebridge1})-(\ref{eqn:dgodebridge3}) with $j$ replaced by $j+1$, 
$\Sigma=0$ and $F=I_d$.  

In the special case of no measurement error and observation of all components 
we have that 
\[
\widehat{\pi}(r_{\tau_{j,k}}| r_{\tau_{j,k-1}},x_{t_{j+1}}) = N\left(r_{\tau_{j,k}}\,;\,r_{\tau_{j,k-1}}+\frac{r_{t_{j+1}}-r_{\tau_{j,k-1}}}{t_{j+1}-\tau_{j,k-1}}\Delta\tau\,,\,
\frac{t_{j+1}-\tau_{j,k}}{t_{j+1}-\tau_{j,k-1}}\beta_{j,k-1}\Delta\tau\right),
\]
which can be seen as a linear approximation of the
Brownian bridge SDE
\begin{equation}\label{SDEbridge}
dR_t=\frac{R_{t_{j+1}}-R_t}{t_{j+1}-t}+\sqrt{\beta(X_t)}\,dW_t.
\end{equation}
We also note that (\ref{SDEbridge}) has the same
diffusion coefficient as the target process and appeal again 
to \cite{Delyon_2006}, to deduce that the distribution 
of the residual process governed by (\ref{eqn:dgode_r}) (conditional on $r_{t_{j+1}}$) 
is absolutely continuous with respect to the distribution of the 
solution to (\ref{eqn:SDEbridge0}).

\subsection{Parameter updates}\label{param}

The full conditional densities of $\Sigma$ and $\psi$ are
\[
\pi(\Sigma| x,y)\propto \pi(\Sigma)\pi(y|\Sigma)
\qquad\text{and}\qquad
\pi(\psi| b)\propto \pi(\psi)\pi(b|\psi).
\]
Often, semi-conjugate priors can be specified for $\Sigma$ and $\psi$
negating the need for Metropolis-within-Gibbs steps. For the remaining
parameters $\theta$ and $b=(b^1,\ldots,b^N)^T$ we have
\[
\pi(\theta| b,x)\propto \pi(\theta)\pi(x|\theta,b)
\qquad\text{and}\qquad
\pi(b|\theta,\psi,x)\propto \pi(b|\psi)\pi(x|\theta,b)
=\prod\limits_{i=1}^{N}\pi(b^i|\psi)\pi(x^i|\theta,b^i)
\]
where the last expression suggests unit-specific updates of the 
components of $b$.

As discussed earlier, since $\theta$ and the components of $b$ enter into 
the diffusion coefficient of (\ref{eqn:sdmem}), sampling the full conditionals 
of $\theta| b,x$ and $b|\theta,\psi,x$ as part of a Gibbs sampler 
will result in a reducible Markov chain as $m\to\infty$ (or $\Delta\tau\to 0$). 
To overcome this problem we use a reparameterisation which is outlined in the next 
section.

\subsubsection{Modified Innovation scheme}

The innovation scheme was first outlined in \cite{Chib_et_al_2004} and
exploits the fact that, given $\theta$ and $b$, under the
Euler-Maruyama approximation there is a one-to-one relationship
between the increments of the process~($\Delta X_t$) and the
increments of the driving Brownian motion~($\Delta W_t$).  Moreover,
whilst the quadratic variation of $X$ determines $\theta$ and $b$ (as
$m\to\infty$), the quadratic variation of the Brownian process is
independent of $\theta$ and $b$ \emph{a priori}. Conditioning on the
Brownian increment innovations in a Gibbs update should therefore be
effective in overcoming the dependence problem. The resulting
algorithm is known as the innovation scheme. Unfortunately, combining
an updated parameter value with the Brownian increments will not
necessarily give an imputed path that is consistent with the
observations. Therefore,
\citet{Golightly_Wilkinson_2008,Golightly_Wilkinson_2010} suggest that
a diffusion bridge (such as the modified diffusion bridge of
\cite{Durham_2001}) be used to determine the innovation process,
leading to a modified innovation scheme.

\cite{Fuchs_2013} considers the modified innovation scheme in a
continuous-time framework. Adapting their innovation process to an
SDMEM, we have, for an interval $[t_j,t_{j+1}]$, an innovation process
$\{Z_t^i,t\in[t_j,t_{j+1}]\}$ satisfying
\begin{align}
dZ_t^i&=\beta(X_t^i,\theta,b^i)^{-\half}
\left(dX_t^i-\frac{x_{t_{j+1}}^i-X_t^i}{t_{j+1}-t}\,dt\right), \label{eqn:innovation} \\
&= \beta(X_t^i,\theta,b^i)^{-\half}
\left\{\alpha(X_t^i,\theta,b^i)-\frac{x_{t_{j+1}}^i-X_t^i}{t_{j+1}-t}\right\}dt+dW_t^i \nonumber
\end{align}
with $Z_{t_{j}}^i=0$. Clearly, each process $Z^i$ has unit diffusion coefficient and 
whilst not Brownian motion processes, the probability measures induced 
by each $Z^i$ are absolutely continuous with respect to Wiener 
measure. A proof of this result can be found in \cite{Fuchs_2013} 
as well as a justification for using this form of innovation 
process as the effective component in a Gibbs sampler.

The aim is to apply a discretisation of (\ref{eqn:innovation}) between
observation times. We therefore define
$x_o^i=(x_{t_{0}}^i,\ldots,x_{t_{n}}^i)^T$ to be the current values of
the (unit-specific) latent process at the observation times, and stack
all $x_o^i$ values into the matrix $x_o$. We have for $k=~1,\ldots,m$
\[
Z_{\tau_{j,k}}^i-Z_{\tau_{j,k-1}}^i=
\beta^*(X_{\tau_{j,k-1}}^i,\theta,b^i)^{-\half}
\left(X_{\tau_{j,k}}^i-X_{\tau_{j,k-1}}^i-\frac{x_{t_{j+1}}^i-X_{\tau_{j,k-1}}^i}{t_{j+1}-\tau_{j,k-1}}\Delta\tau\right), 
\]
where $Z_{\tau_{j,0}}=0$ and
\[
\beta^*(X_{\tau_{j,k-1}}^i,\theta,b^i)=
\frac{t_{j+1}-\tau_{j,k}}{t_{j+1}-\tau_{j,k-1}}\beta(X_{\tau_{j,k-1}}^i,\theta,b^i).
\]
Note that our discretisation of (\ref{eqn:innovation}) follows 
\cite{Golightly_Wilkinson_2008} by using the modified diffusion bridge to 
construct the innovation process. Now define a function $f$ so that \hbox{$X_{\tau_{j,k}}^i=f(Z_{\tau_{j,k}}^i,\theta,b^i)$} and 
$Z_{\tau_{j,k}}^i=f^{-1}(X_{\tau_{j,k}}^i,\theta,b^i)$. Let $z_{imp}^i$ denote the (unit-specific) 
innovation values over $[t_{0},t_{n}]$ and stack all $z_{imp}^i$ values into the matrix $z_{imp}$. 
Define $x_{imp}^i$ and $x_{imp}$ similarly. The modified innovation scheme samples 
$\theta| b,z_{imp},x_o$ and $b^i|\theta,\psi,z_{imp}^i,x_o^i$, $i=1,\ldots,N$. 
Note that for an updated value of $b^i$, say $b^{i*}$, a new $x_{imp}^{i*}$ is updated 
deterministically through $x_{imp}^{i*}=f(z_{imp}^{i*},\theta,b^{i*})$. Likewise, for a new 
$\theta^*$, a new $x_{imp}^*$ is updated deterministically through 
$x_{imp}^{i*}=f(z_{imp}^{i*},\theta^*,b^i)$, $i=1,\ldots,N$. The full conditional 
density of $\theta$ is
\begin{align}
& \pi(\theta| b,z_{imp},x_o) \propto \pi(\theta)\prod\limits_{i=1}^{N}\prod\limits_{j=1}^{n-1}
\left[\prod\limits_{k=1}^{m}\pi(x^i_{\tau_{j,k}}| x^i_{\tau_{j,k-1}},\theta,b^i)
\prod\limits_{k=1}^{m-1}J\{f(z_{\tau_{j,k}}^i,\theta,b^i)\} \right], \label{eqn:innovationFC1}
\end{align}
where
\[
J\{f(z_{\tau_{j,k}}^i,\theta,b^i)\} =
\left|\beta^*(x^i_{\tau_{j,k-1}},\theta,b^i)\right|^{-\half}
\]
is the Jacobian determinant of $f$. Similarly, the full conditional 
density of $b^i$, $i=1,\ldots,N$ is
\begin{align}
&\pi(b^i|\theta,\psi,z_{imp}^i,x_o^i)\propto \pi(b^i|\psi)\prod\limits_{j=1}^{n-1}\left[\prod\limits_{k=1}^{m}
\pi(x^i_{\tau_{j,k}}|x^i_{\tau_{j,k-1}},\theta,b^i)
\prod\limits_{k=1}^{m-1}J\{f(z_{\tau_{j,k}}^i,\theta,b^i)\} \right]. \label{eqn:innovationFC2}
\end{align}
Naturally, the full conditionals in (\ref{eqn:innovationFC1}) and (\ref{eqn:innovationFC2}) will 
typically be intractable, requiring the use of Metropolis-within-Gibbs updates. 

\section{Linear noise approximation}\label{lna}

In this section we outline a competing solution which uses an
inference scheme based on a linear noise approximation (LNA) to the
SDMEM. The LNA typically refers to an approximation to the solution of
the forward Kolmogorov equation governing the transition probability
of a Markov jump process
\citep{Kurtz_1970,Ferm_et_al_2008,Komorowski_et_al_2009,Finkenstadt_2013}. Specifically,
the forward Kolmogorov equation is approximated by a Fokker-Planck
equation with linear coefficients. Equivalently, a general Fokker-Planck 
equation can be deduced and then linearised. In this context, therefore, 
the LNA aims to replace intractable transition densities 
with Gaussian approximations. In what follows, we give a brief 
informal derivation of the LNA and refer the reader to 
\cite{Fearnhead_2014} and the references therein for further details.

\subsection{Setup}\label{setup}

For notational simplicity and clarity of exposition, 
we suppress parameter dependence and the unit-specific 
$i$ for the remainder of this sub-section.

Without loss of generality, consider a time $t\in[t_{j},t_{j+1}]$ 
at which we wish to approximate the intractable transition density 
associated with $X_{t}|X_{t_{j}}=x_{t_{j}}$. The LNA uses the same 
partition of $X_t$ given in (\ref{eqn:partx}), that is $X_t=\eta_t+R_t$ 
where the deterministic process $\eta_t$ satisfies (\ref{eqn:dgode_eta}) 
and the residual stochastic process satisfies (\ref{eqn:dgode_r}). The 
key assumption underpinning the LNA is that the residual stochastic 
perturbation is ``small'' relative to the deterministic process, 
allowing suitable truncation of a Taylor series expansion of 
$\alpha(X_t)$ and $\beta(X_t)$ about $\eta_t$. Taking the first 
two terms in the expansion of $\alpha(X_t)$, 
and the first term in the expansion of $\beta(X_t)$ gives an 
SDE satisfied by an approximate residual process 
$\{\tilde{R}_t,t\in[t_{j},t_{j+1}]\}$ of the form
\begin{equation}\label{eqn:tilder}
d\tilde{R}_t = H_t\tilde{R}_t\,dt + \sqrt{\beta(\eta_t)}\,dW_t,
\end{equation}
where $H_t$ is the Jacobian matrix with $(i,j)$th element 
$(H_t)_{i,j}=\partial\alpha_i(\eta_t)/\partial\eta_{j,t}$.

Assuming fixed or Gaussian initial conditions $\tilde{R}_{t_j}\sim N(m_{t_j},V_{t_j})$ 
gives $\tilde{R}_t\sim N(m_t,V_t)$, where $m_t$ and $V_t$ satisfy the ODE system
\begin{align}
\dfrac{dm_t}{dt} &= H_tm_t, \label{eqn:lna_m} \\
\dfrac{dV_t}{dt} &=  H_tV_t +  \beta(\eta_t,\theta,b)  +  V_tH_t^T. \label{eqn:lna_v}
\end{align}
In the absence of an analytic solution, the system of coupled ODEs
\eqref{eqn:dgode_eta} and \hbox{\eqref{eqn:lna_m}--\eqref{eqn:lna_v}} which characterise the LNA,
must be solved numerically. For initial conditions
$\eta_{t_j}=~x_{t_j}$, we have $m_{t_j}=0$ and $V_{t_j}=0$ so that
\eqref{eqn:lna_m} does not need to be solved, and the approximating
transition distribution is $X_{t}|X_{t_j}=x_{t_{j}}\sim N(\eta_t,V_t)$.

It is worth noting here that the linear form of the SDE (\ref{eqn:tilder}) 
satisfied by the approximate residual process coupled with the additive 
Gaussian observation regime admits a closed form expression for densities 
of the form $\widehat{\pi}(\tilde{r}_{\tau_{j,k}}| \tilde{r}_{\tau_{j,k-1}},y_{t_{j+1}})$, 
suggesting use of the LNA as a proposal mechanism inside the Bayesian imputation approach 
of Section~\ref{bayes}. Whilst the LNA could in principle be used to directly approximate the conditioned 
residual process governed by the SDE in (\ref{eqn:dgode_r}) we note that the SDEs 
in (\ref{eqn:dgode_r}) and (\ref{eqn:tilder}) have different diffusion coefficients. 
Consequently, the probability law governing $\tilde{R}_{t}$ is not absolutely continuous 
with respect to the law of $R_t$. We therefore do not advocate use of the LNA in this way.

In the next section we outline an inference scheme for SDMEMs of the
form \eqref{eqn:sdmem} based on the LNA. It exploits the computational
efficiency of a filtering algorithm proposed by \cite{Fearnhead_2014}
that allows closed-form calculation of the marginal likelihood
$\pi(y|\theta,b,\Sigma)$ under our Gaussian observation
regime~\eqref{eqn:obs}; see the supplementary material for further
details.

\subsection{Application to SDMEMs}\label{lna-app}

Under the linear noise approximation of \eqref{eqn:sdmem} the marginal
posterior for all parameters is given by 
\begin{align*}
\pi(\theta,\psi,\Sigma,b| y)&\propto\pi(\theta)\pi(\psi)\pi(\Sigma)\pi(b|\psi)\pi(y|\theta,\Sigma,b)\\
&\propto  \pi(\theta)\pi(\psi)\pi(\Sigma)\prod\limits_{i=1}^N\pi(b^i|\psi)\pi(y^i|\theta,\Sigma,b^i). 
\end{align*}
This factorisation suggests a Gibbs sampler with blocking that
sequentially takes draws from the full conditionals 
$\pi(\Sigma|\theta,\psi,b,y)=\pi(\Sigma|y)$,
$\pi(\theta|\psi,\Sigma,b,y)=\pi(\theta| b,y)$,
$\pi(b|\theta,\psi,\Sigma,y)=\pi(b|\theta,\psi,y)$ and 
$\pi(\psi|\theta,\Sigma,b,y)=\pi(\psi|b)$.
A Metropolis-Hastings step can be used when a full conditional density is
intractable. An algorithm for computing the marginal likelihood
$\pi(y^i|\theta,\Sigma,b^i)$ for each experimental unit is given in
the supplementary material. Interest may also lie in the joint
posterior $\pi(\theta,\psi,\Sigma,b,x| y)$ where, since no imputation
is required for the LNA, $x^i=(x_{t_0},\ldots,x_{t_n})^T$ and
$x=(x^1,\ldots,x^N)$. Realisations from this posterior can be obtained
using the above Gibbs sampler with an extra step that draws from
$\pi(x^i|\theta,\psi,\Sigma,b^i,y^i)=\pi(x^i|\theta,\Sigma,b^i,y^i)$
for $i=1,\ldots,N$. An efficient mechanism for making such draws can also
be found in the supplementary material. The method uses a forward
filter, backward sampling (FFBS) algorithm. 

\section{Applications}\label{apps}
We now compare the accuracy and efficiency of our Bayesian imputation
approach (coupled with the modified innovation scheme) with an
LNA-based solution. We consider two scenarios: one in which the ODEs
governing the LNA are tractable and one in which numerical solvers are
required. In the first we use synthetic data generated from a simple
univariate SDE description of orange tree growth. The second example
uses real data taken from \cite{Matis_2008} to fit an SDMEM driven by
the bivariate diffusion approximation of a stochastic kinetic model of
aphid dynamics. The resulting SDMEM is particularly challenging to fit
as both the drift and diffusion functions are nonlinear and also only
one component of the model is observed (with error). We also include
(in the supplementary material) a simulation study based on synthetic
data generated from the model of aphid dynamics, to explore further
any differences between the Bayesian imputation and LNA-based
approaches.

\subsection{Orange tree growth}\label{orange}

The SDMEM developed by \cite{Picchini_2010} and \cite{Picchini_2011}
to model orange tree growth describes the dynamics of the
circumference ($X_t^i$) of individual trees (mm) by
\[
dX_t^i = \dfrac{1}{\phi_1^i\phi_2^i}X_t^i(\phi_1^i-X_t^i)\,dt 
+ \sigma\sqrt{X_t^i}\,dW_t^i,\quad X_0^i=x_0^i, \quad i=1,\ldots,N
\]
with $\phi_1^i\sim N(\phi_1,\sigma_{\phi_1}^2)$ and $\phi_2^i\sim
N(\phi_2,\sigma_{\phi_2}^2)$ independently. Here $\theta=\sigma$ is
common to all trees, the random effects are
$b^i=(\phi_1^i,\phi_2^i)^T$, $i=1,\ldots,N$ and the parameter vector
governing the random effects distributions is
$\psi=(\phi_1,\phi_2,\sigma_{\phi_1},\sigma_{\phi_2})^T$. Note that
the $\phi_1^i$ can be interpreted as asymptotic circumferences and the
$\phi_2^i$ as the time-distance between the inflection point of the
model obtained by ignoring stochasticity and the point where
$X_t^i=\phi_1^i / (1+e^{-1})$.

To allow identifiability of all model parameters we generated 16
observations for the circumference of $N=100$ trees at intervals of
100 days. Following \cite{Picchini_2011} we gave each tree the same
initial condition ($x_0^i=30$) and took
$(\phi_1,\phi_2,\sigma_{\phi_1},\sigma_{\phi_2},\sigma)=(195,350,25,52.5,0.08)$,
which gives random effects distributions $\phi_1^i\sim N(195,25^2)$
and $\phi_2^i\sim N(350,52.5^2)$. For our analysis of these data we
assumed the parameters to be independent \emph{a priori} with $\phi_1$
and $\phi_2$ having weak $N(0,100^2)$ priors and
$1/\sigma_{\phi_1}^2$, $1/\sigma_{\phi_2}^2$ and $1/\sigma^2$ having
weak gamma $Ga(1,0.01)$ priors. In this example we assume there is no
measurement error and therefore the target posterior is given by
\[
\pi(\theta,\psi,b|x)\propto\pi(\theta)\pi(\psi)\pi(b|\psi)\pi(x|\theta,b).
\]
In the Bayesian imputation approach, $\pi(x|\theta,b)$ is as
in~\eqref{eqn:xdens} whereas for the LNA--based solution
\[
\pi(x|\theta,b)=
\prod\limits_{i=1}^{N}\prod\limits_{j=0}^{n-1}N(x^{i}_{t_{j+1}}\,;\, \eta^{i}_{t_{j+1}},V^{i}_{t_{j+1}}),
\] 
where, for each interval $[t_j,t_{j+1}]$ and each tree~$i$, the $\eta_t^i$
and $V_t^i$ satisfy the ODE system
\begin{align*}
\dfrac{d\eta_t^i}{dt} &= \dfrac{1}{\phi_1^i\phi_2^i}\eta_t^i(\phi_1^i-\eta_t^i),\qquad \eta_{t_j}^i=x_{t_j}^i, \\
\dfrac{dV_t^i}{dt} &= \dfrac{2}{\phi_1^i\phi_2^i}(\phi_1^i-2\eta_t^i)V_t^i + \sigma^2\eta_t^i,\qquad V_{t_j}^i=0.
\end{align*}
Fortunately this ODE system can be solved analytically giving
$\eta^i_t=A\phi_1^ie^{t/\phi_2^i}/(1+ Ae^{t/\phi_2^i})$ and
\[
V_t^i=B\left(\frac{1}{2}A^3\phi_2^i e^{2t/\phi_2^i} + 3A^2\phi_2^ie^{t/\phi_2^i}-\phi_2^ie^{-t/\phi_2^i}
+3At - \dfrac{1}{2}A^3\phi_2^i - 3A^2\phi_2^i +\phi_2^i \right)
\]
where $A=x^i_{t_0}/(\phi_1^i-x^i_{t_0})$ and
$B=\sigma^2A\phi_1^ie^{2t/\phi_2^i}/(1+Ae^{t/\phi_2^i})^4$. 

The MCMC scheme can make use of simple semi-conjugate updates for
$\phi_1$, $\phi_2$, $\sigma_{\phi_1}$ and $\sigma_{\phi_2}$. However
the remaining parameters ($\sigma$ and the $b^i$) require 
Metropolis-within-Gibbs updates and we have found that componentwise normal random
walk updates (so-called random walk Metropolis) on the log scale work
particularly well. Also, for the modified innovation scheme, the
dynamics of the SDMEM permit the use of the modified diffusion bridge
construct to update the latent trajectories between observation times:
the improved bridge construct of Section~\ref{dgode} is not needed.

\begin{figure}[t]
\begin{center}
\begin{minipage}[b]{0.28\linewidth}
        \centering
        \includegraphics[scale=0.22]{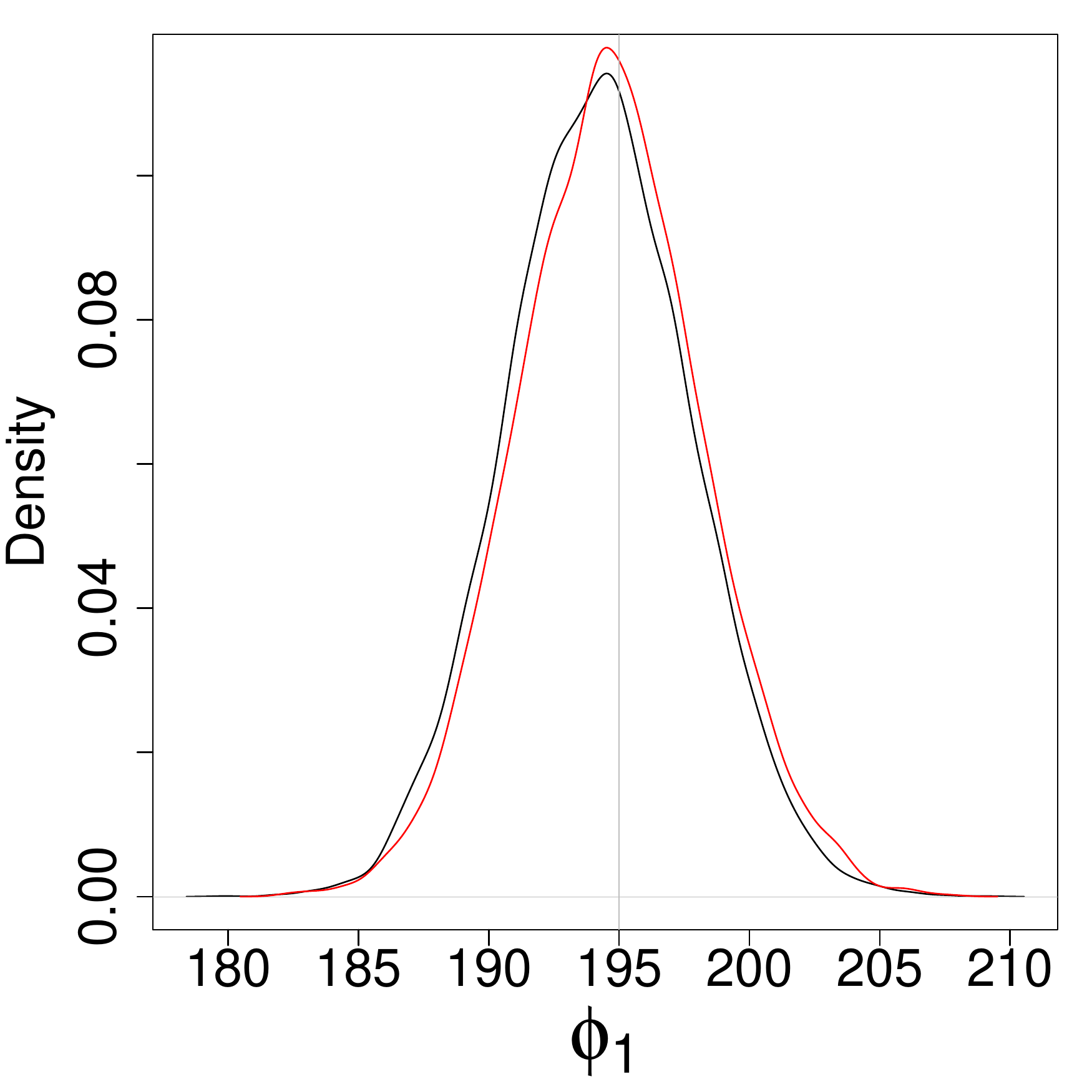}
\end{minipage} 
\hspace{-0.25cm}
\begin{minipage}[b]{0.28\linewidth}
        \centering
        \includegraphics[scale=0.22]{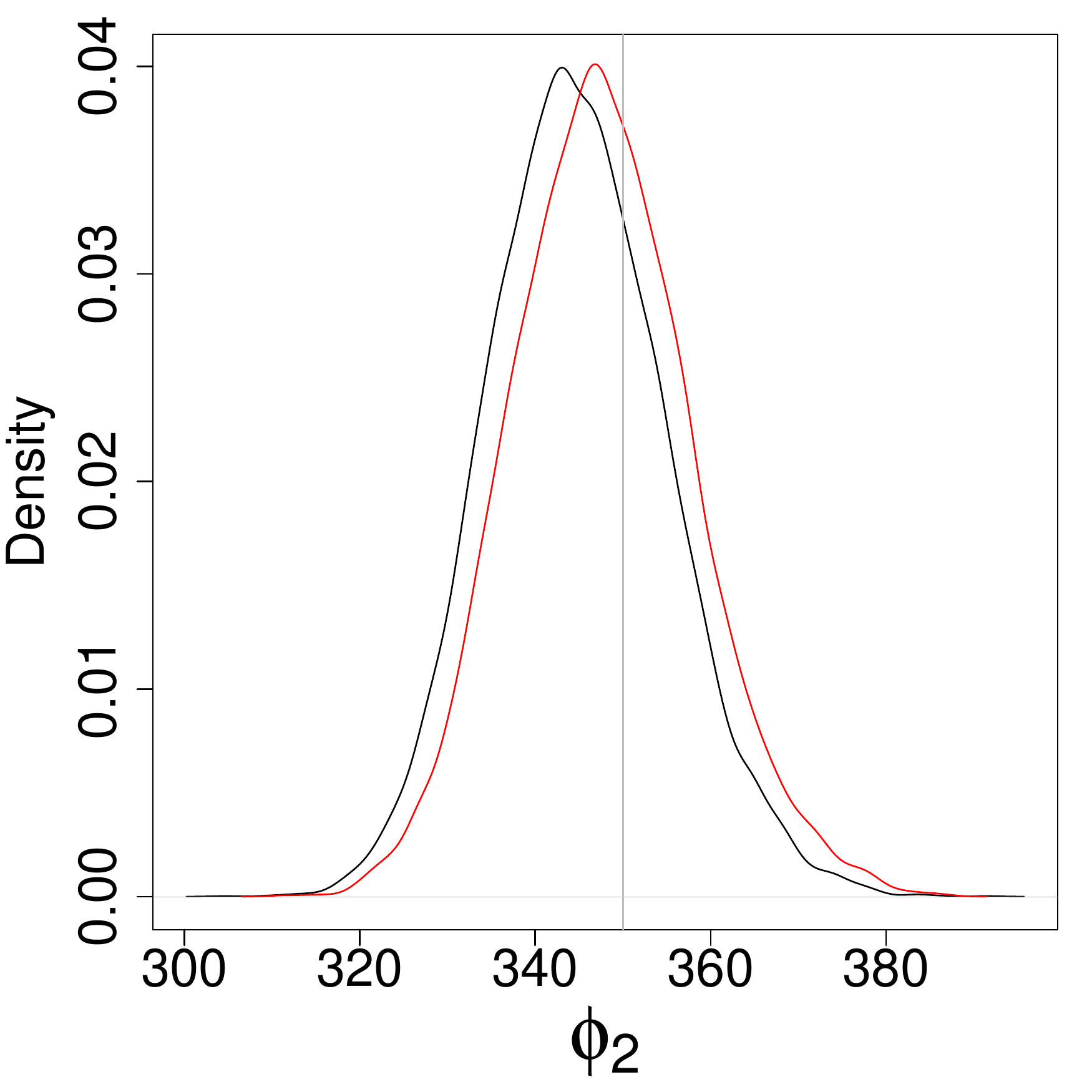}
\end{minipage} 
\hspace{-0.25cm}
\begin{minipage}[b]{0.28\linewidth}
	\centering
        \includegraphics[scale=0.22]{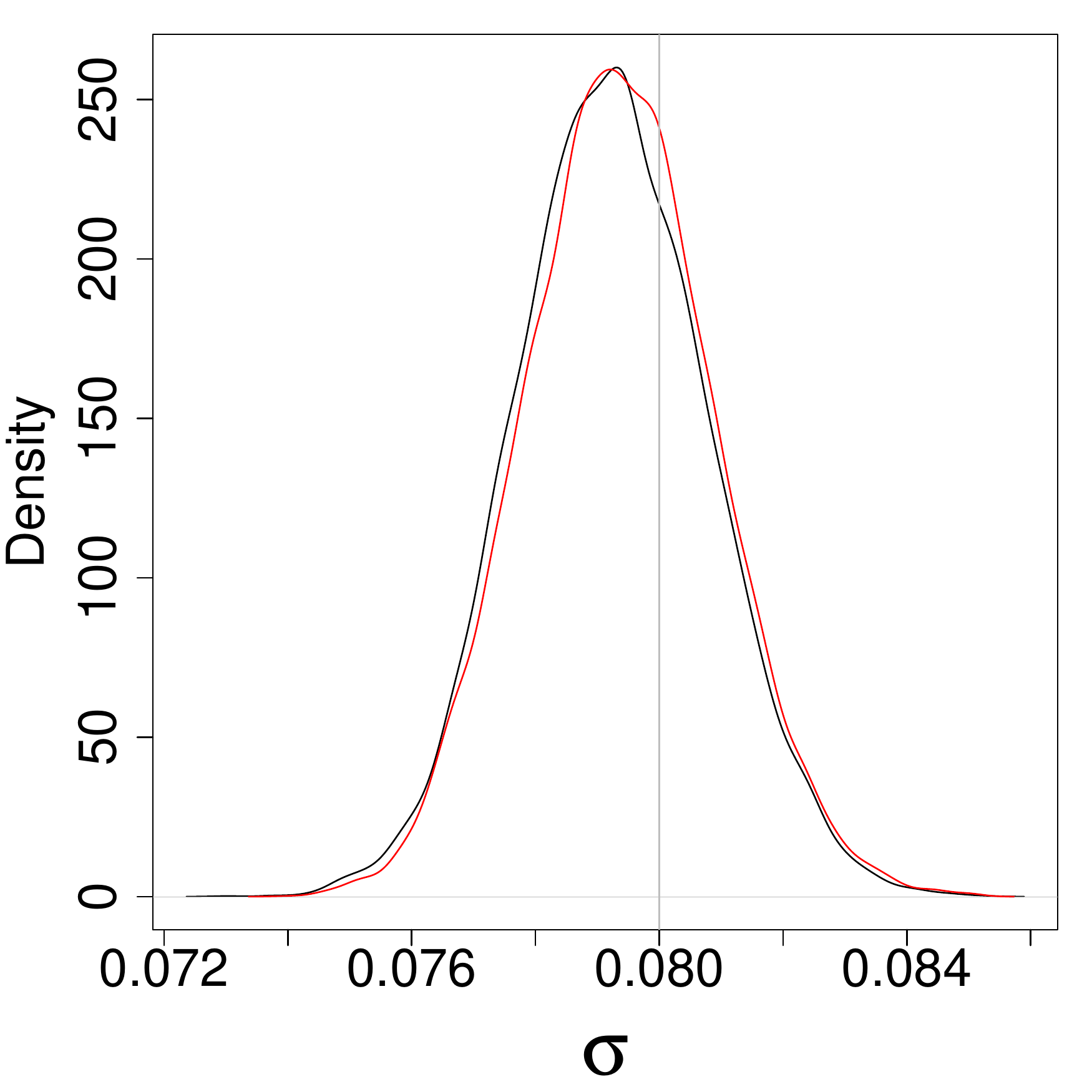}
\end{minipage}\\
\vspace{0.2cm}
\begin{minipage}[b]{0.28\linewidth}
	\centering
        \includegraphics[scale=0.22]{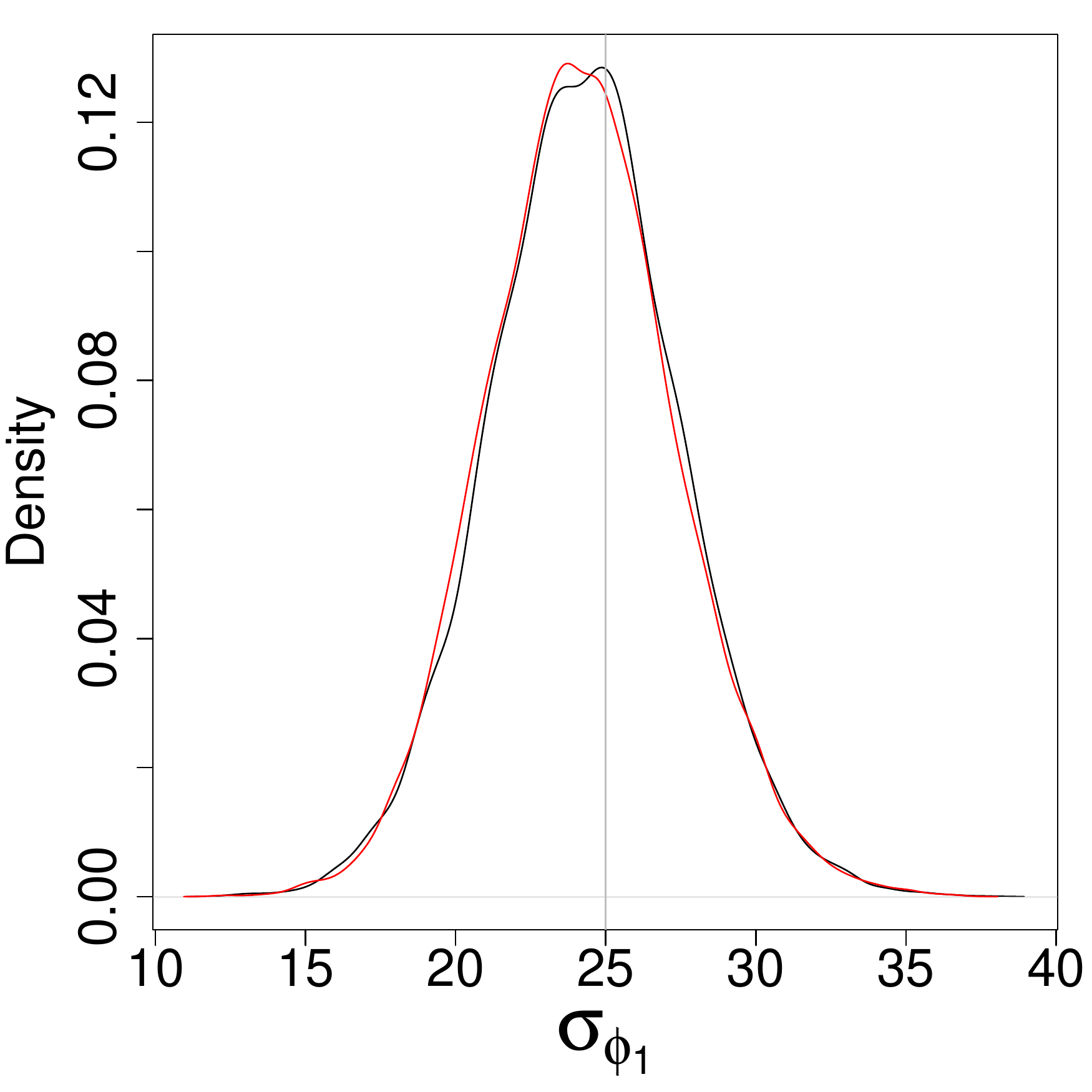}
\end{minipage}
\hspace{-0.25cm}
\begin{minipage}[b]{0.28\linewidth}
        \centering
        \includegraphics[scale=0.22]{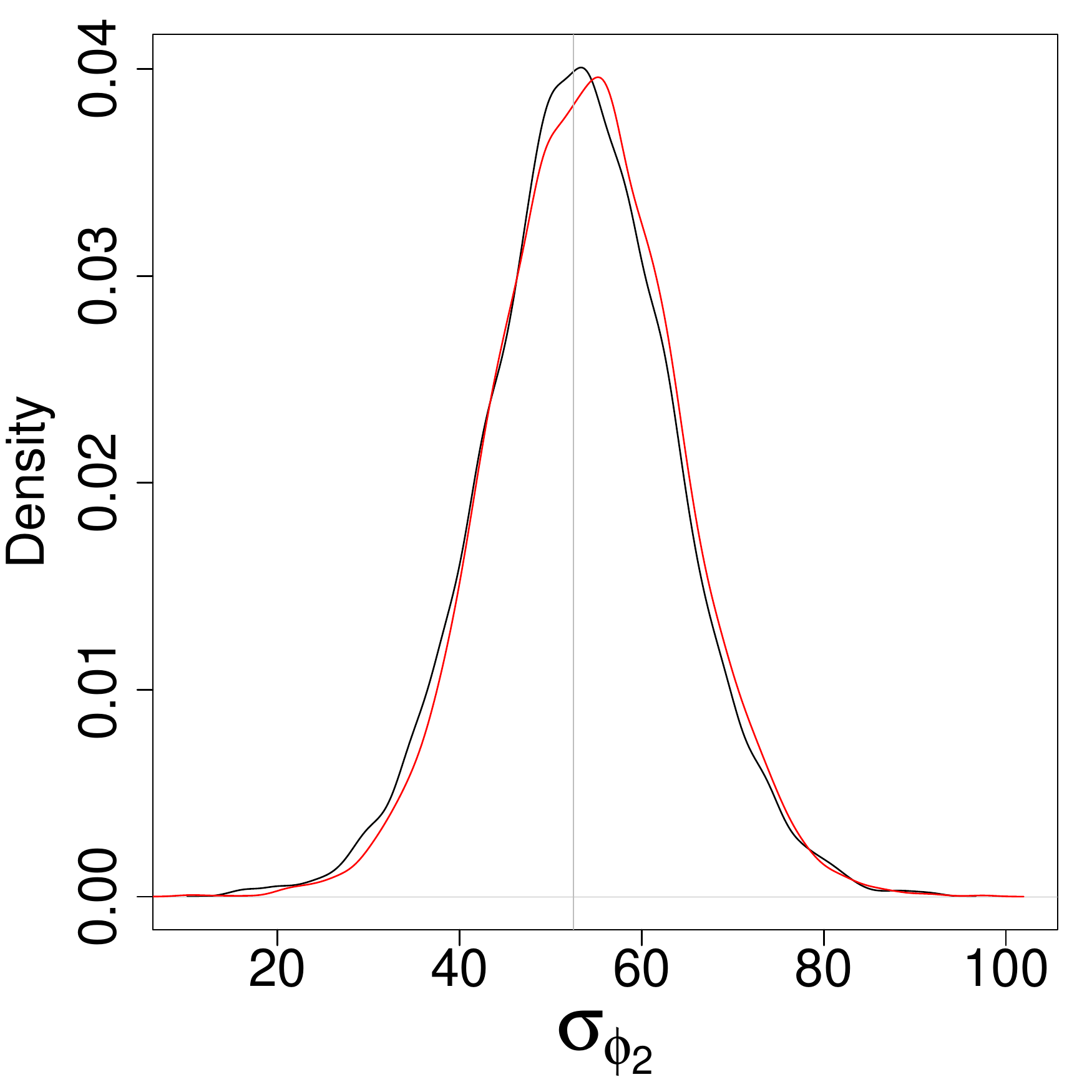}
\end{minipage}
\hspace{-0.25cm}
\begin{minipage}[b]{0.28\linewidth}
        \centering
        \includegraphics[scale=0.22]{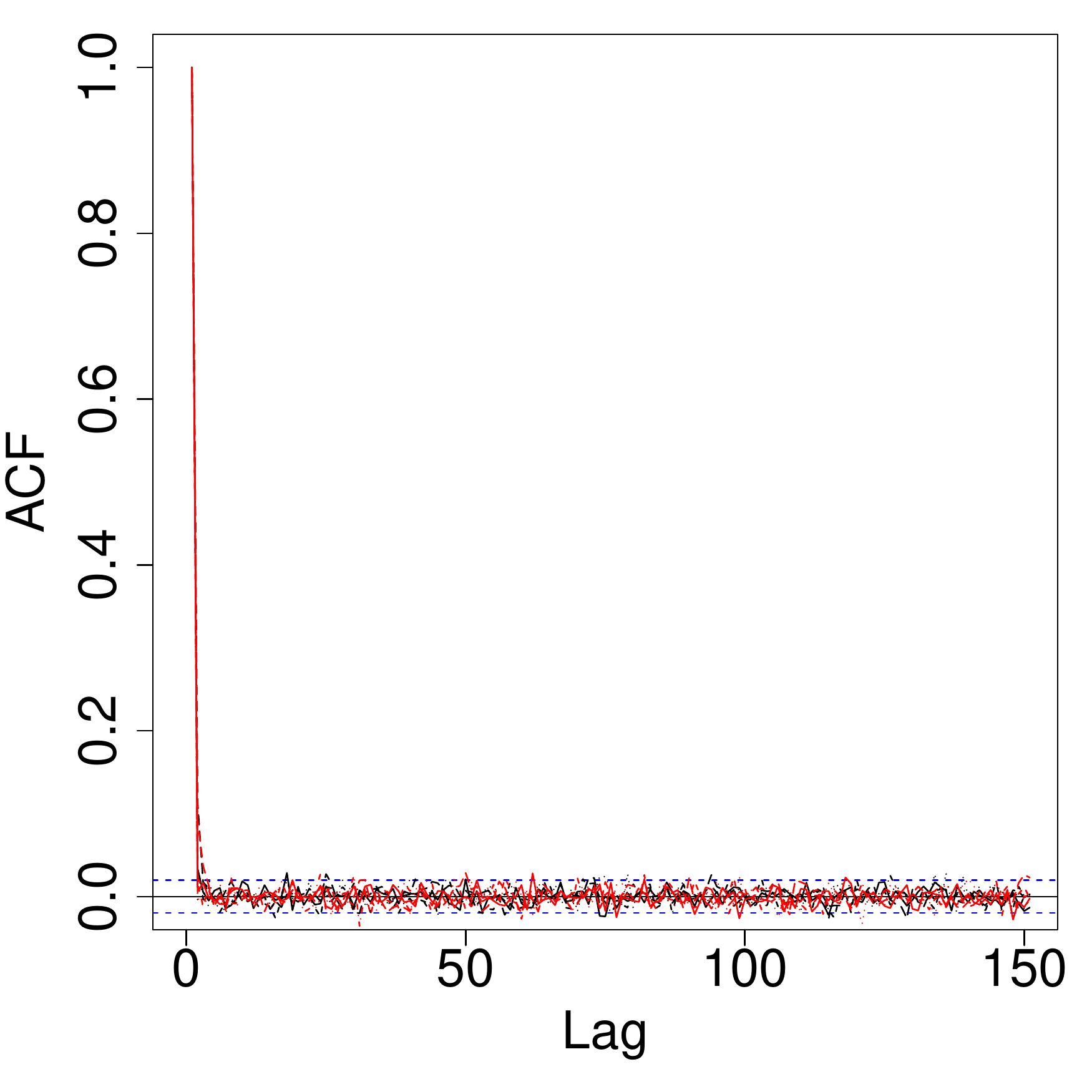}
\end{minipage}
\caption{Marginal posterior densities for the random effects
  hyper-parameters ($\phi_1$, $\phi_2$, $\sigma_{\phi_1}$,
  $\sigma_{\phi_2}$) and common parameter $\sigma$ in the orange tree
  growth SDMEM, together with their (overlayed) autocorrelation
  functions. Black: Bayesian imputation. Red:~LNA. The vertical grey lines
  indicate the ground truth.} \label{orange fig_params}
\end{center}
\end{figure}

The modified innovation scheme requires specification of the level
of discretisation~$m$. We performed several short pilot runs of
the scheme with $m\in\{5,10,20,40\}$ and found no discernible
difference in posterior output for $m\geq 10$. We therefore took
$m=10$. The sample output was also used to estimate the marginal
posterior variances of $\sigma$ and the $b^i$, to provide sensible
innovation variances in the random walk Metropolis updates. Both the
modified innovation scheme and the LNA--based scheme required a burn
in of 500 iterations, a thin of 100 iterates and were run long enough
to yield a sample of approximately 10K independent posterior draws.
Figure~\ref{orange fig_params} shows the marginal posterior densities
and autocorrelations for the common parameter $\sigma$ and the
parameters governing the random effects distributions. 
The marginal posterior means and standard deviations of
$(\phi_1,\phi_2,\sigma_{\phi_1},\sigma_{\phi_2},\sigma)$ are given in
Table~\ref{tab orange_param}. The figures and table show that for these parameters 
both the imputation approach and LNA--based approach generally give similar
output and are consistent with the true values from which the data
were simulated. Similar results are obtained for the random effects 
parameters (see the supplementary material). 

Both schemes were coded in C and run on an Intel Xeon 3.0GHz
processor; the modified innovation scheme took 43504 seconds to run
whilst the LNA inference scheme took 2483 seconds. We use the
minimum (over each parameter chain) effective sample size (minESS) to
measure the statistical efficiency of each scheme. The modified
innovation scheme produced a minESS of 7949 and the LNA--based
approach gave 7821. Therefore, in terms of minESS/sec, using the
LNA outperforms the imputation approach in this example by a factor of
approximately 17. It should be noted, however, that for most nonlinear
SDMEMs the ODEs governing the LNA solution will rarely be tractable
and the consequent use of numerical schemes will degrade its
performance. 

In the next section we consider an example in which the LNA ODEs are
intractable.

\begin{table}[h]
\begin{center}
\begin{tabular}{c|ccccc}
	\hline
  & $\phi_1$ & $\phi_2$ & $\sigma_{\phi_1}$ & $\sigma_{\phi_2}$ & $\sigma$ \\
  \hline
  \multirow{2}{*}{Imputation} & 194.229 & 344.799 & 24.316 & 53.219 & 0.079 \\
                    & (3.509) & (10.098) & (3.149) & (10.410) & (0.002) \\ 
  \rule{0pt}{3ex} \multirow{2}{*}{LNA} 							& 194.634 & 347.631 & 24.207 & 53.960 & 0.079 \\
                    & (4.025) & (10.844) & (3.154) & (10.193) & (0.002) \\
	\hline
\end{tabular}
\caption{Marginal posterior means (standard deviations) of the random effects hyper-parameters 
($\phi_1$, $\phi_2$, $\sigma_{\phi_1}$, $\sigma_{\phi_2}$) and common parameter $\sigma$ in the 
orange tree growth SDMEM. The synthetic data used $\phi_1=195$, $\phi_2=350$, $\sigma_{\phi_1}=25$, 
$\sigma_{\phi_2}=52.5$ and $\sigma=0.08$.} \label{tab orange_param}
\end{center} 
\end{table}

\subsection{Cotton aphid dynamics}\label{aphid}

\subsubsection{Model and data}

Aphids (also known as plant lice or greenfly) are small sap sucking
insects which live on the leaves of plants. As they suck the sap they
also secrete honey-dew which forms a protective cover over the leaf,
ultimately resulting in aphid starvation. \cite{Matis_2006} describe a
model for aphid dynamics in terms of population size ($N_t$) and
cumulative population size ($C_t$).  The model is a stochastic
birth-death model with linear birth rate $\lambda N_t$ and death rate
$\mu N_t C_t$.  The key probabilistic laws governing the time-evolution of
the process over a small interval $(t,t+dt]$ are
\begin{equation}\begin{split}
\Pr(N_{t+dt}=n_t +1,\,C_{t+dt}=c_t +1\,|\,n_t,c_t)&=\lambda n_t\,dt+o(dt), \label{eqn:aphid1and2} \\
\Pr(N_{t+dt}=n_t -1,\,C_{t+dt}=c_t\,|\,n_t,c_t)&=\mu n_t c_t\,dt+o(dt).
\end{split}\end{equation}
The diffusion approximation of the Markov jump process defined by
\eqref{eqn:aphid1and2} is
\begin{equation}\label{eqn:aphidSDE}
\begin{pmatrix}
dN_t\\
dC_t\\ \end{pmatrix}= \begin{pmatrix}
\lambda N_t-\mu N_tC_t \\
\lambda N_t \\ \end{pmatrix}dt+ \begin{pmatrix}
\lambda N_t+\mu N_tC_t & \lambda N_t  \\
\lambda N_t & \lambda N_t \\ \end{pmatrix}^\half dW_t.
\end{equation}
\cite{Matis_2008} also provide a dataset of cotton aphid counts
collected from three blocks (1/2/3) and using treatments constructed
from two factors: water irrigation (low/medium/high) and nitrogen
(blanket/variable/none). The data were collected in July 2004 in
Lamesa, Texas and consist of five observations of aphid counts
aggregated over twenty randomly chosen leaves in each plot for the
twenty-seven treatment-block combinations. The data were recorded at
times $t=0, 1.14, 2.29, 3.57$ and $4.57$ weeks (approximately every
7/8 days).

We now formulate an appropriate SDMEM model driven by
\eqref{eqn:aphidSDE} for these data and then fit the model.  For
notational simplicity, let $i,j,k$ denote the level of water, nitrogen
and block number respectively with $i,j,k\in\{1,2,3\}$, where 1
represents low water/blanket nitrogen, 2 represents medium
water/variable nitrogen and 3 represents high water/zero nitrogen. Let
$N_t^{ijk}$ denote the number of aphids at time~$t$ for combination
$ijk$ and $C_t^{ijk}$ the corresponding cumulative population size. We
write $X_t^{ijk}=(N_t^{ijk},C_t^{ijk})^T$ and consider the SDMEM
\[
dX_t^{ijk}=\alpha(X_t^{ijk},b^{ijk})\,dt+\sqrt{\beta(X_t^{ijk},b^{ijk})}\,dW_t^{ijk}, \quad i,j,k\in\{1,2,3\},
\]
where
\begin{align*}
\alpha(X_t^{ijk},b^{ijk})&=\begin{pmatrix}
\lambda^{ijk} N_t^{ijk}-\mu^{ijk} N_t^{ijk}C_t^{ijk} \\
\lambda^{ijk} N_t^{ijk} \\ \end{pmatrix},\\
\beta(X_t^{ijk},b^{ijk})&=\begin{pmatrix}
\lambda^{ijk} N_t^{ijk}+\mu^{ijk} N_t^{ijk}C_t^{ijk} & \lambda^{ijk} N_t^{ijk}  \\
\lambda^{ijk} N_t^{ijk} & \lambda^{ijk} N_t^{ijk} \\ \end{pmatrix}.
\end{align*}
The fixed effects $b^{ijk}=(\lambda^{ijk},\mu^{ijk})^T$ have a
standard structure which allows for main factor and block effects and single
factor-block interactions, with
\begin{equation}\begin{split}
\lambda^{ijk}&=\lambda + \lambda_{W_i} + \lambda_{N_j} + \lambda_{B_k} + \lambda_{WN_{ij}} + \lambda_{WB_{ik}} + \lambda_{NB_{jk}} \label{aphid_lambdamu}\\
\mu^{ijk}&=\mu + \mu_{W_i} + \mu_{N_j} + \mu_{B_k} + \mu_{WN_{ij}} + \mu_{WB_{ik}} + \mu_{NB_{jk}}. 
\end{split}\end{equation}
Also for identifiability we use the corner constraints $\lambda_{W_1}
= \lambda_{N_1} = \lambda_{B_1} = 0$,
$\lambda_{WN_{ij}}=\lambda_{WN_{ij}}(1-\kappa_{ij})$,
$\lambda_{WB_{ik}}=\lambda_{WB_{ik}}(1-\kappa_{ik})$ and
$\lambda_{NB_{jk}}=\lambda_{NB_{jk}}(1-\kappa_{jk})$, where
$\kappa_{rs}=\max(\delta_{1r},\delta_{1s})$ and $\delta_{1\cdot}$ is
the Kronecker delta, with equivalent constraints on the death rates.
The interpretation of \eqref{aphid_lambdamu} is straightforward. For
example, $\lambda^{111}=\lambda$ and $\mu^{111}=\mu$ are the baseline
birth and death rates inferred using all $5\times 3^3=135$
observations, and correspond to the treatment combination low water,
blanket nitrogen and block 1. Likewise, all $5\times 3^2=45$
observations taken from block 2 inform the main effects of block 2
($\lambda_{B_2}$ and $\mu_{B_2}$) relative to the baseline.

A related approach can be found in \cite{Gillespie_Golightly_2010},
where the diffusion approximation is eschewed in favour of a further
approximation via moment closure. Our approach further differs from
theirs by allowing for measurement error and leads to a much improved
predictive fit. The measurement error model is in part motivated by an
over-dispersed Poisson error structure which we then approximate by a
Gaussian distribution. Specifically, we assume that aphid population
size $N_t$ is observed with Gaussian error and that the error variance
is proportional to the latent aphid numbers, giving
\begin{equation}\label{eqn:obsAphid}
Y_t^{ijk}|N_t^{ijk},\sigma \indep N(N_t^{ijk},\sigma^{2}N_t^{ijk}),\quad t=0, 1.14, 2.29, 3.57, 4.57.
\end{equation} 
   
\subsubsection{Implementation}

Our prior beliefs for $1/\sigma^2$ are described by a $Ga(a,a)$
distribution. We found little difference in results for
$a\in\{0.01,0.1,1\}$ and so here we report results for $a=1$.  The
prior for the elements in \eqref{aphid_lambdamu} consists of
independent components subject to the birth and death rates for each
treatment combination $(\lambda^{ijk},\mu^{ijk})$ being positive. The
baseline rates $\lambda$ and $\mu$ must be positive and so, following
\cite{Gillespie_Golightly_2010}, we assign weak $U(-10,10)$ priors to
$\log\lambda$ and $\log\mu$ and also to the remaining parameters. We
also take a fairly weak $N(24,90)$ prior for each $N_{t_0}^{ijk}$ and
use a proposal of the form $N(N_{t_0}$,$\sigma^2N_{t_0})$ for
updates. The cumulative population sizes must be at least as large as
their equivalent population size. However, we do not expect them to be
greatly different \emph{a priori}.  We investigated using a truncated
distribution of the form $C_{t_0}|N_{t_0}\sim N(N_{t_0},d_c^2)$,
$C_{t_0}>N_{t_0}$ as the prior and found that this led to little
difference in posterior output for $d_{c}\in\{1,10,100\}$. We have,
therefore, chosen to fix $C_{t_0}^{ijk}=N_{t_0}^{ijk}$ in our
analysis. Note that the form of the prior for $\sigma$ gives a semi-conjugate update.  The
remaining parameters in \eqref{aphid_lambdamu} are updated using
random walk Metropolis on the pairwise $\lambda,\mu$ component blocks
$(\lambda,\mu),(\lambda_{W_2},\mu_{W_2}),(\lambda_{W_3},\mu_{W_3}),
\ldots,(\lambda_{NB_{33}},\mu_{NB_{33}})$.

The nonlinear form of the observation model \eqref{eqn:obsAphid} can
be problematic for the modified innovation scheme. In particular, the
proposal mechanism for the path update requires an observation model
that is linear in $N_t$. Therefore, when proposing from the bridge
construct in Section~\ref{dgode}, we replace $\Sigma$ in
\eqref{eqn:dgodebridge2} and \eqref{eqn:dgodebridge3} with $\sigma^2
\eta_{N,t_{j+1}}$, where
$\eta_{t_{j+1}}=(\eta_{N,t_{j+1}},\eta_{C,t_{j+1}})^T$ is the solution
of \eqref{eqn:dgode_eta}. Since the proposal mechanism is corrected
for via the Metropolis-Hastings step, no additional approximations to
the target distribution are needed. 

\begin{figure}[t!]
\begin{center}
\begin{minipage}[b]{0.49\linewidth}
				\centering
				\caption*{\qquad Improved bridge construct}\vspace{-0.25cm}
				\includegraphics[scale=0.3]{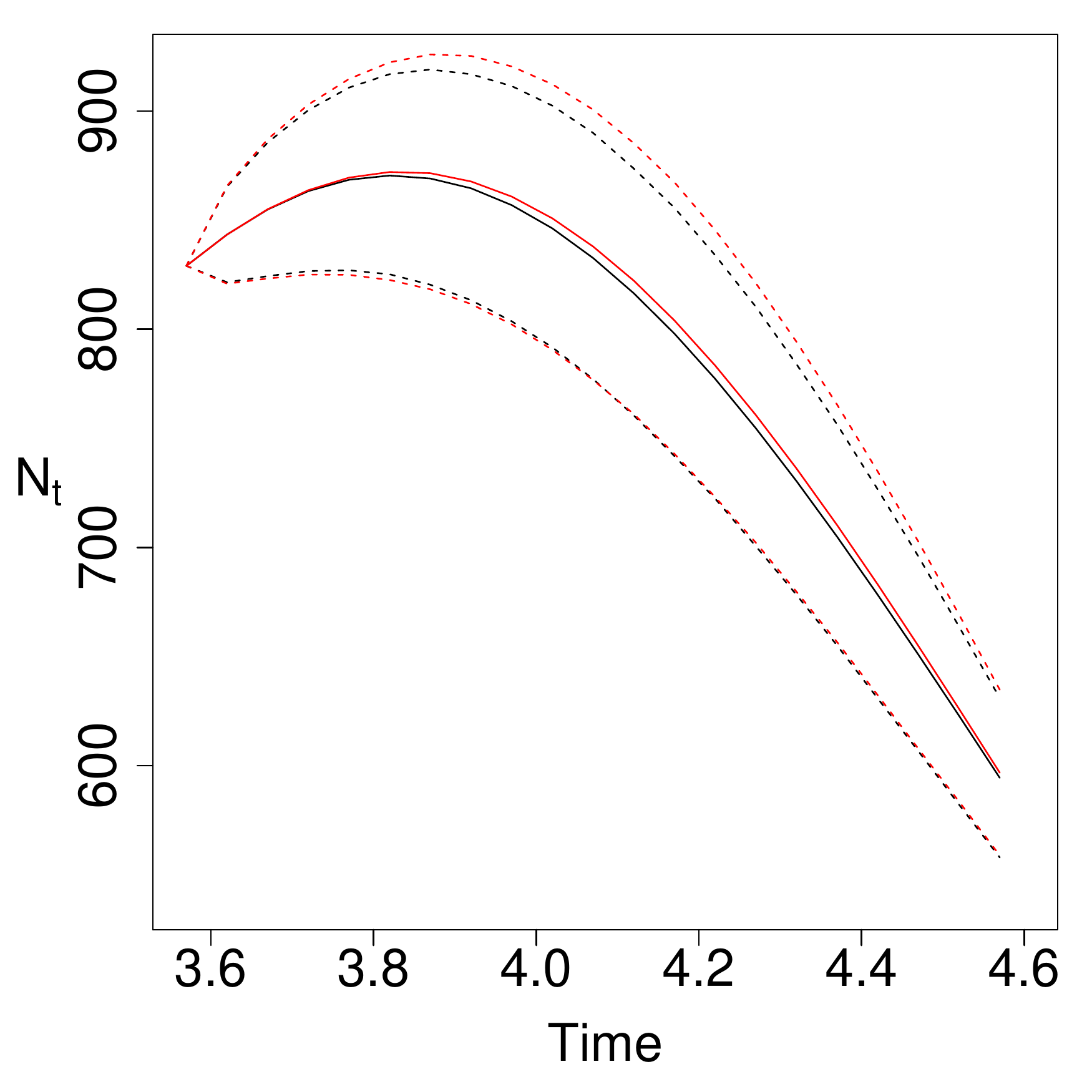}
\end{minipage} 
\hspace{-2cm}
\begin{minipage}[b]{0.49\linewidth}
				\centering
				\caption*{\qquad Modified diffusion bridge}\vspace{-0.25cm}
				\includegraphics[scale=0.3]{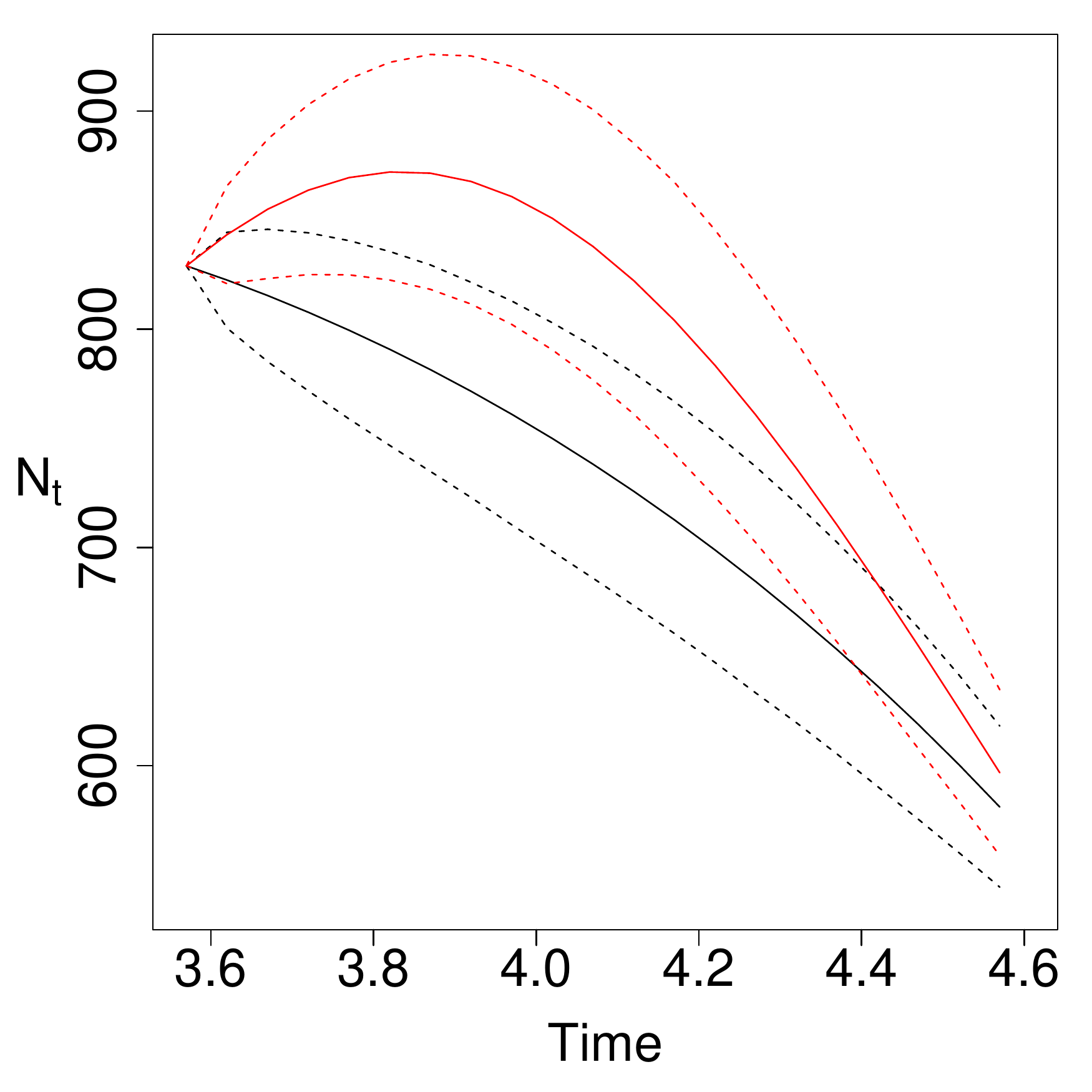}
\end{minipage}
				\caption{95\% credible region (dashed line) and mean (solid line) of the true conditioned 
aphid population component $N_t|x_{3.57},y_{4.57}$ (red) and two competing bridge constructs (black).} \label{aphid fig_bridges}
\end{center}
\end{figure}

In order to obtain a statistically efficient implementation 
of the modified innovation scheme, we investigate the 
performance of the modified diffusion bridge construct of \cite{Durham_2001} 
and our improved bridge construct of Section~\ref{dgode} in a scenario typical of the real 
dataset. Using the simulation study of \cite{Gillespie_Golightly_2010}, 
we take $(\lambda,\mu)^T=(1.75, 0.00095)^T$, \hbox{$x_0=(28,28)^T$} 
and recursively apply the Euler-Maruyama approximation to 
give \linebreak$x_{3.57}=(829.08,1406.07)^T$. We then compare the performance 
of each bridge construct over the final observation interval 
$[3.57,4.57]$ by taking $y_{4.57}$ as the median of 
(\ref{eqn:obsAphid}) with $\sigma=1$. Figure~\ref{aphid fig_bridges} shows 95\% credible regions 
of the true conditioned process $N_t|x_{3.57},y_{4.57}$ 
(found via Monte Carlo simulation) with 95\% credible regions 
obtained by repeatedly simulating from the modified 
diffusion bridge and our improved construct. 
It is clear that the modified diffusion bridge fails to adequately 
account for the nonlinear behaviour of the conditioned process. 
Use of each construct as a proposal mechanism inside a 
Metropolis-Hastings independence sampler (100K iterations) results in 
an acceptance rate of around 58\% for the improved bridge 
construct and just 1\% for the modified diffusion bridge. It is for these 
reasons that the modified diffusion bridge is eschewed 
in favour of our improved bridge construct when applying the Bayesian 
imputation approach.

Finally, fitting the LNA requires the
solution of an ODE system given by \eqref{eqn:dgode_eta} and
\eqref{eqn:lna_v} where the Jacobian matrix is
\[
H_t= \begin{pmatrix}
\lambda - \mu\eta_{C,t}  &  -\mu\eta_{N,t} \\
\lambda  &  0 \end{pmatrix}.
\]
This ODE system is intractable and so our C implementation uses a
standard ODE solver from the GNU scientific library, namely the
explicit embedded Runge-Kutta-Fehlberg $(4, 5)$ method. Note that the
tractability of the marginal likelihood under the LNA requires a
linear Gaussian observation model. Therefore, when applying the FFBS
algorithm in the supplementary material, we make an approximation to
the marginal likelihood calculation by replacing $\Sigma$ with
$\sigma^2 \eta_{N,t_{j+1}}$.

\subsubsection{Results}
The time between observations is almost but not quite constant and so
we have allowed each interval to have its own discretisation level,
$m$.  That said, the interval-specific values vary very little, and by
at most two for the larger $m$ values. Several short pilot runs of the
modified innovation scheme were performed with typical
$m\in\{5,10,20,40,50\}$. These gave no discernible difference in
posterior output for $m\geq 20$ and so we took $m=20$. The sample
output was also used to estimate the marginal posterior variances of
the $\lambda,\mu$ component blocks of the parameters in
\eqref{aphid_lambdamu}, to be used in the random walk Metropolis
updates. Both the modified innovation scheme and MCMC scheme under
the LNA were run for 40M iterations with the output thinned by taking
every 4Kth iterate to give a final sample of size 10K.

\begin{figure}[t!]
\begin{center}
\begin{minipage}[b]{0.28\linewidth}
        \centering
        \includegraphics[scale=0.22]{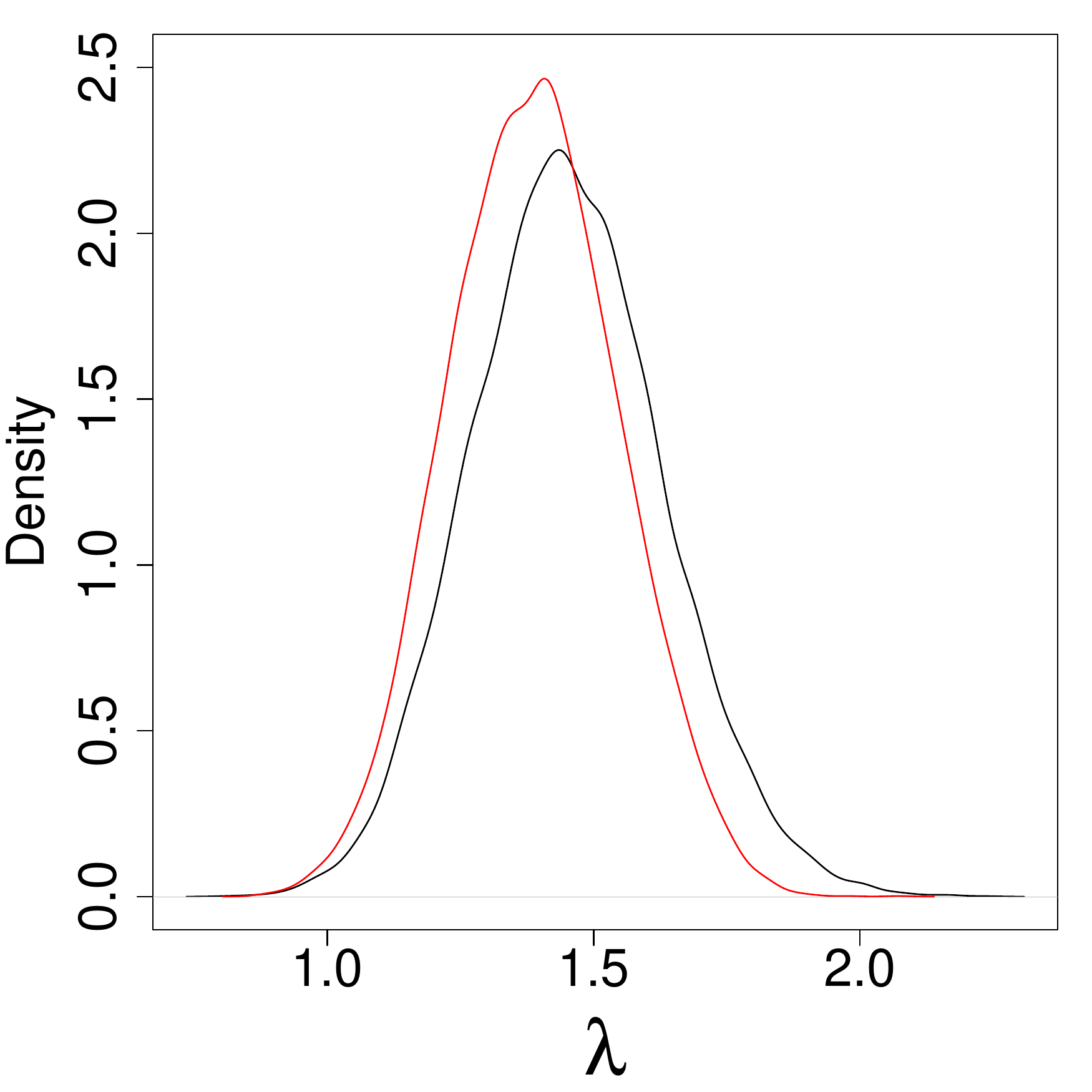}
\end{minipage} 
\hspace{-0.25cm}
\begin{minipage}[b]{0.28\linewidth}
        \centering
        \includegraphics[scale=0.22]{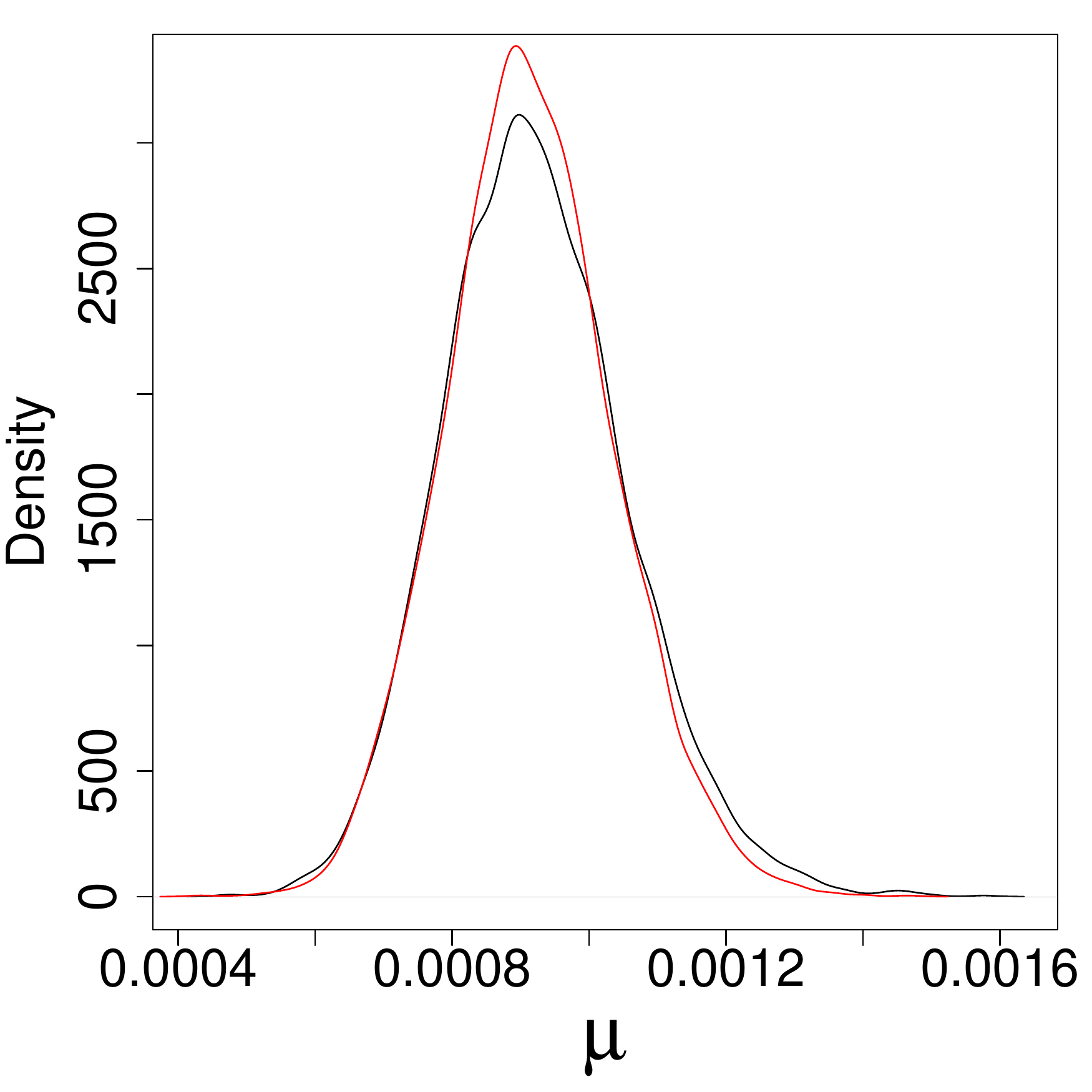}
\end{minipage} 
\hspace{-0.25cm}
\begin{minipage}[b]{0.28\linewidth}
				\centering
        \includegraphics[scale=0.22]{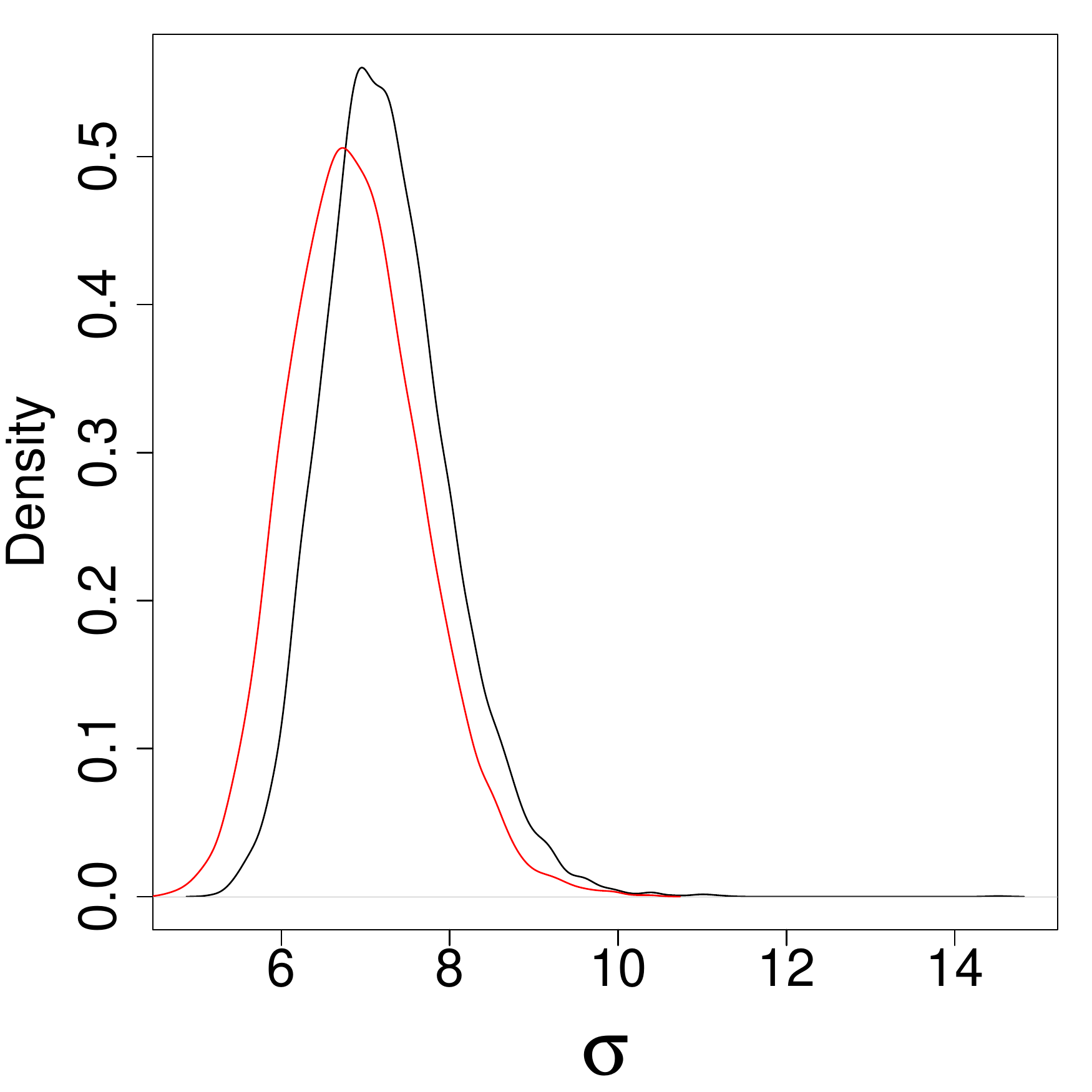}
\end{minipage}\\
\vspace{0.2cm}
\begin{minipage}[b]{0.28\linewidth}
				\centering
        \includegraphics[scale=0.22]{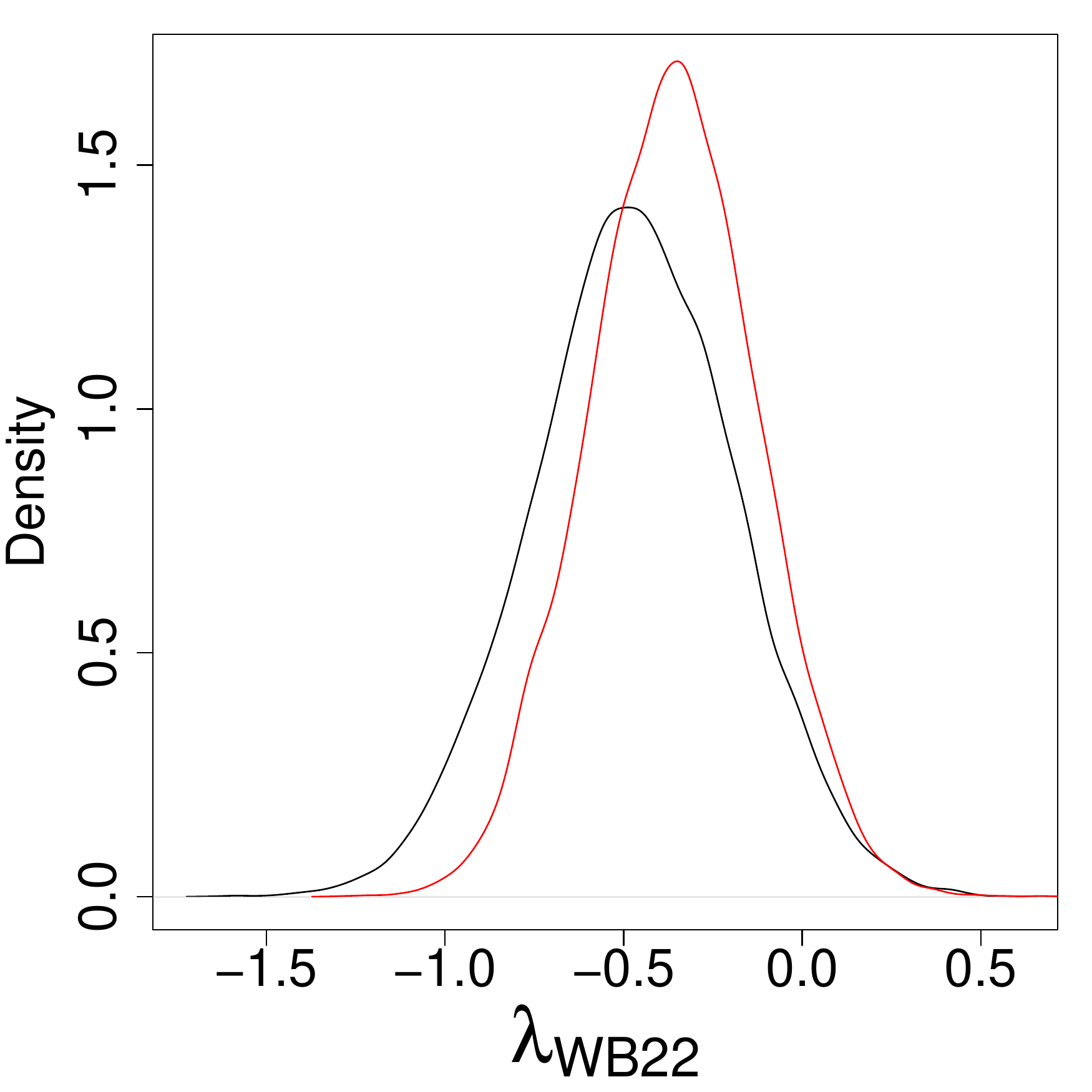}
\end{minipage}
\hspace{-0.25cm}
\begin{minipage}[b]{0.28\linewidth}
        \centering
        \includegraphics[scale=0.22]{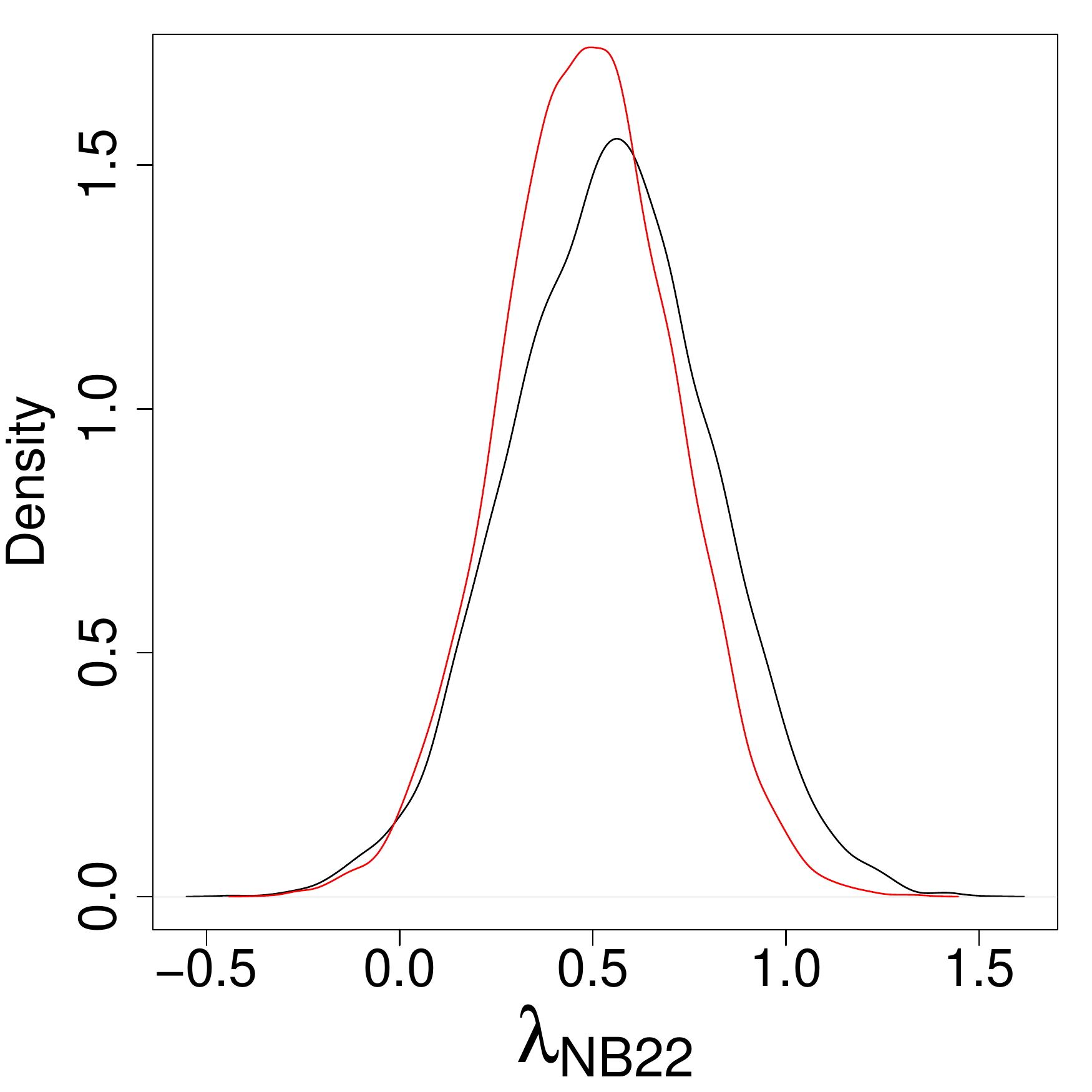}
\end{minipage}
\hspace{-0.25cm}
\begin{minipage}[b]{0.28\linewidth}
        \centering
        \includegraphics[scale=0.22]{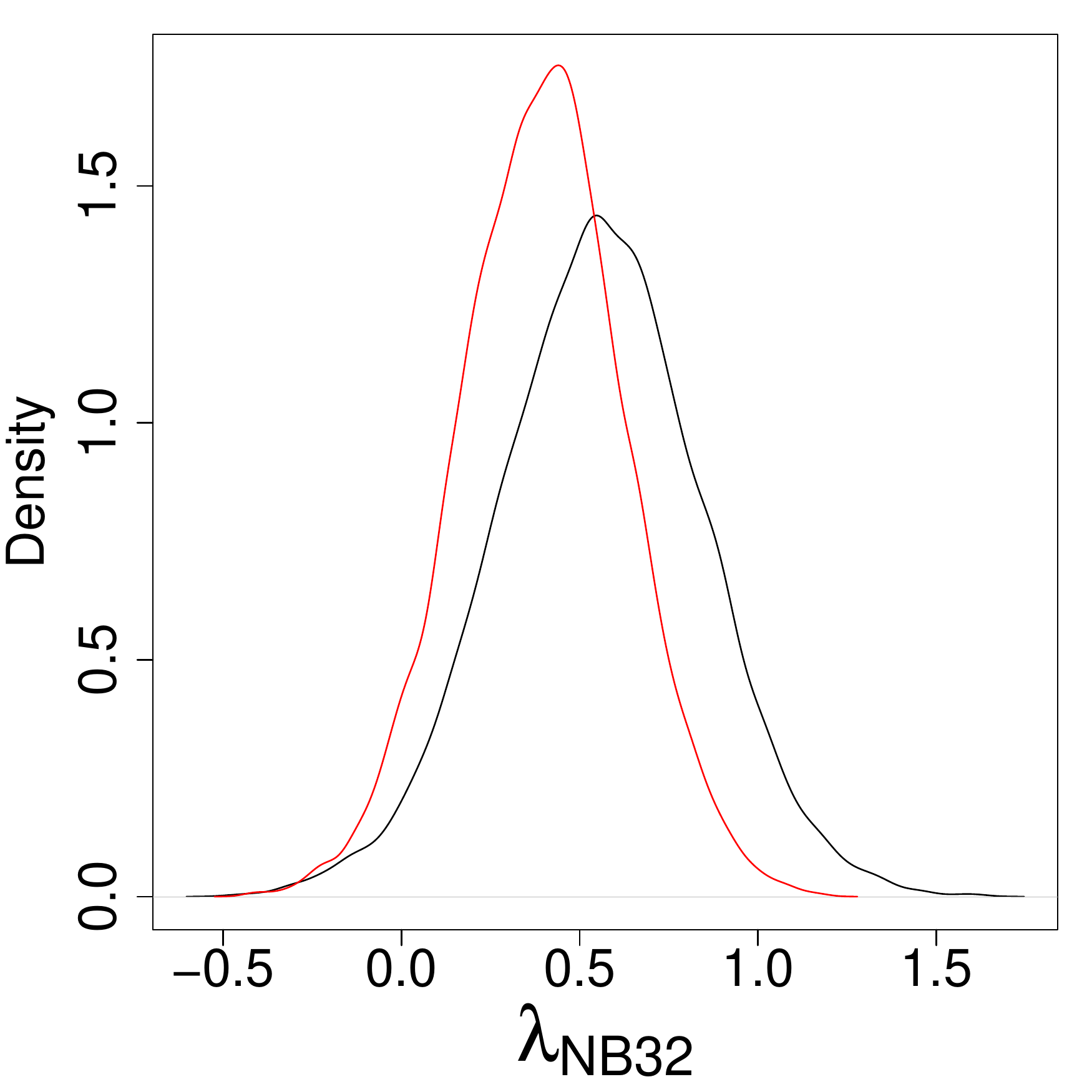}
\end{minipage}\\
\vspace{0.2cm}
\begin{minipage}[b]{0.28\linewidth}
        \centering
        \includegraphics[scale=0.22]{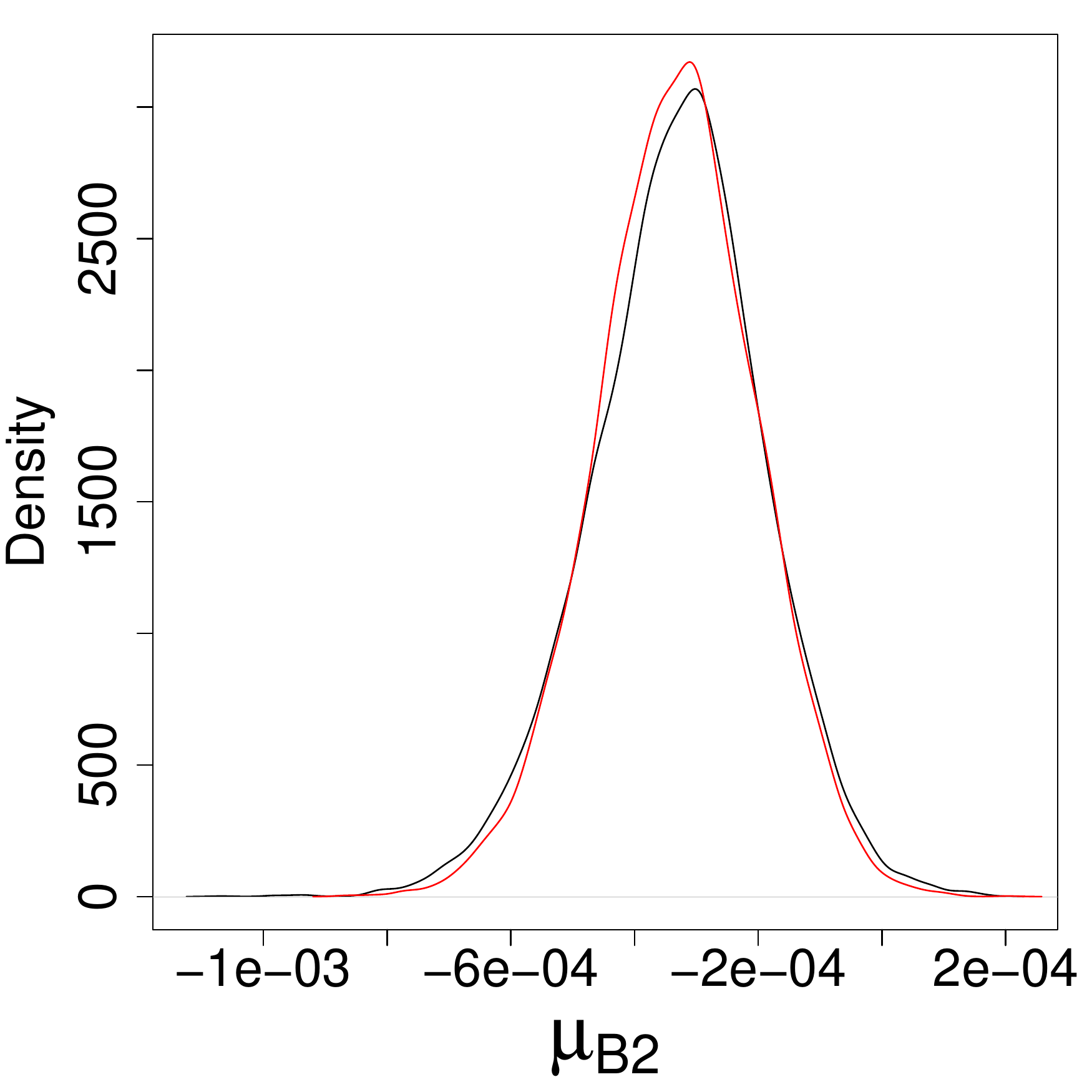}
\end{minipage} 
\hspace{-0.25cm}
\begin{minipage}[b]{0.28\linewidth}
        \centering
        \includegraphics[scale=0.22]{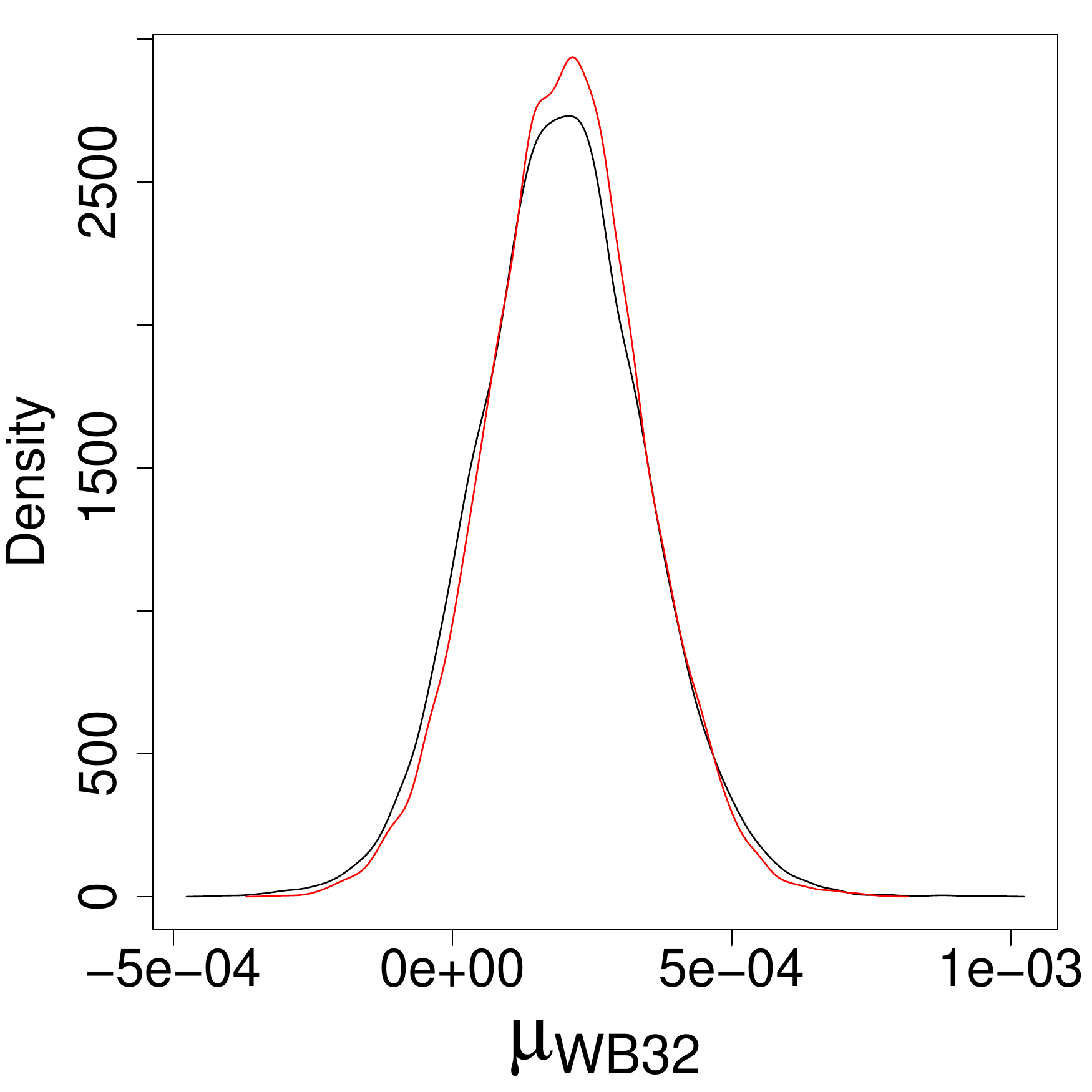}
\end{minipage} 
\hspace{-0.25cm}
\begin{minipage}[b]{0.28\linewidth}
				\centering
        \includegraphics[scale=0.22]{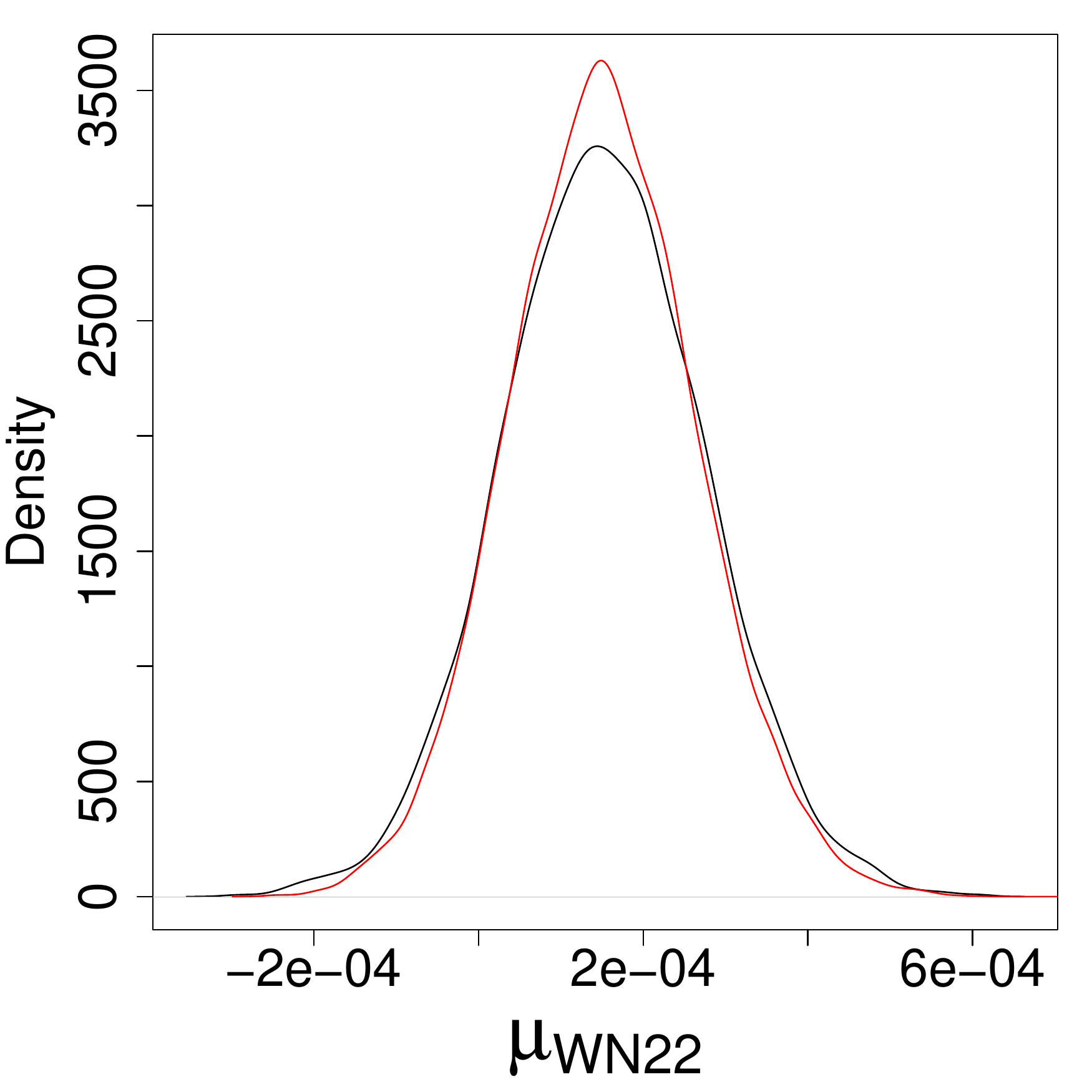}
\end{minipage}
\caption{Marginal posterior densities for a selection of the aphid model parameters. 
Black: Bayesian imputation. Red: LNA.} \label{aphid fig_baselines}
\end{center}
\end{figure}

Figure~\ref{aphid fig_baselines} shows the marginal posterior
densities of the baseline parameters, the parameter $\sigma$
controlling the observation error variance and a selection of the
remaining parameters. As in \cite{Gillespie_Golightly_2010} we find
that block 2 plays an important role. The 95\% credible regions for
$\mu_{B_2}$, the main block 2 death rate, and $\lambda_{NB_{22}}$, the
birth rate characterising the interaction with nitrogen, are plausibly
non-zero. Whilst the imputation approach and LNA generally give
consistent output, there are some notable differences. For example, we find, in
general, that the LNA tends to underestimate parameter values 
(and slightly exaggerates the confidence in these estimates) compared
to those obtained under the modified innovation scheme.

\begin{figure}
\begin{center}
\begin{minipage}[b]{0.28\linewidth}
        \centering
				\caption*{\quad$ijk=122$}\vspace{-0.25cm}
        \includegraphics[scale=0.22]{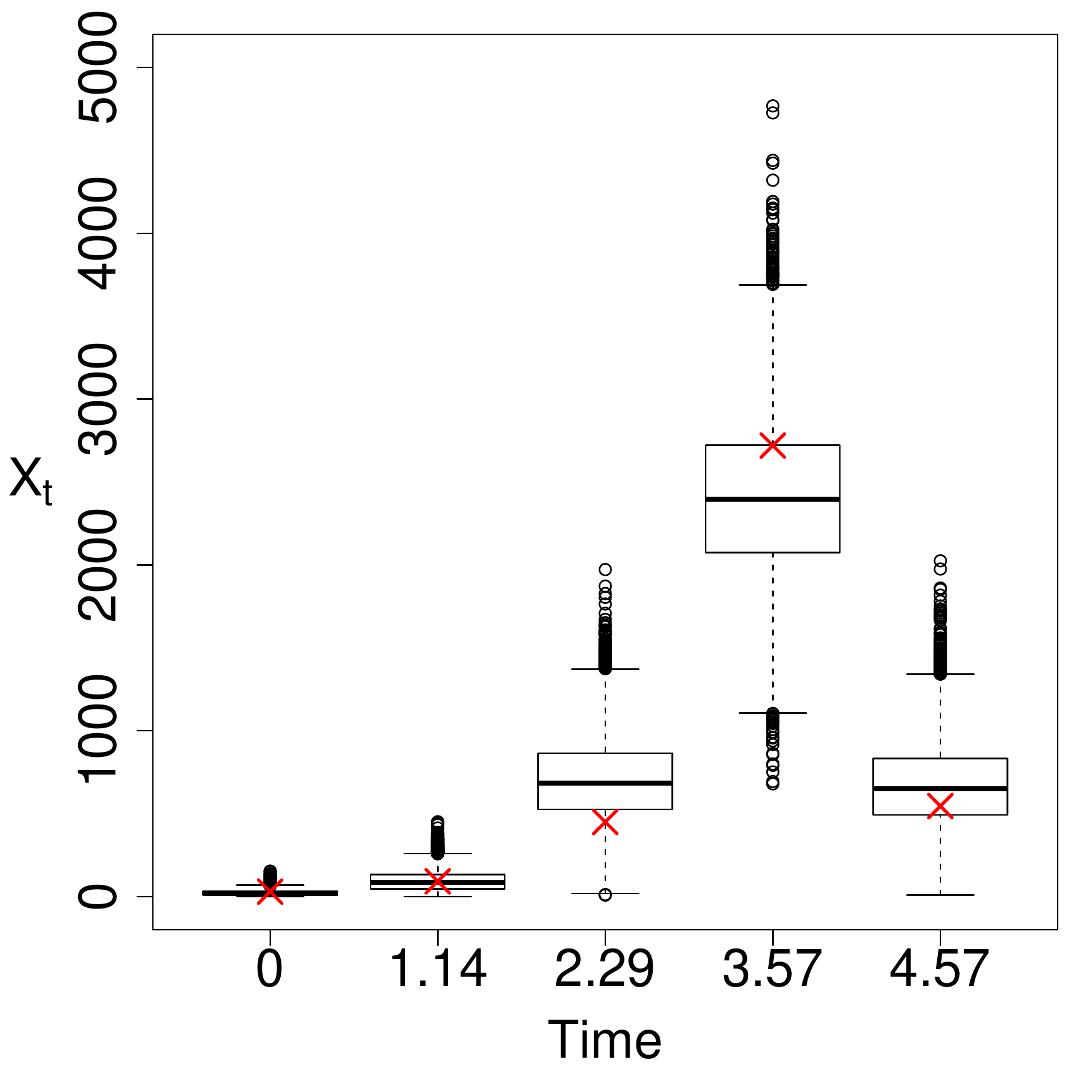}
\end{minipage} 
\hspace{-0.25cm}
\begin{minipage}[b]{0.28\linewidth}
        \centering
				\caption*{\quad$ijk=133$}\vspace{-0.25cm}
        \includegraphics[scale=0.22]{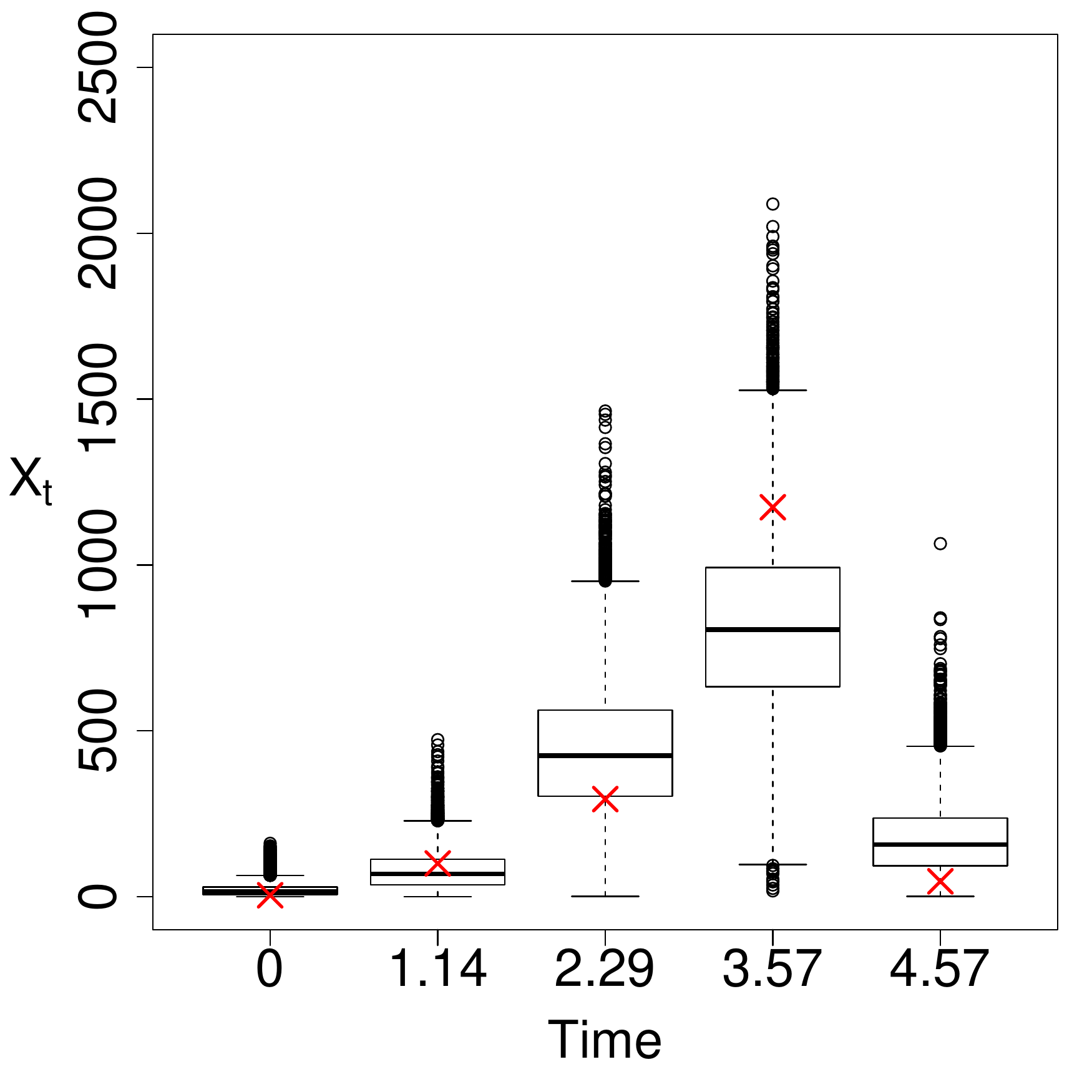}
\end{minipage} 
\hspace{-0.25cm}
\begin{minipage}[b]{0.28\linewidth}
				\centering
				\caption*{\quad$ijk=212$}\vspace{-0.25cm}
        \includegraphics[scale=0.22]{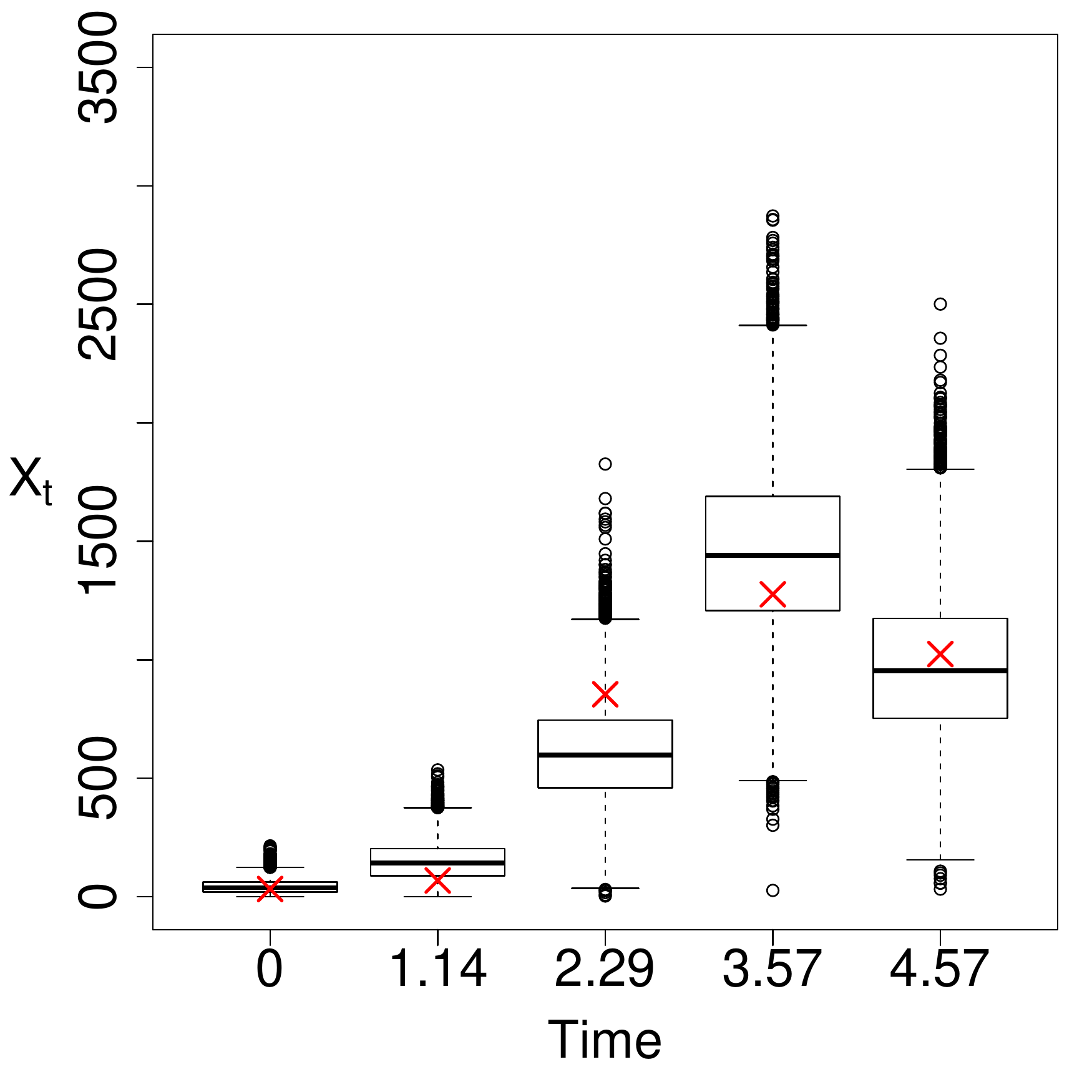}
\end{minipage}\\
\vspace{0.2cm}
\begin{minipage}[b]{0.28\linewidth}
				\centering
				\caption*{\quad$ijk=222$}\vspace{-0.25cm}
        \includegraphics[scale=0.22]{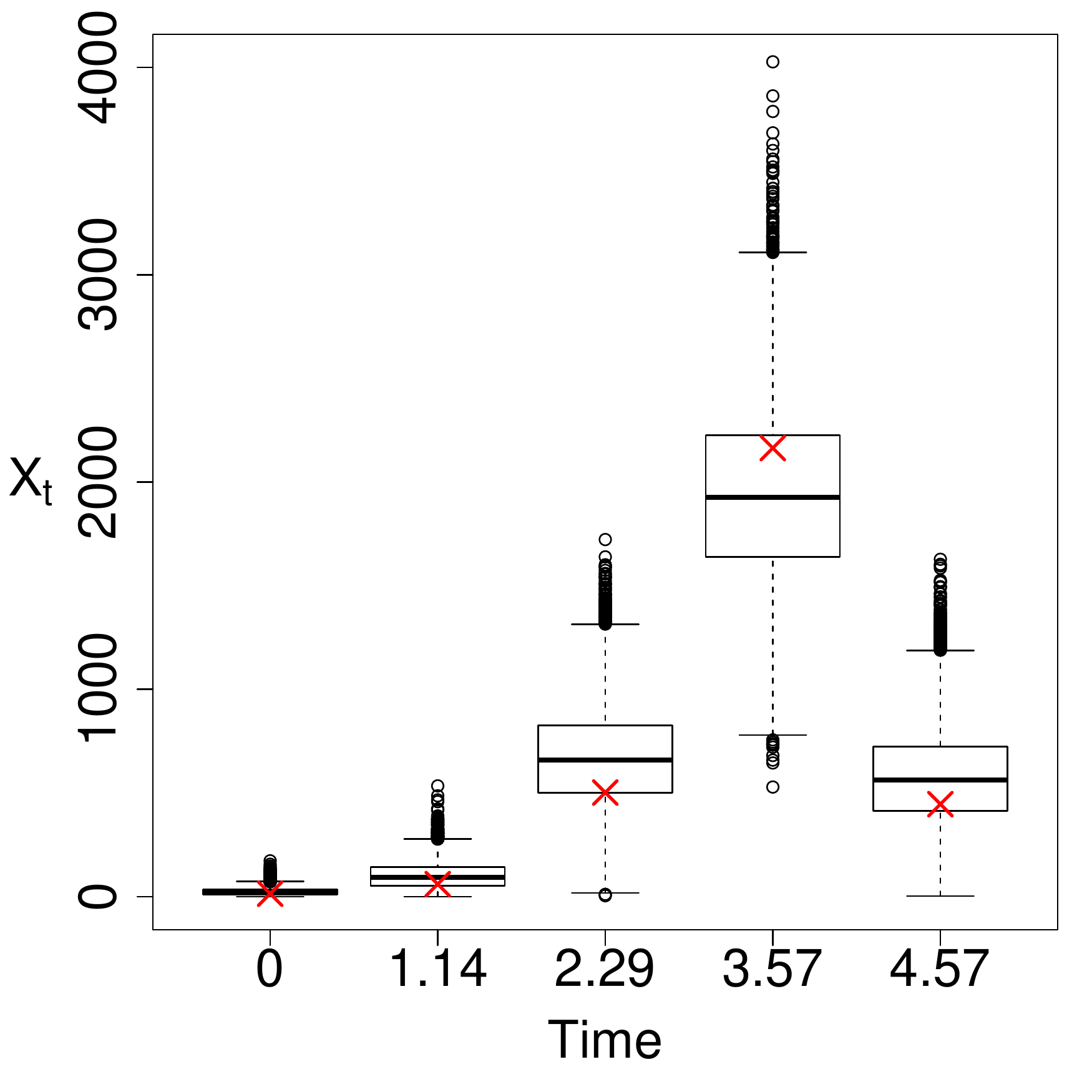}
\end{minipage}
\hspace{-0.25cm}
\begin{minipage}[b]{0.28\linewidth}
        \centering
				\caption*{\quad$ijk=311$}\vspace{-0.25cm}
        \includegraphics[scale=0.22]{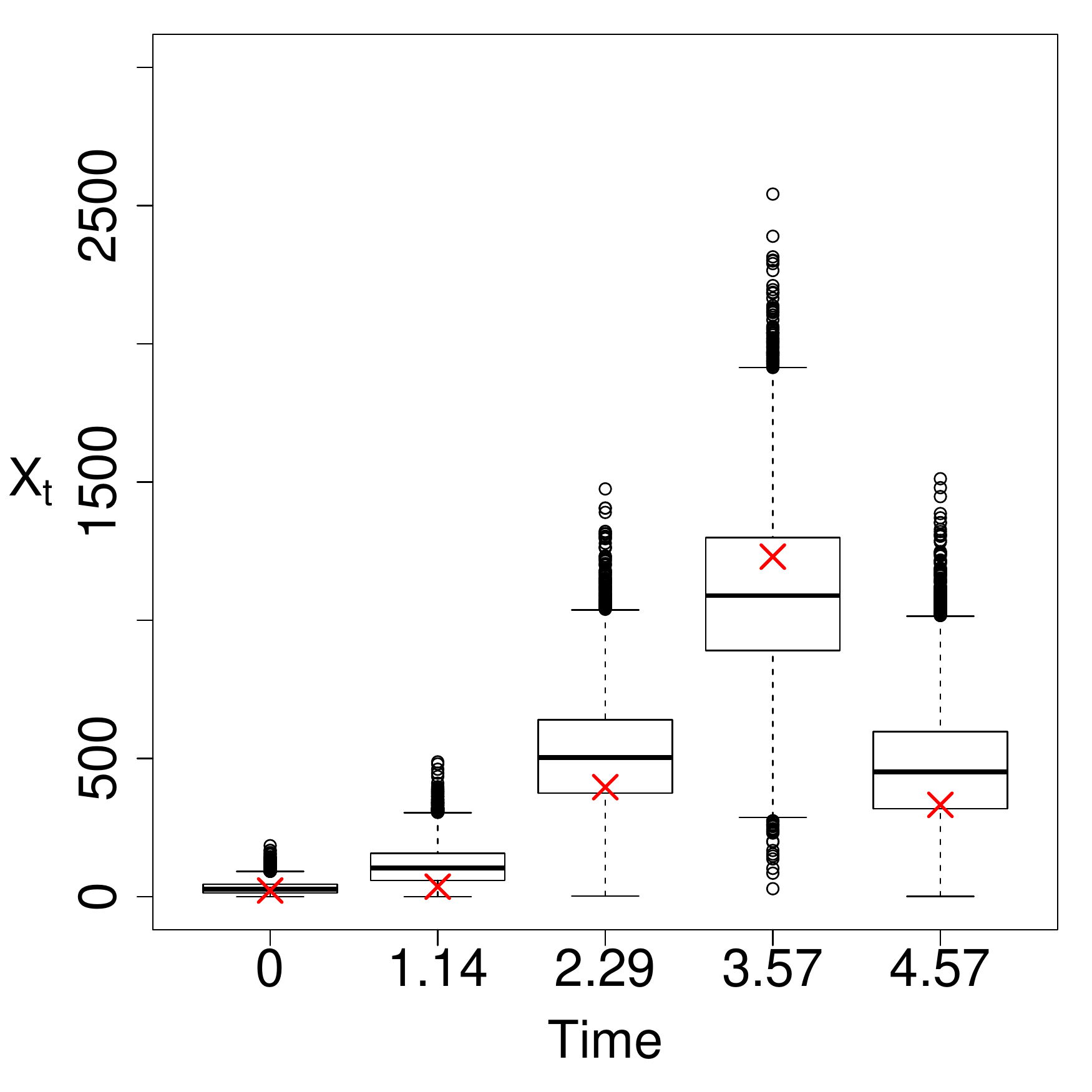}
\end{minipage}
\hspace{-0.25cm}
\begin{minipage}[b]{0.28\linewidth}
        \centering
				\caption*{\quad$ijk=332$}\vspace{-0.25cm}
        \includegraphics[scale=0.22]{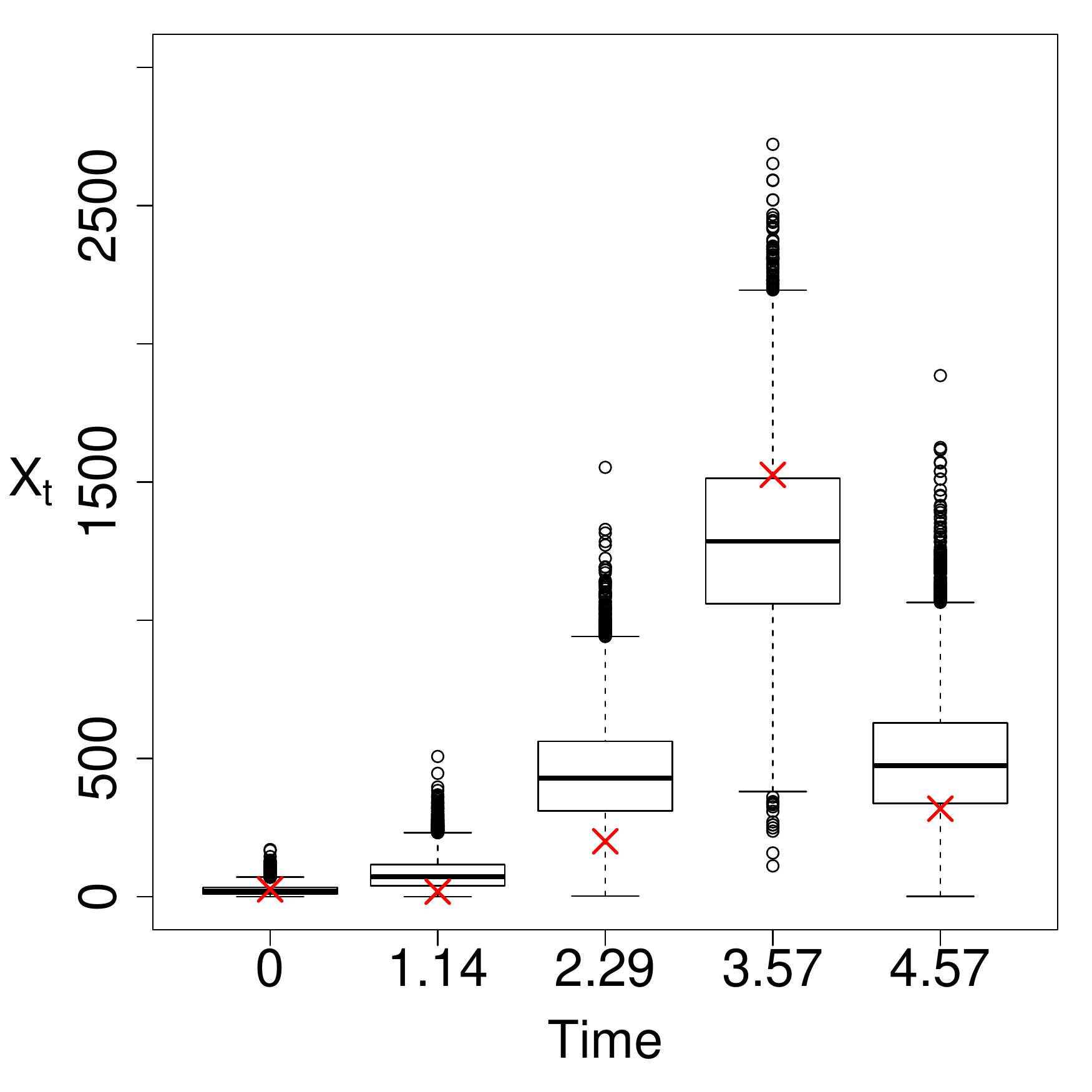}
\end{minipage}\\
\vspace{0.2cm}
\begin{minipage}[b]{0.28\linewidth}
        \centering
				\caption*{\quad$ijk=122$}\vspace{-0.25cm}
        \includegraphics[scale=0.22]{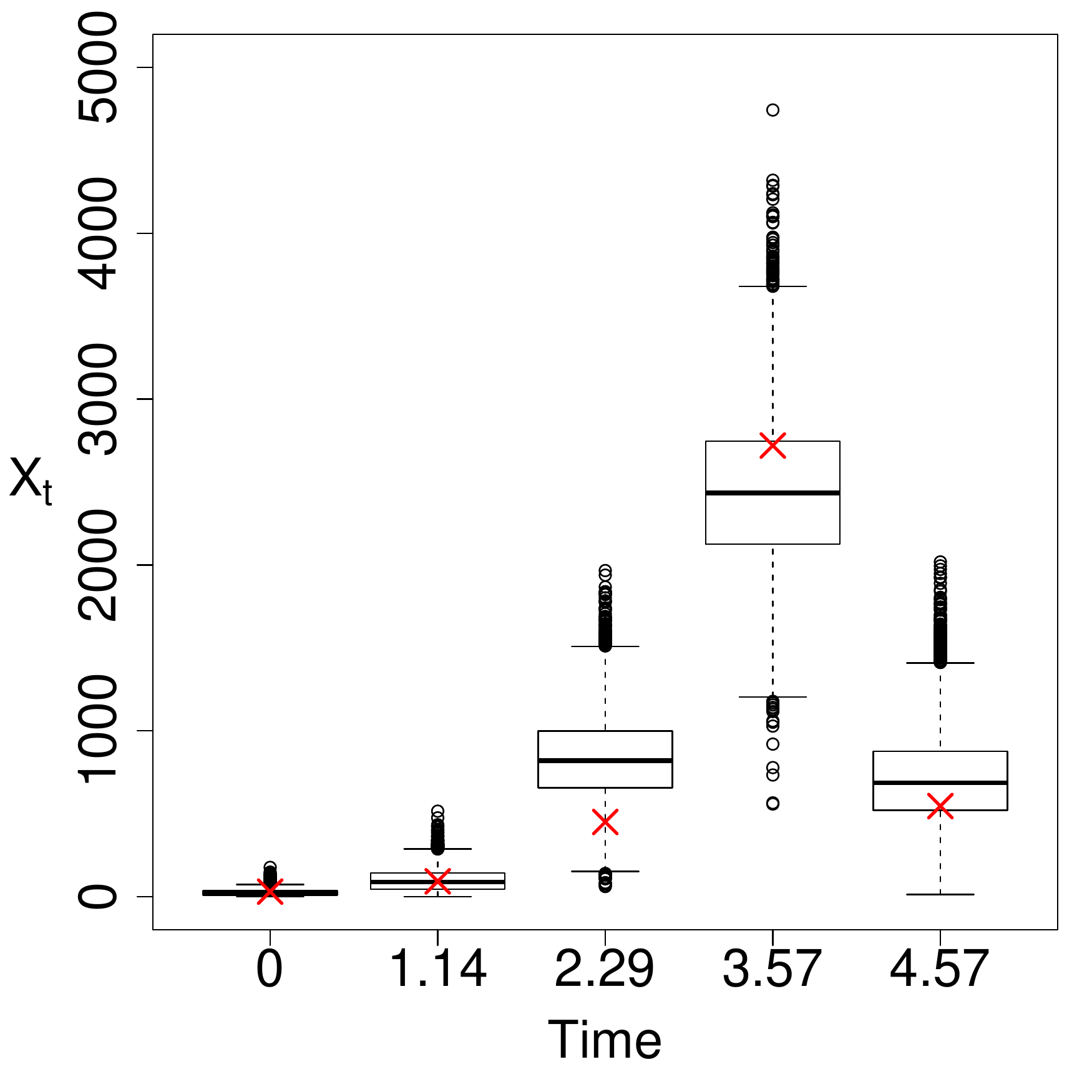}
\end{minipage} 
\hspace{-0.25cm}
\begin{minipage}[b]{0.28\linewidth}
        \centering
				\caption*{\quad$ijk=133$}\vspace{-0.25cm}
        \includegraphics[scale=0.22]{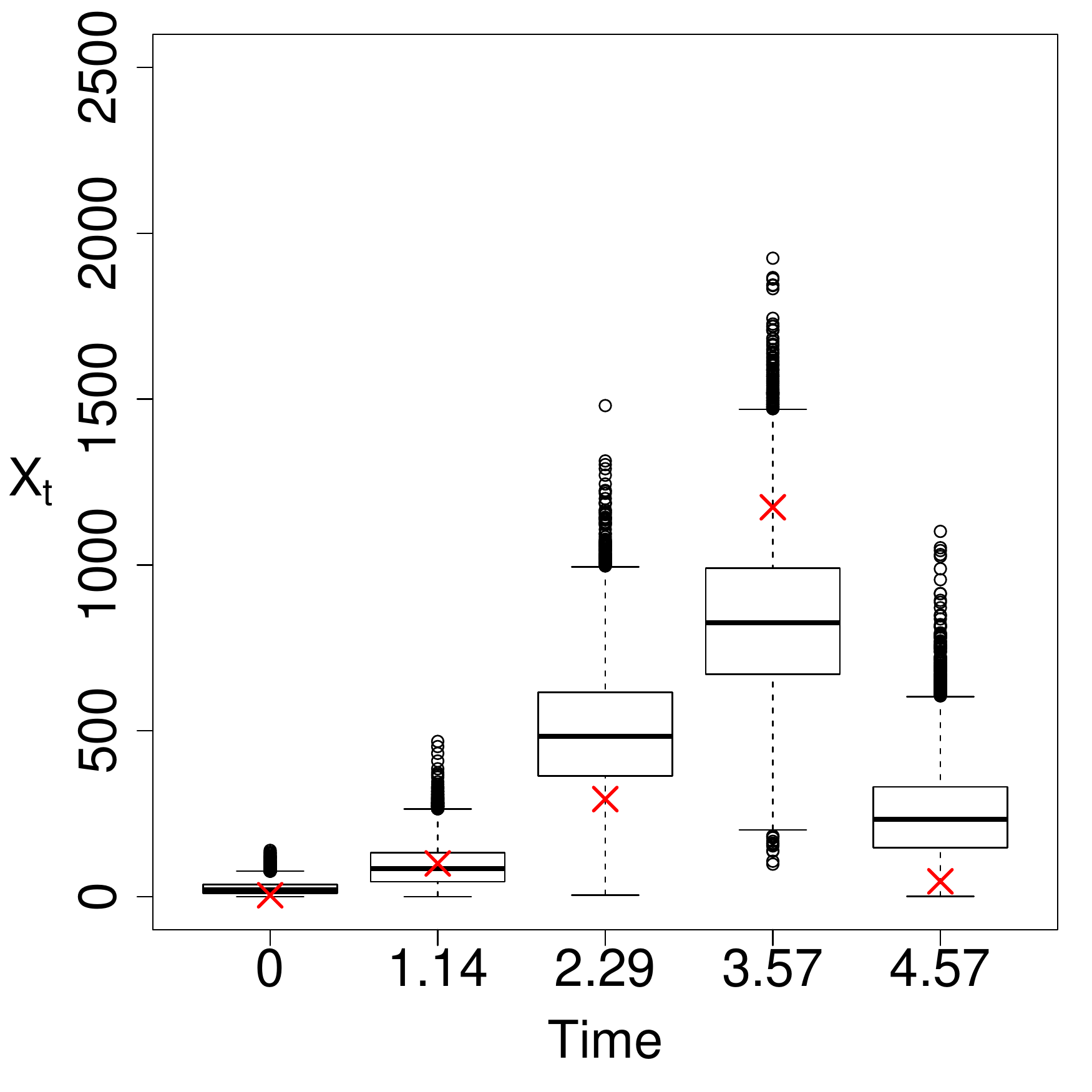}
\end{minipage} 
\hspace{-0.25cm}
\begin{minipage}[b]{0.28\linewidth}
				\centering
				\caption*{\quad$ijk=212$}\vspace{-0.25cm}
        \includegraphics[scale=0.22]{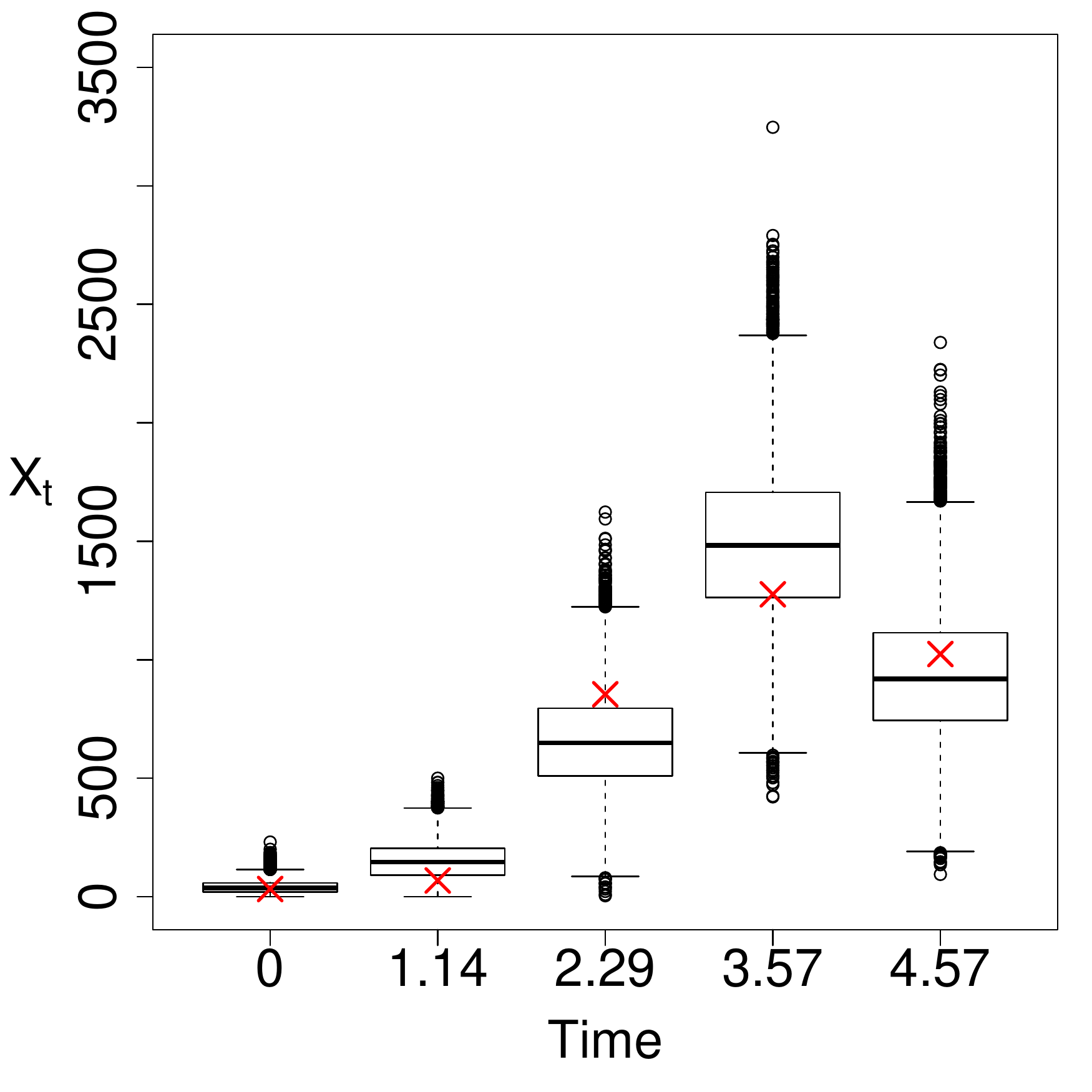}
\end{minipage}\\
\vspace{0.2cm}
\begin{minipage}[b]{0.28\linewidth}
				\centering
				\caption*{\quad$ijk=222$}\vspace{-0.25cm}
        \includegraphics[scale=0.22]{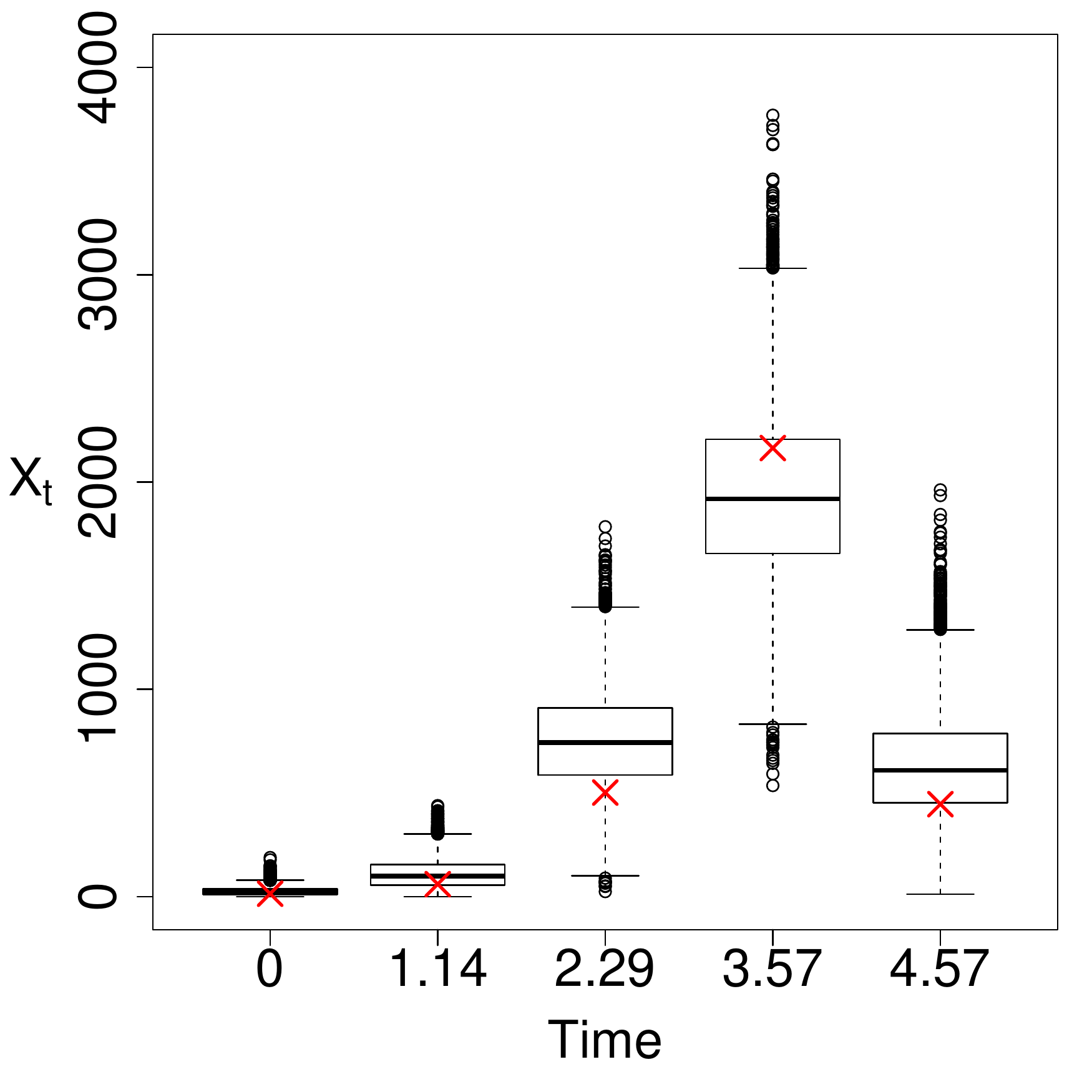}
\end{minipage}
\hspace{-0.25cm}
\begin{minipage}[b]{0.28\linewidth}
        \centering
				\caption*{\quad$ijk=311$}\vspace{-0.25cm}
        \includegraphics[scale=0.22]{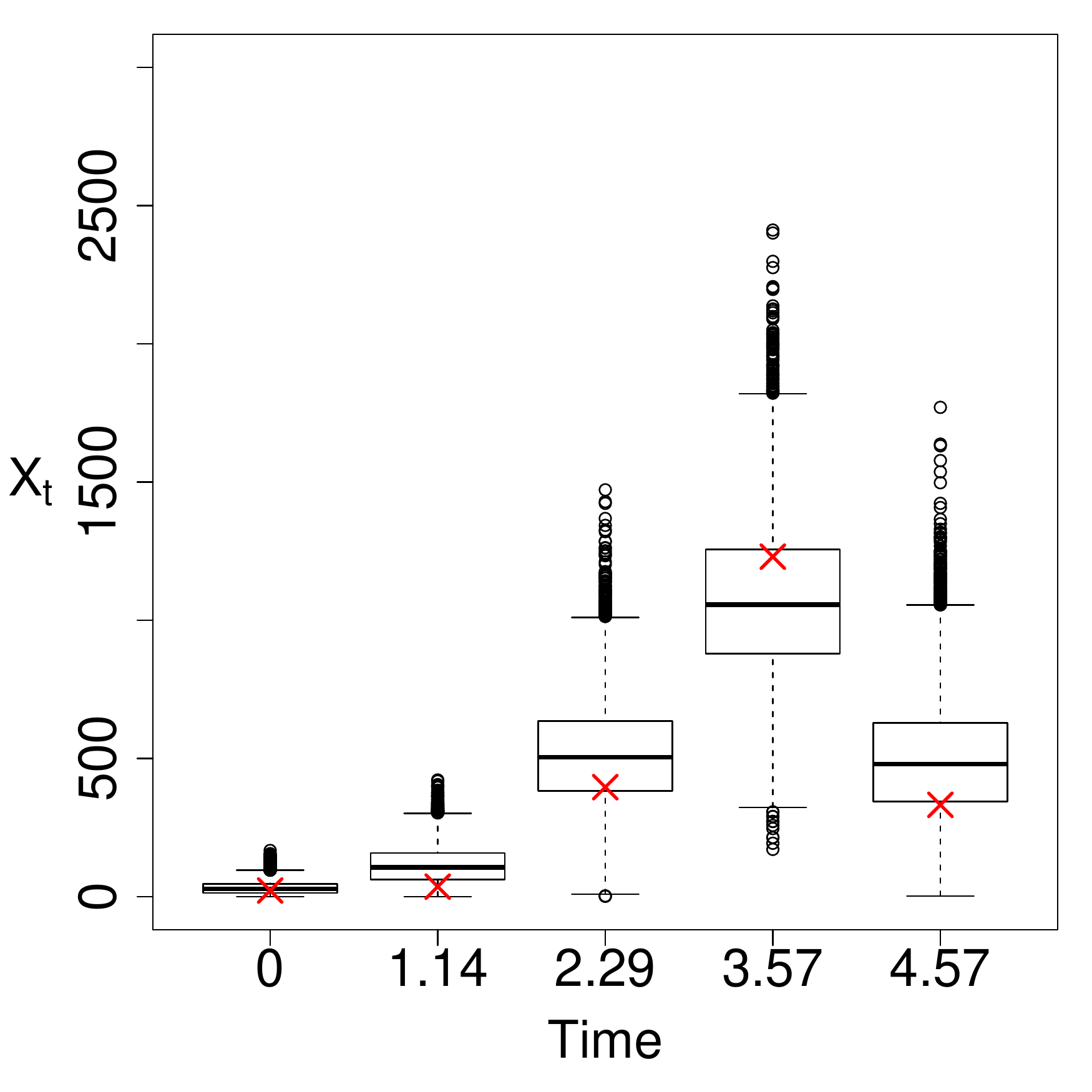}
\end{minipage}
\hspace{-0.25cm}
\begin{minipage}[b]{0.28\linewidth}
        \centering
				\caption*{\quad$ijk=332$}\vspace{-0.25cm}
        \includegraphics[scale=0.22]{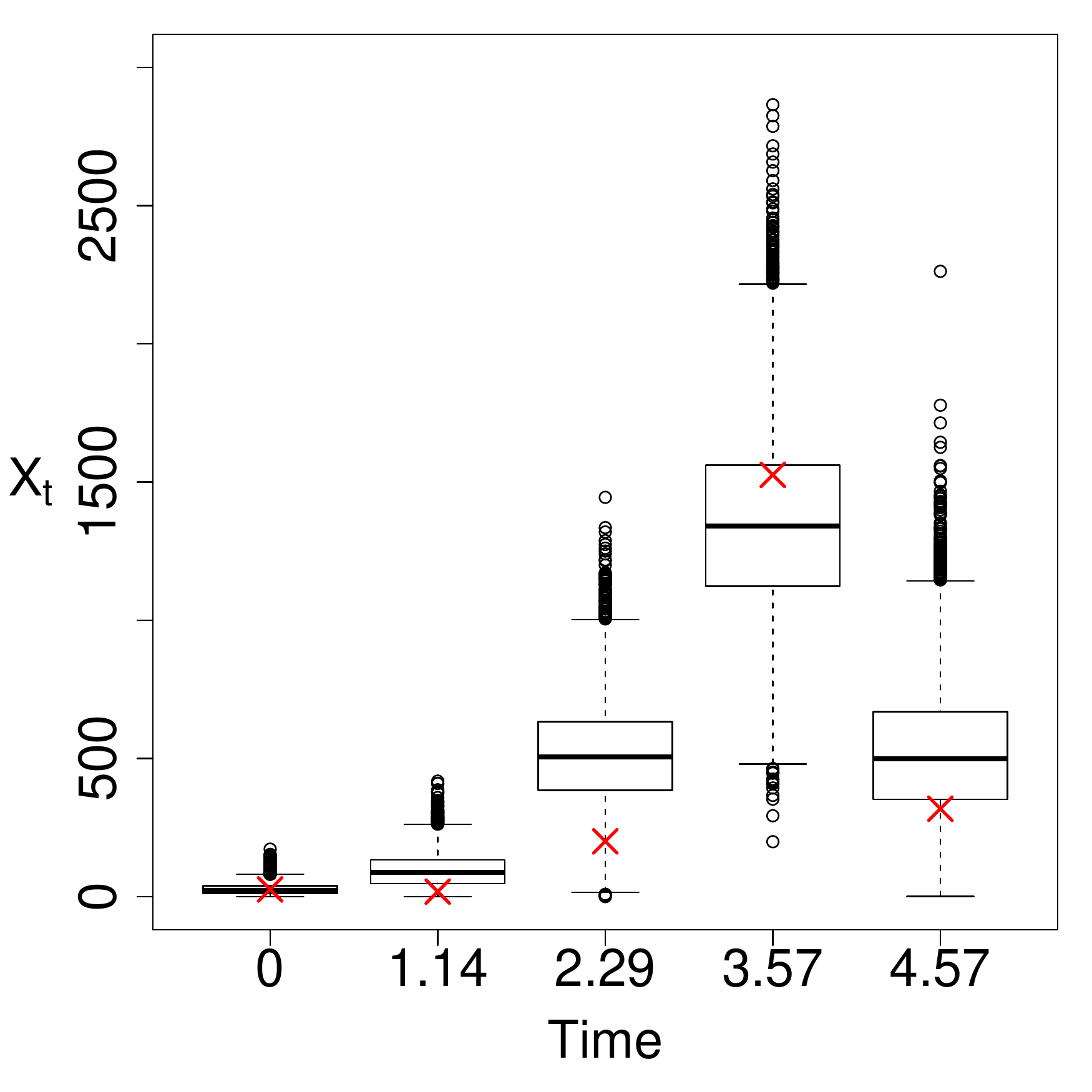}
\end{minipage}
\caption{Within sample predictive distributions for Bayesian imputation (top 2 rows) 
and LNA (bottom 2 rows). The red crosses indicate the observed values.} \label{aphid fig_fit}
\end{center}
\end{figure}

\begin{figure}
\begin{center}
\begin{minipage}[b]{0.28\linewidth}
        \centering
				\caption*{\quad $ijk=122$}\vspace{-0.25cm}
        \includegraphics[scale=0.22]{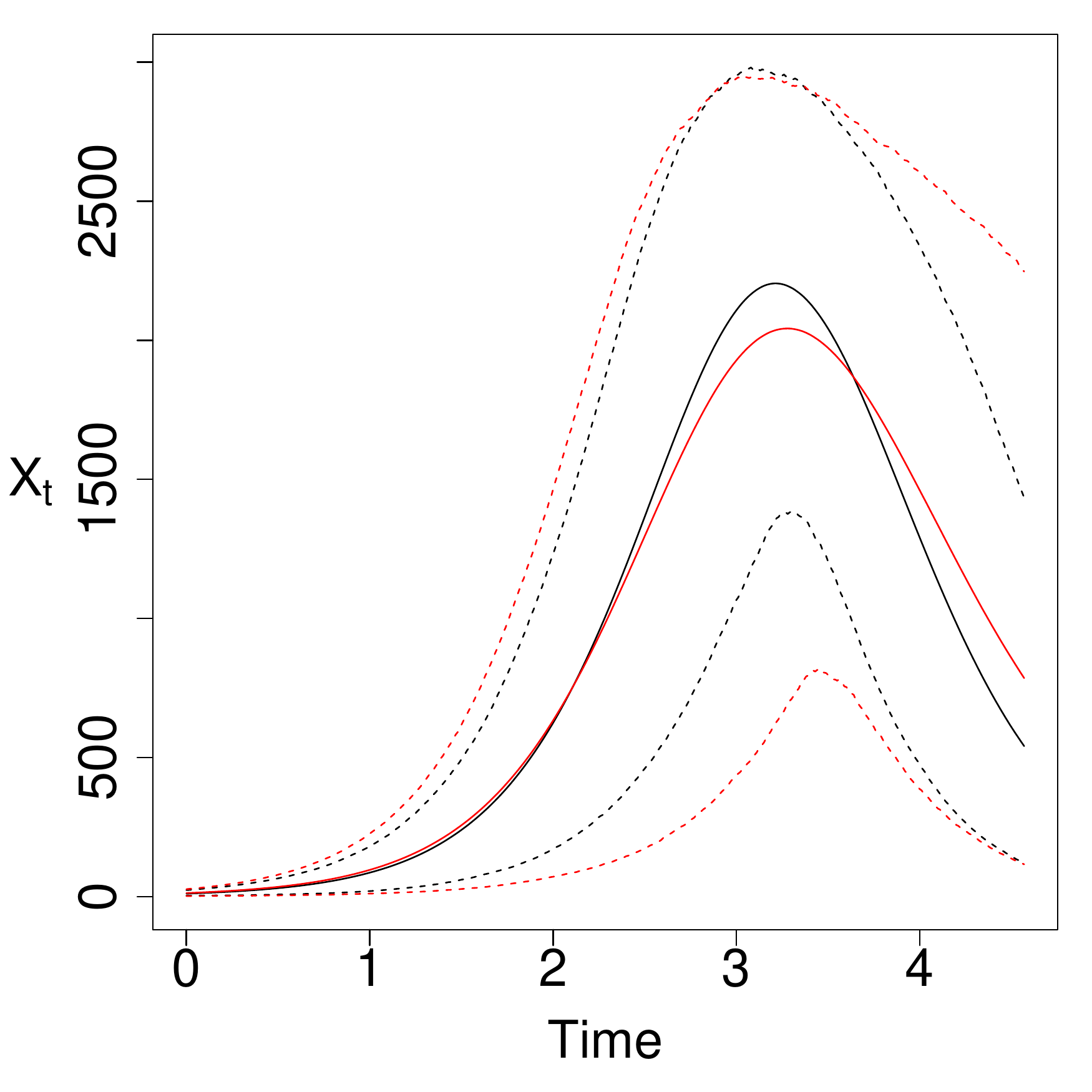}
\end{minipage} 
\hspace{-0.25cm}
\begin{minipage}[b]{0.28\linewidth}
        \centering
				\caption*{\quad $ijk=133$}\vspace{-0.25cm}
        \includegraphics[scale=0.22]{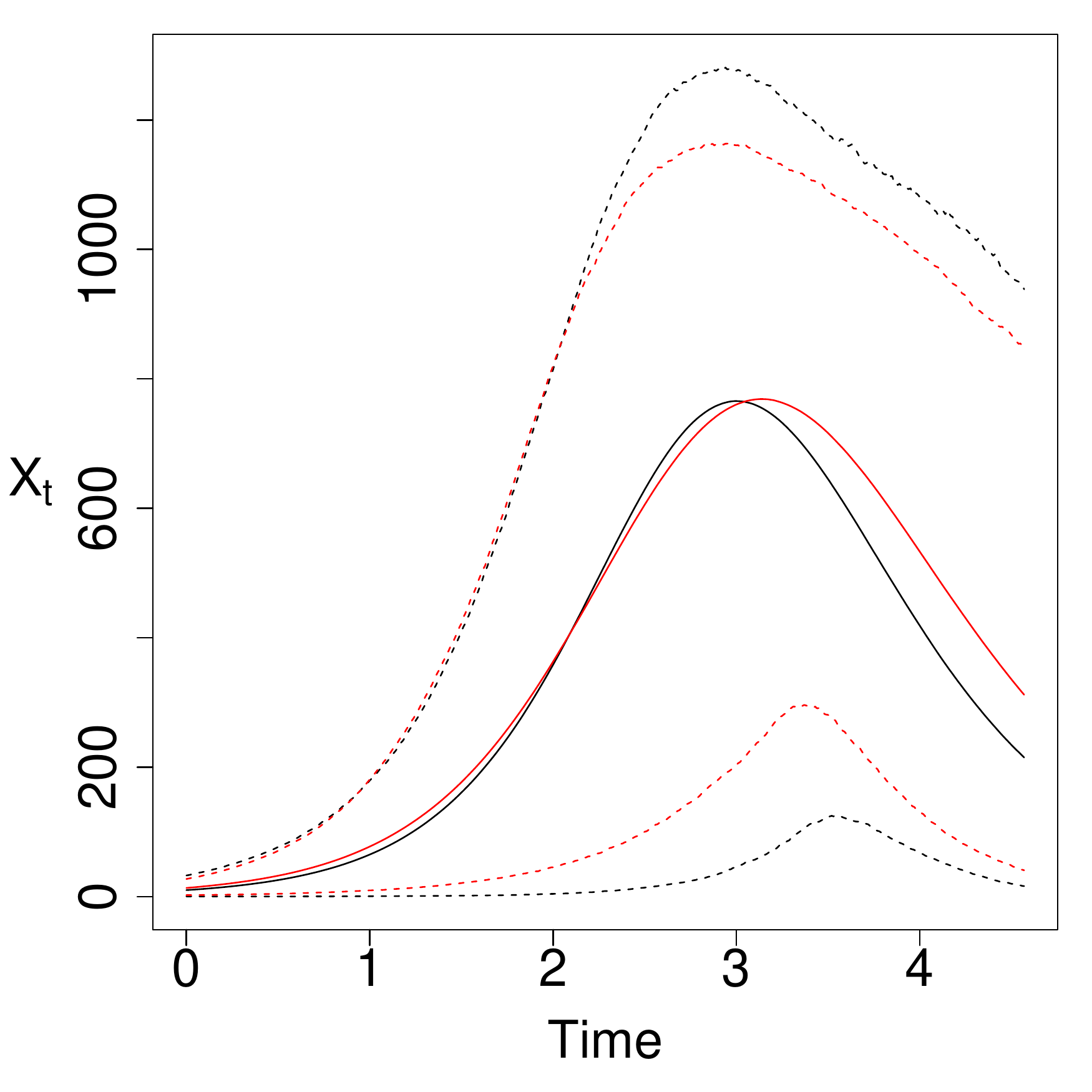}
\end{minipage} 
\hspace{-0.25cm}
\begin{minipage}[b]{0.28\linewidth}
				\centering
				\caption*{\quad $ijk=212$}\vspace{-0.25cm}
        \includegraphics[scale=0.22]{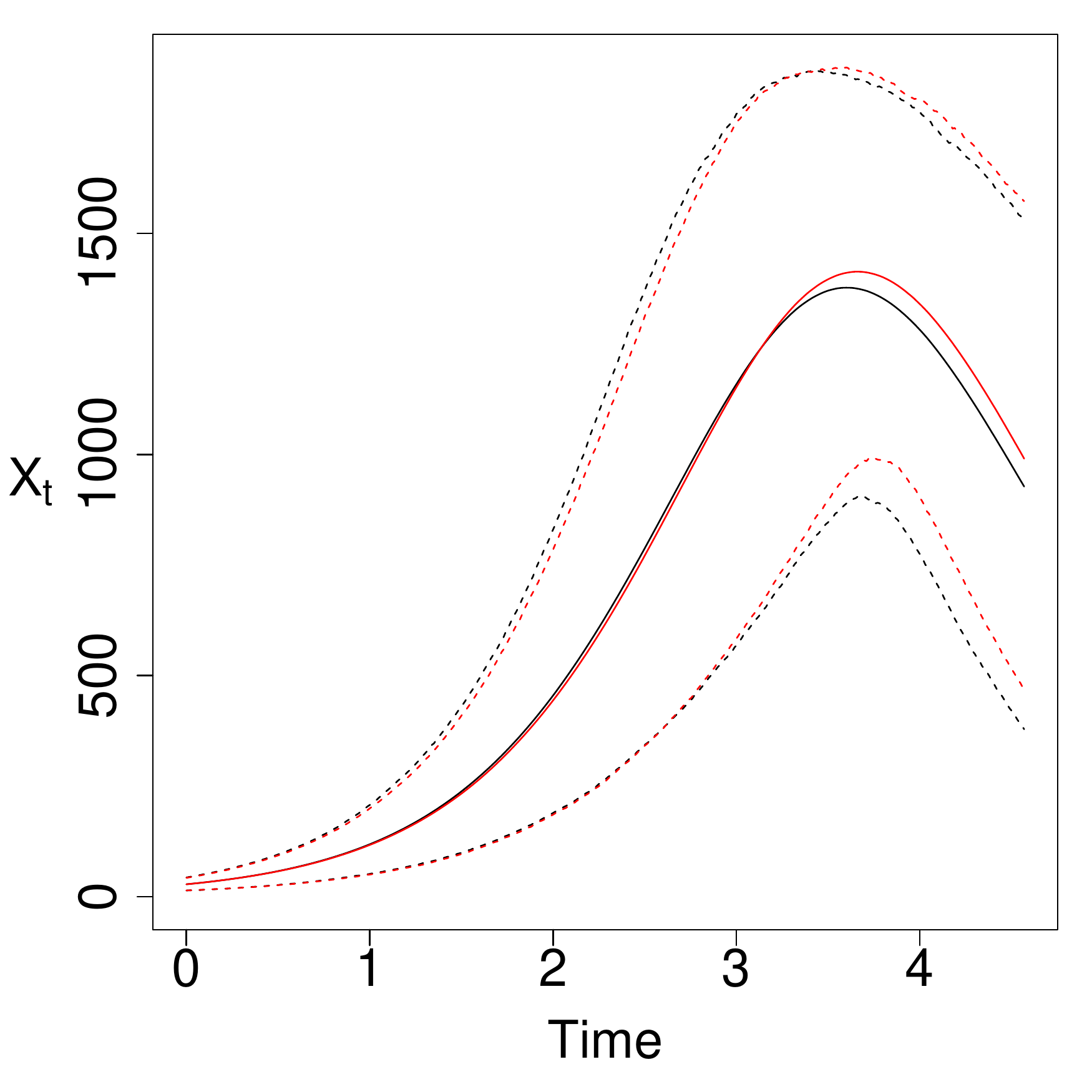}
\end{minipage}\\
\vspace{0.2cm}
\begin{minipage}[b]{0.28\linewidth}
				\centering
				\caption*{\quad $ijk=222$}\vspace{-0.25cm}
        \includegraphics[scale=0.22]{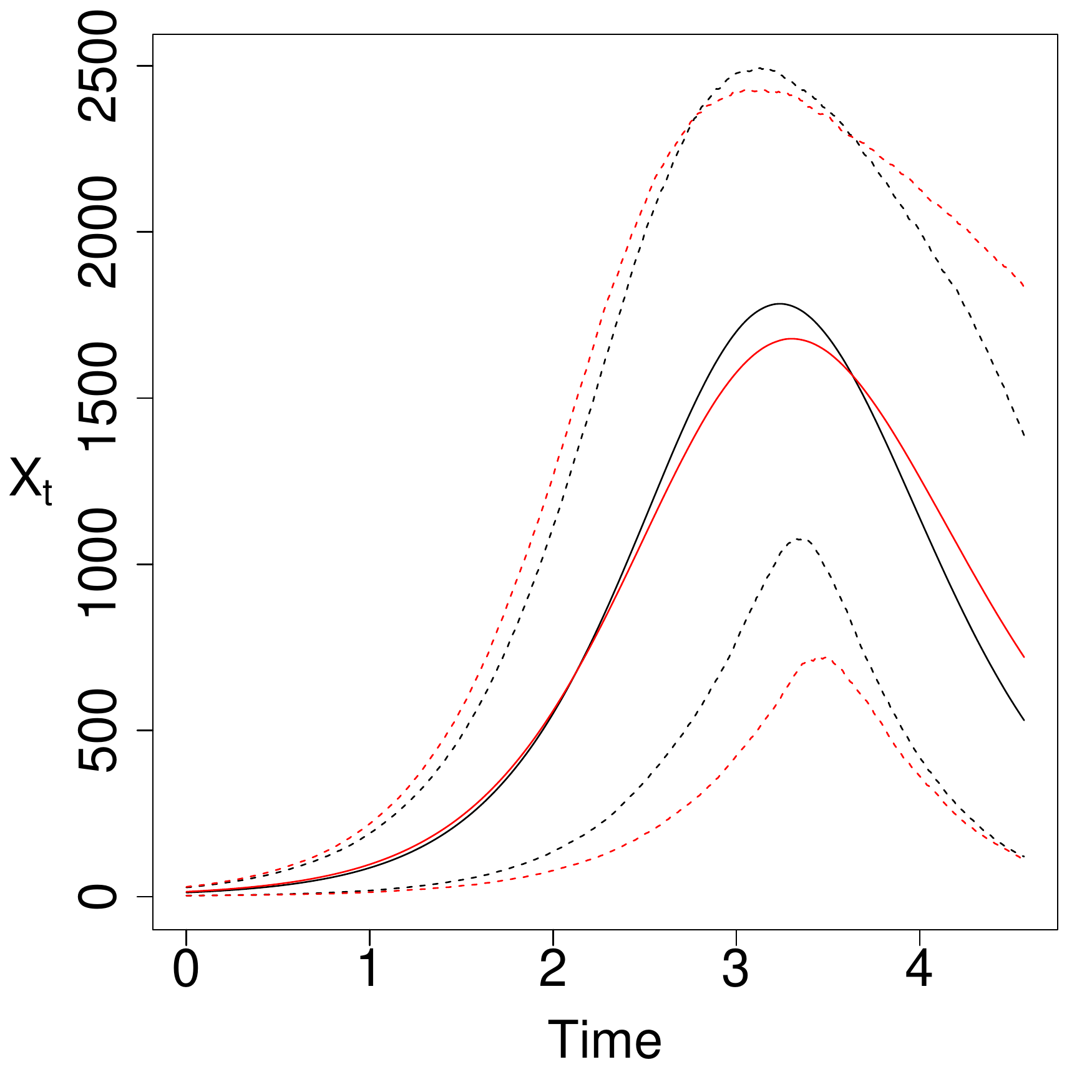}
\end{minipage}
\hspace{-0.25cm}
\begin{minipage}[b]{0.28\linewidth}
        \centering
				\caption*{\quad $ijk=311$}\vspace{-0.25cm}
        \includegraphics[scale=0.22]{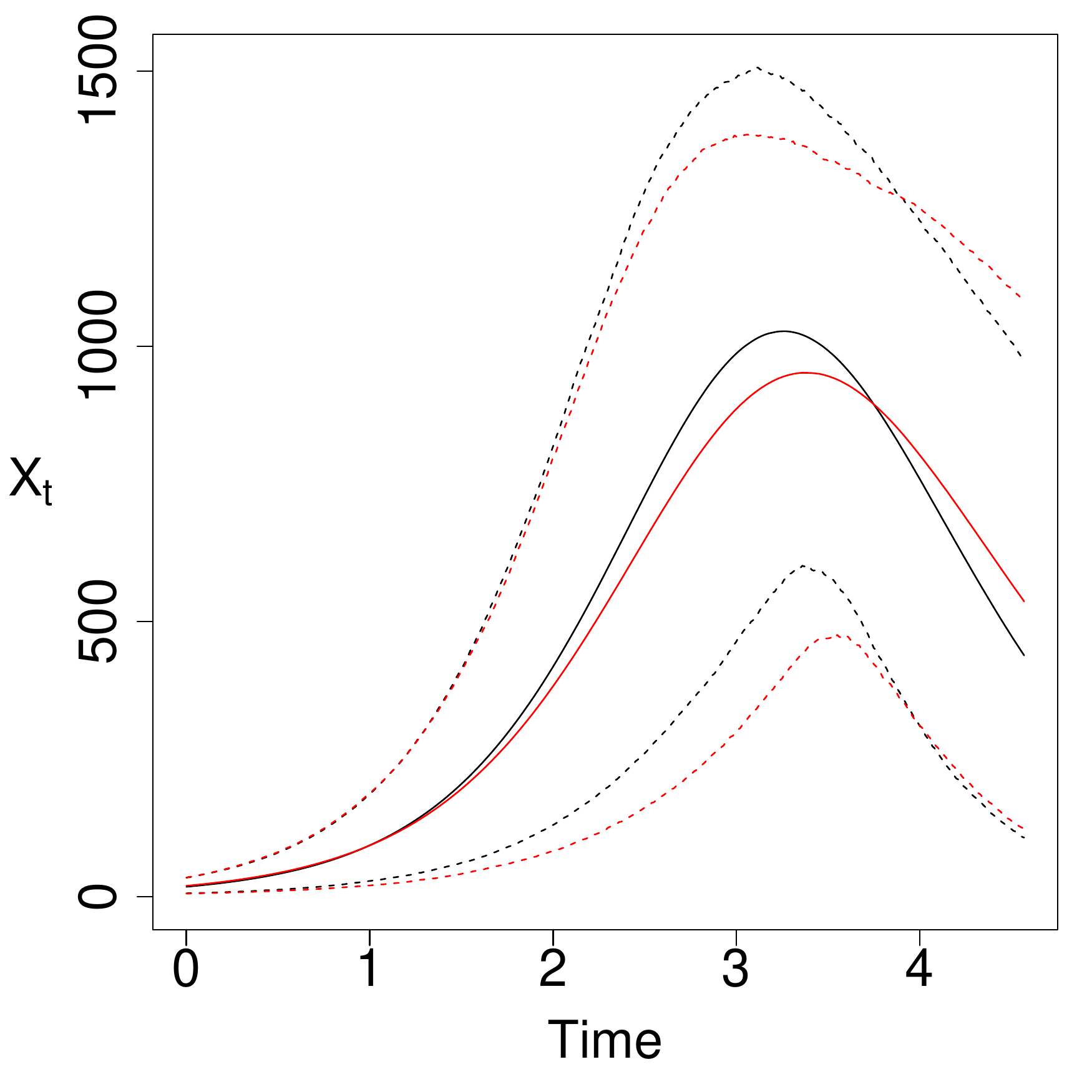}
\end{minipage}
\hspace{-0.25cm}
\begin{minipage}[b]{0.28\linewidth}
        \centering
				\caption*{\quad $ijk=332$}\vspace{-0.25cm}
        \includegraphics[scale=0.22]{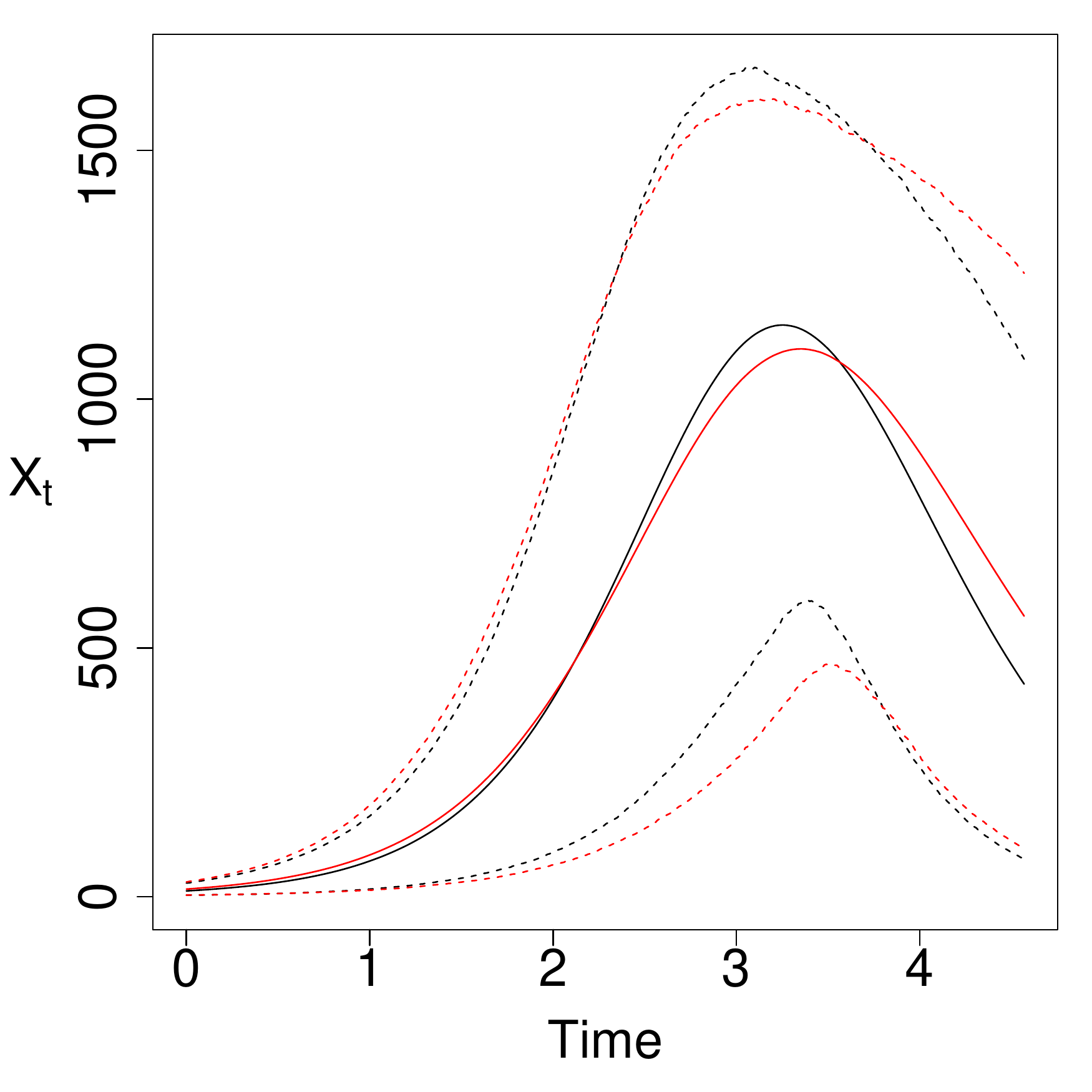}
\end{minipage}
\caption{Out-of-sample predictive intervals for the aphid population size $(N_t^{ijk})$ against time 
for a random selection of treatment combinations. The mean is depicted by the solid line with 
the dashed representing a 95\% credible region. Black: Bayesian imputation. Red: LNA.} \label{aphid fig_preds}
\end{center}
\end{figure}

\begin{figure}
\begin{center}
\begin{minipage}[b]{0.28\linewidth}
        \centering
        \includegraphics[scale=0.22]{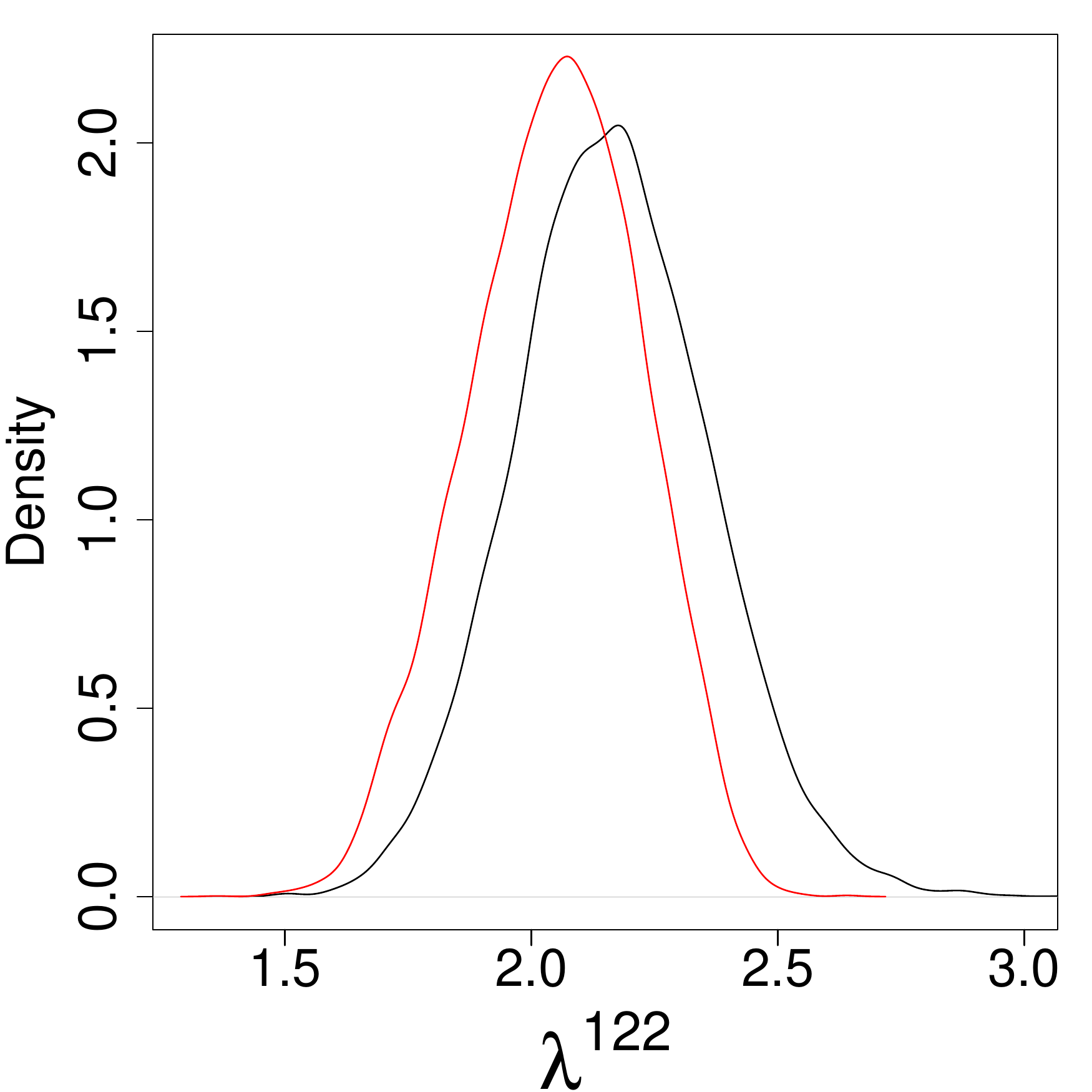}
\end{minipage} 
\hspace{-0.25cm}
\begin{minipage}[b]{0.28\linewidth}
        \centering
        \includegraphics[scale=0.22]{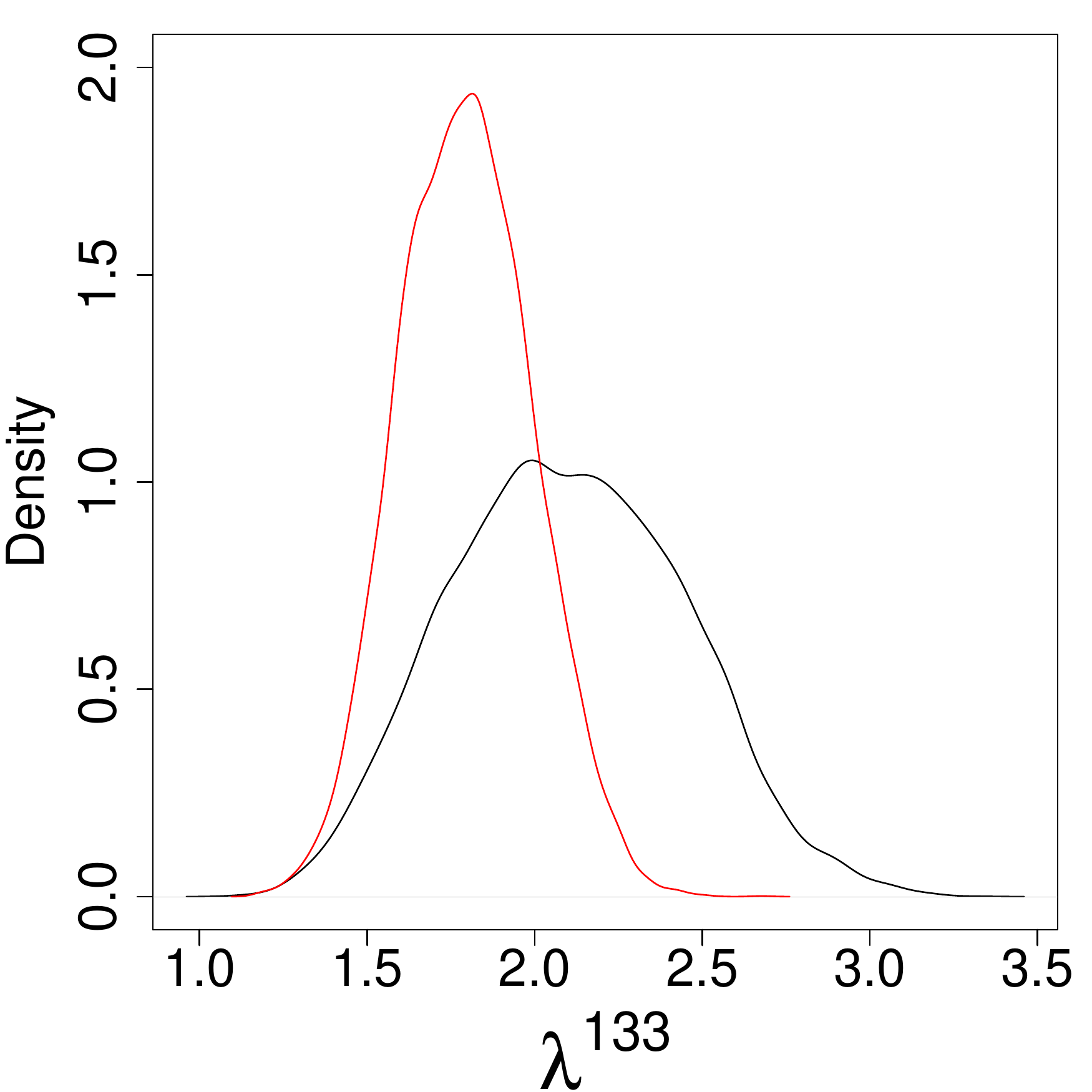}
\end{minipage} 
\hspace{-0.25cm}
\begin{minipage}[b]{0.28\linewidth}
				\centering
        \includegraphics[scale=0.22]{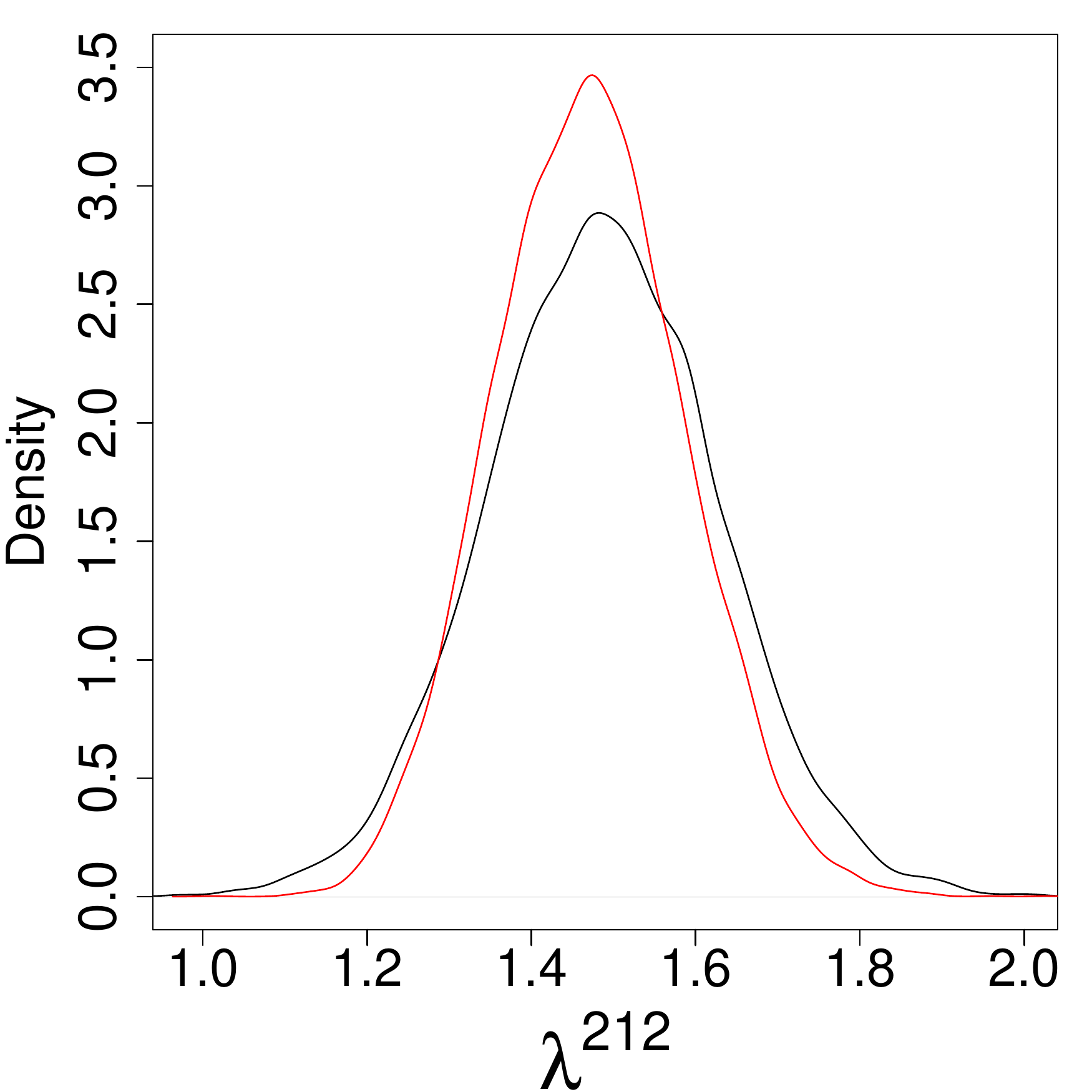}
\end{minipage}\\
\vspace{0.2cm}
\begin{minipage}[b]{0.28\linewidth}
				\centering
        \includegraphics[scale=0.22]{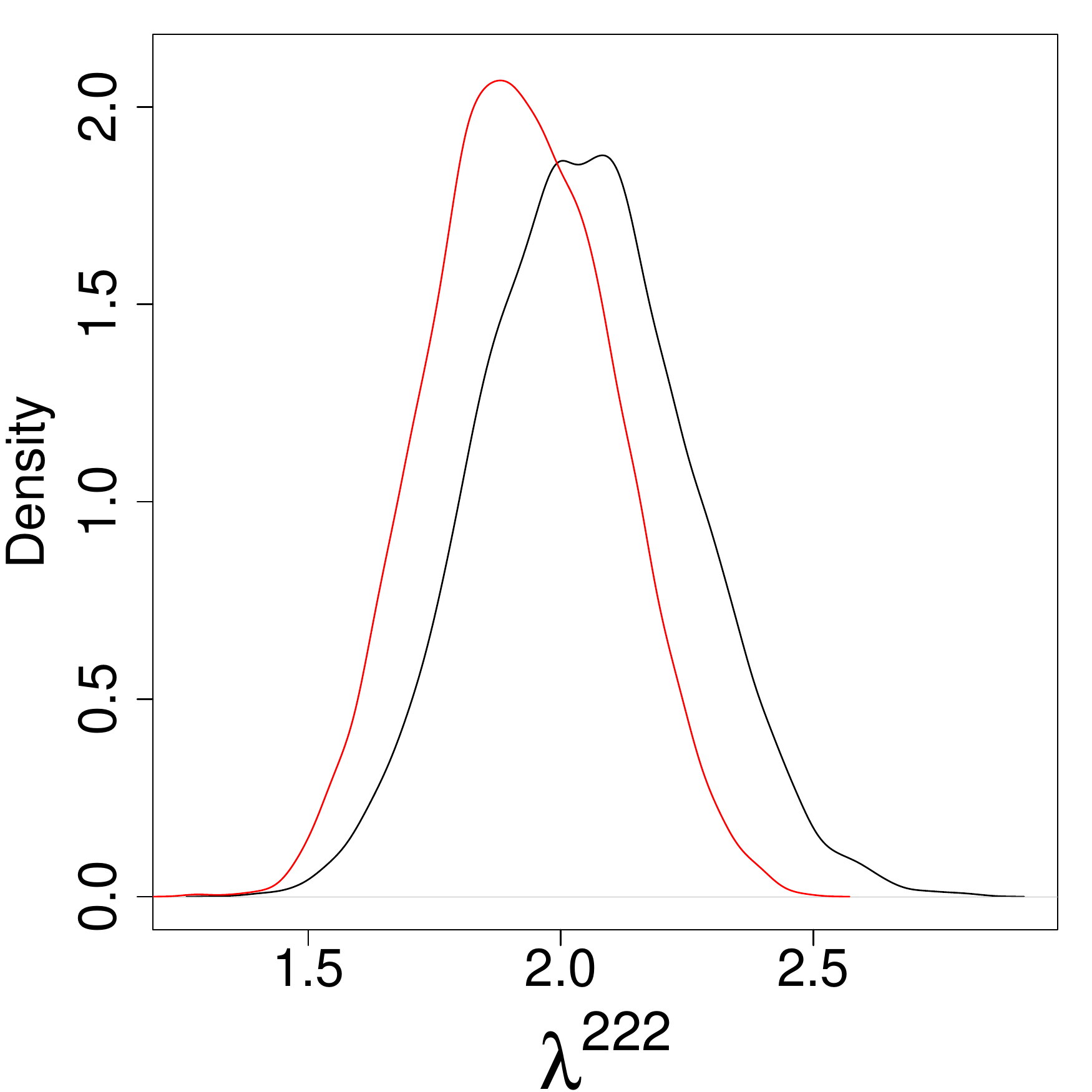}
\end{minipage}
\hspace{-0.25cm}
\begin{minipage}[b]{0.28\linewidth}
        \centering
        \includegraphics[scale=0.22]{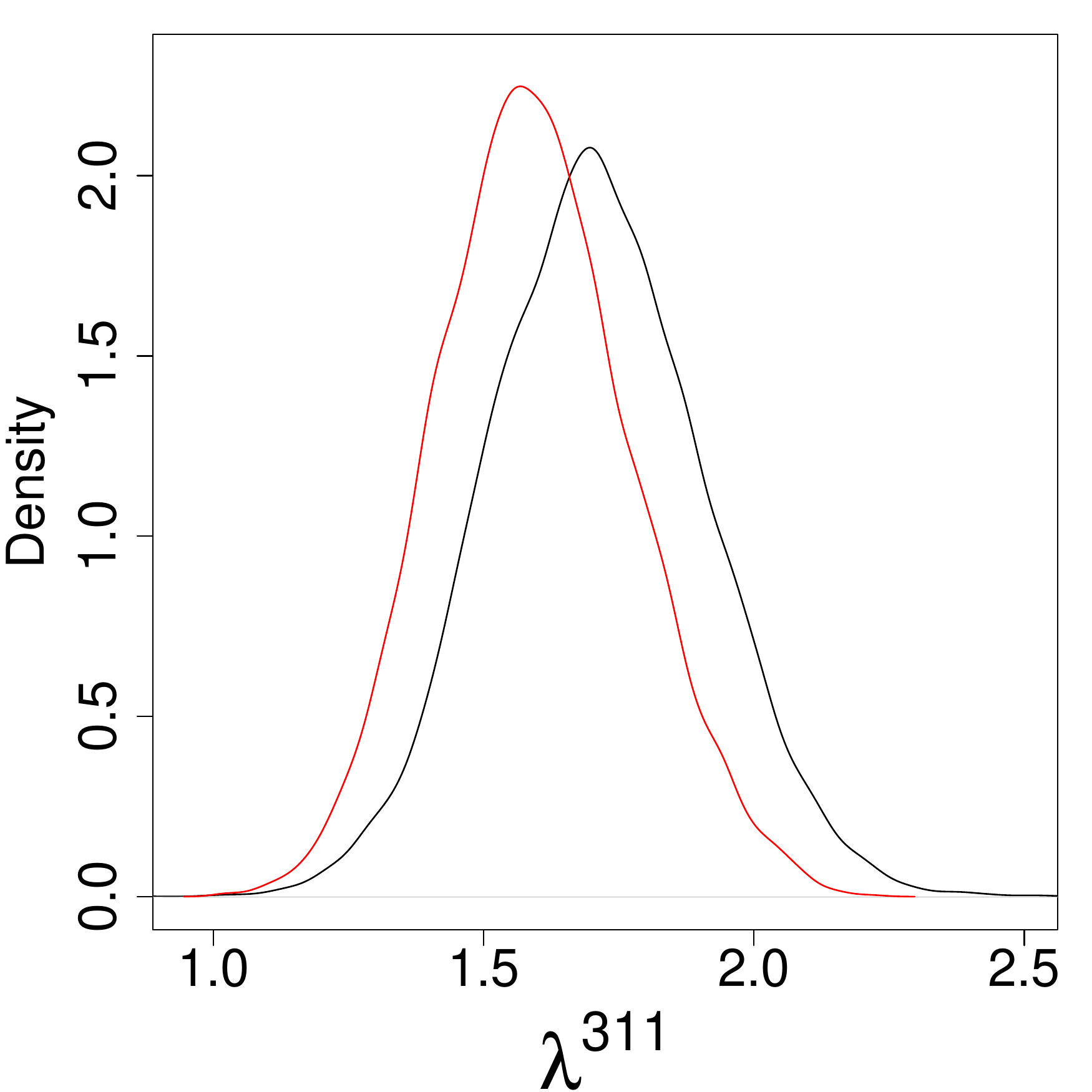}
\end{minipage}
\hspace{-0.25cm}
\begin{minipage}[b]{0.28\linewidth}
        \centering
        \includegraphics[scale=0.22]{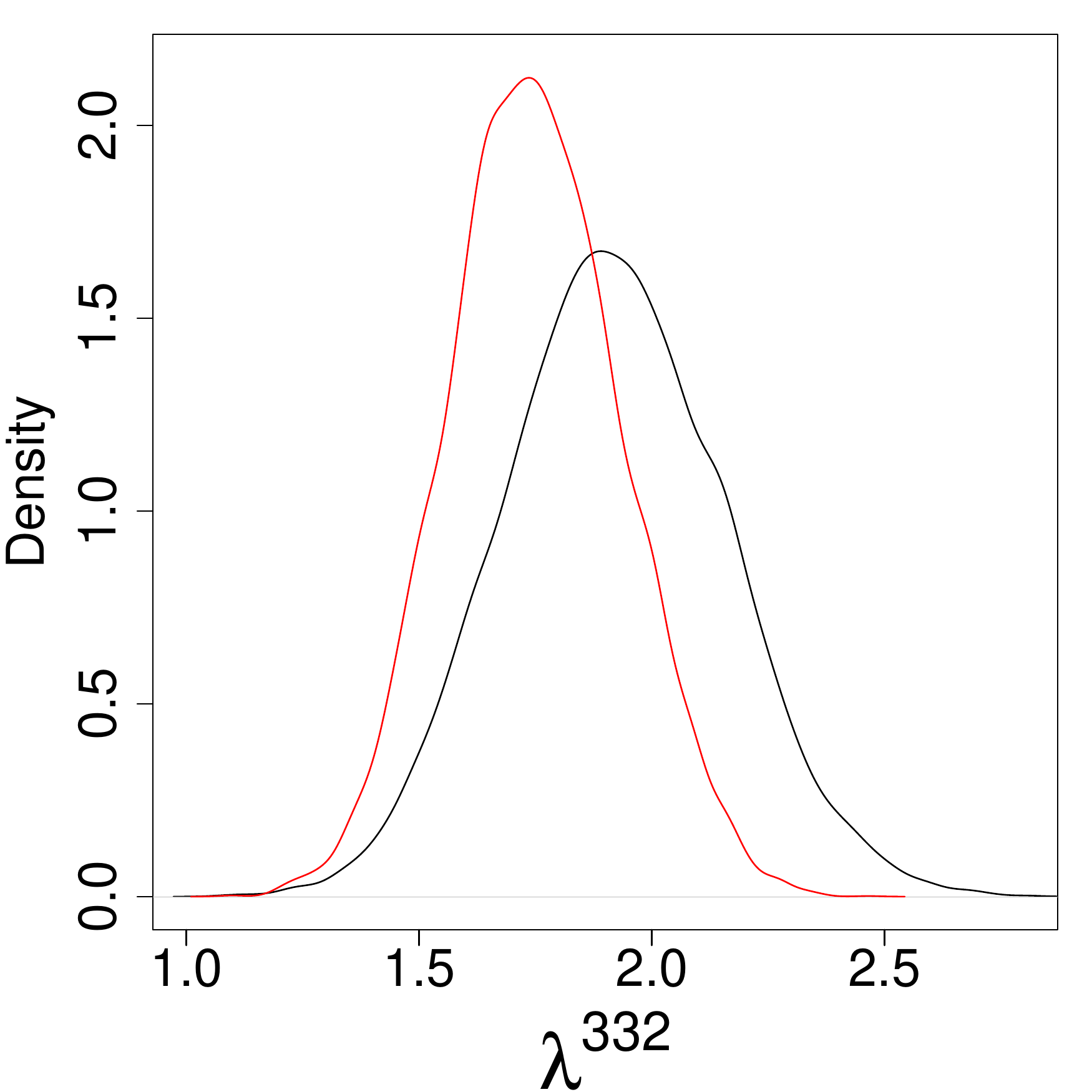}
\end{minipage}
\caption{Marginal posterior densities for a random selection of the birth rates 
associated with specific treatment combinations in the aphid model. 
Black: Bayesian imputation. Red: LNA.} \label{aphid fig_re}
\end{center}
\end{figure}

We also compared the predictive distributions obtained under each
inferential model. The within-sample predictive distribution for the
observation process $\{Y_t,t=0,\ldots,4.57\}$ can be obtained by
integrating over the posterior uncertainty of the latent process and
parameter values in the observation model
\eqref{eqn:obsAphid}. Specifically, given samples
$\{(n_t^{ijk(l)},\sigma^{(l)}),$ $l=1,\ldots,L\}$ from the marginal
posterior $\pi(n_t^{ijk},\sigma|y)$, the predictive density at time $t$
can be estimated by
\[
\frac{1}{L}\sum_{l=1}^{L}N\left(y_t\,;\,n_t^{ijk(l)},(\sigma^{(l)})^2 n_t^{ijk(l)}\right).
\]
Likewise, for a new experiment repeated under the same conditions, the
out-of-sample predictive distribution for the aphid population size
can be determined for each treatment combination. This is estimated by
averaging realisations of $N_t$ (obtained by applying the
Euler-Maruyama approximation to (\ref{eqn:aphidSDE})) over draws from
the marginal posterior $\pi(n_0^{ijk},b^{ijk}|y)$ obtained using
either Bayesian imputation or the LNA. Figures~\ref{aphid fig_fit} and
\ref{aphid fig_preds} summarise these predictive distributions for a
random selection of treatment combinations. Both the SDMEM and LNA
give a satisfactory fit to the observed data, with all observations
within or close to the central $50\%$ of the distribution, and no
observation outside the equi-tailed $95\%$ credible intervals.  As
expected, the SDMEM gives a better fit over the LNA, although there is
little difference between the two. There are however noticeable
differences in the out-of-sample predictives, especially in the lower 
credible bound (in Figure~\ref{aphid fig_preds}) 
suggesting that in some situations, using the inferences
made under the LNA to predict the outcome of future experiments can
give misleading results. These differences lead us to examine the 
marginal posterior densities of the treatment-block specific birth and death 
rates, $\lambda^{ijk}$ and $\mu^{ijk}$, over whose uncertainty we average. 
Samples from these posteriors are straightforward to obtain, using the 
posterior samples of the constituent parameters in (\ref{aphid_lambdamu}). 
Figure~\ref{aphid fig_re} shows marginal posterior densities of 
the overall birth rates ($\lambda^{ijk}$) associated with the six 
treatment-block combinations for which predictives are presented in 
Figure~\ref{aphid fig_preds}. We see distinct differences between 
posteriors obtained under the Bayesian imputation approach and the LNA approach. 
The posteriors displayed are indicative of those obtained for all 
treatment combinations. Moreover, similar patterns are evident in the overall 
death rates ($\mu^{ijk}$).

We obtained a minESS of 1039 under the modified innovation
scheme. The LNA, however, clearly benefits from analytically
integrating out the latent process and gave a minESS of 8908.  For
this example, we found that significant gains in computational
efficiency were possible by performing the parameter updates and, for
the modified innovation scheme, the path updates, in parallel. For
example, updating $\lambda_{B_2}$ and $\mu_{B_2}$ involves calculating
a product of likelihoods (or marginal likelihoods for the LNA) over
all $3^2=9$ treatment combinations that include block 2. These
constituent likelihoods can be calculated in parallel. Similarly, for
the modified innovation scheme, the treatment specific path updates
can be performed in parallel. Both the modified innovation scheme and
the LNA--based scheme were again coded in C and run on a high
performance computing cluster with 14 cores (made up of Intel Xeon
3.0GHz processors). The modified innovation scheme took 
approximately 18 days to run whereas the LNA--based scheme
required only 
approximately 4.3 days. Note that here the speed advantage of the
LNA--based scheme has reduced, now being roughly~4 times faster than
the modified innovation scheme, whereas in Section~\ref{orange}, the
LNA was approximately 20 times faster. The intractability of the ODEs
driving the LNA clearly plays a significant role in computational
efficiency. In terms of overall efficiency (as measured by minESS/sec)
the LNA--based scheme outperforms the Bayesian imputation approach by
a factor of around 36. These computational advantages of the LNA must 
be weighted against the inaccuracies of the resulting posterior and 
predictive distributions, inaccuracies which can at times be substantial, 
as demonstrated by the simulation study in the supplementary material.

\section{Discussion}\label{disc}

We have provided a framework that permits (simulation-based) Bayesian
inference for a large class of multivariate SDMEMs using discrete-time
observations that may be incomplete and subject to measurement
error. By adopting a Bayesian imputation approach, we have shown how
the modified innovation scheme of \cite{Golightly_Wilkinson_2008},
which is necessary for overcoming the problematic dependence between
the latent process and any parameters that enter the diffusion
coefficient, can be applied to SDMEMs.  Fundamental to our approach is
the development of a novel bridge construct that can be used to sample
a discretisation of a conditioned diffusion process, and does not
break down when the process exhibits strong nonlinearity over 
inter-observation times of interest. The computational cost of the 
Bayesian imputation scheme is dictated by the number of 
imputed points (characterised by $m$) between observation times. 
In the examples considered here we see little difference in 
posterior output under the Bayesian imputation scheme for 
$m\geq 20$. 

We also considered a tractable approximation to the SDMEM, the linear
noise approximation, and provided a systematic comparison using two
applications. The computational efficiency of the LNA depends on the
dimension of the SDE driving the SDMEM. For a $d$-dimensional SDE
system, the LNA requires the solution of a system of order $d^2$
coupled ODEs. In our first application, the resulting ODE system can
be solved analytically, leading to increases in both computational and
overall efficiency (as measured by minimum ESS per second) of around a
factor of 20. Moreover, we found little difference in the accuracy of
inferences made under the LNA and imputation approaches. In our second
application, we fitted the diffusion approximation of a Markov jump
process description of aphid dynamics using data from
\cite{Matis_2008}. In this example, the ODE system governing the LNA
is intractable and the computational advantage of using the LNA over
an imputation approach reduced to around a factor of 4.  However, the
benefit of using the LNA to analytically integrate over the latent
process is clear, giving an overall increase in efficiency of around a
factor of 36. It is important to note that whilst the LNA is preferred
in terms of overall efficiency for the examples considered here, as
the dimension $d$ of the SDE is increased, the LNA is likely to become
infeasible. Moreover, whilst both the imputation and LNA approaches
provided a reasonable fit to the aphid data, differences were found
between the parameter posteriors, leading to differences in the
out-of-sample predictive distributions. A simulation study (given in the supplementary material) 
highlighted further differences between the LNA and Bayesian imputation approaches. 
Care must therefore be taken in trying to fit the SDMEM by 
using an LNA--based inference approach.

\subsection*{Acknowledgements}
The authors would like to thank the
associate editor and two anonymous referees for their suggestions which 
improved the paper.

\bibliographystyle{apalike}
\bibliography{arxiv_paper_final}

\newpage

\title{Supplementary material for Bayesian inference for diffusion-driven mixed-effects models}
\author{Gavin A.~Whitaker$^1$ \and\ Andrew Golightly$^1$ \and\ Richard J.~Boys$^1$ \and\ Chris Sherlock$^2$}
\date{\small $^{1}$School of Mathematics \& Statistics, Newcastle University,\\
  Newcastle-upon-Tyne, NE1 7RU, UK \\
$^{2}$Department of Mathematics and Statistics, Lancaster University, Lancaster, LA1 4YF}

\maketitle


\appendix
\numberwithin{equation}{section}
\setcounter{figure}{0}

\section{LNA FFBS algorithm}\label{a-lna-ffbs}

To ease the notation, consider a single experimental unit and drop $i$
from the notation. Since the parameters $\theta$, $\psi$, $b$ and
$\Sigma$ remain fixed throughout this section, we also drop them from
the notation where possible. Define $y_{0:j}=(y_{t_0},
\ldots,y_{t_j})^T$. Now suppose that $X_0\sim N(a,C)$ \emph{a priori}.
The marginal likelihood under the LNA, $\pi(y|\theta,\Sigma,b)$, can
be obtained from the forward filter described below. After execution of the 
forward filter, realisations of $\pi(x| y,\theta,\Sigma,b )$ can be generated using a backward sampler. 
Note that the backward sweep requires
\begin{align*}
Cov(X_{t_{j+1}},X_{t_j})&=Cov(\tilde{R}_{t_{j+1}},\tilde{R}_{t_j})
=Cov(P_{t_{j+1}}\tilde{R}_{t_j},\tilde{R}_{t_j})
= P_{t_{j+1}}Var(\tilde{R}_{t_j}).
\end{align*}
Here $P_t$ is a $d\times d$ matrix that can be shown to satisfy the
ODE
\begin{equation}
\dfrac{dP_t}{dt} = H_tP_t, \label{eqn:lna_p}
\end{equation}
with initial condition $P_0=I_d$, the $d\times d$ identity matrix. 

\begin{enumerate}
\item Forward filter. Initialisation. Compute
  $\pi(y_{t_0})=N(y_{t_{0}}\,;\, F^Ta\,,\,F^TCF+\Sigma)$. The
  posterior at time $t_0=0$ is therefore $X_{t_0}|y_{t_0}\sim
  N(a_0,C_0)$, where
\begin{align*}
a_0 &= a+CF(F^TCF+\Sigma)^{-1}(y_{t_{0}}-F^Ta) \\
C_0 &= C-CF(F^TCF+\Sigma)^{-1}F^TC\,.
\end{align*}
Store the values of $a_0$ and $C_0$.

\item For $j=0,1,\ldots,n-1$,
\begin{itemize}
\item[(a)] Prior at $t_{j+1}$. Initialise the LNA with $\eta_{t_j}=a_{t_j}$, 
$V_{t_j}=C_{t_j}$ and $P_{t_j}=I_d$. Integrate the ODEs (11), 
(22) and \eqref{eqn:lna_p} forward to $t_{j+1}$ to obtain 
$\eta_{t_{j+1}}$, $V_{t_{j+1}}$ and $P_{t_{j+1}}$. Hence
$X_{t_{j+1}}|y_{0:j+1}\sim N(\eta_{t_{j+1}},V_{t_{j+1}})$.
\item[(b)] One step forecast. Using the observation equation, we have that 
\[
Y_{t_{j+1}}|y_{0:j}\sim N(F^T\eta_{t_{j+1}},F^TV_{t_{j+1}}F+\Sigma).
\]
Compute the updated marginal likelihood
\begin{align*}
\pi(y_{0:j+1})&=\pi(y_{0:j})\pi(y_{t_{j+1}}|y_{0:j})
=\pi(y_{0:j})N(y_{t_{j+1}}\,;\, F^T\eta_{t_{j+1}}\,,\,F^TV_{t_{j+1}}F+\Sigma).
\end{align*}
\item[(c)] Posterior at $t_{j+1}$. Combining the distributions in (a) and (b) gives the joint 
distribution of $X_{t_{j+1}}$ and $Y_{t_{j+1}}$ (conditional on $y_{0:j}$) as
\[
\begin{pmatrix}
	X_{t_{j+1}} \\	
	Y_{t_{j+1}}
	\end{pmatrix}\sim N\left\{\begin{pmatrix}
	\eta_{t_{j+1}} \\
	F^T\eta_{t_{j+1}} 	
	\end{pmatrix}\,,\, \begin{pmatrix}
	V_{t_{j+1}} & V_{t_{j+1}}F  \\
	F^TV_{t_{j+1}} & F^TV_{t_{j+1}}F+\Sigma  	 
	\end{pmatrix} \right \} 
\]
and therefore $X_{t_{j+1}}|y_{0:j+1}\sim N(a_{t_{j+1}},C_{t_{j+1}})$, where
\begin{align*}
a_{t_{j+1}} &= \eta_{t_{j+1}}+V_{t_{j+1}}F(F^TV_{t_{j+1}}F+\Sigma)^{-1}(y_{t_{j+1}}-F^T\eta_{t_{j+1}}) \\
C_{t_{j+1}} &= V_{t_{j+1}}-V_{t_{j+1}}F(F^TV_{t_{j+1}}F+\Sigma)^{-1}F^TV_{t_{j+1}}\,.
\end{align*}
Store the values of $a_{t_{j+1}}$, $C_{t_{j+1}}$, $\eta_{t_{j+1}}$, $V_{t_{j+1}}$ and $P_{t_{j+1}}$.
\end{itemize}
\end{enumerate}
We sample $\pi(x| y)$ using a backward sampler. The algorithm is as follows. 
\begin{enumerate}
\item Backward sampler. First draw $x_{t_n}$ from $X_{t_n}|y\sim
  N(a_{t_n},C_{t_n})$.
\item For $j=n-1,n-2,\ldots,0$,
\begin{itemize}
\item[(a)] Joint distribution of $X_{t_j}$ and $X_{t_{j+1}}$. Note
  that $X_{t_j}|y_{0:j}\sim N(a_{t_j},C_{t_j})$. The joint
  distribution of $X_{t_j}$ and $X_{t_{j+1}}$ (conditional on
  $y_{0:j}$) is 
\[
\begin{pmatrix}
	X_{t_{j}} \\	
	X_{t_{j+1}}
	\end{pmatrix}\sim N\left\{\begin{pmatrix}
	a_{t_{j}} \\
	\eta_{t_{j+1}} 	
	\end{pmatrix}\,,\, \begin{pmatrix}
	C_{t_{j}} & C_{t_{j}}P_{t_{j+1}}^T  \\
	P_{t_{j+1}}C_{t_{j}} & V_{t_{j+1}}  	 
	\end{pmatrix} \right \}. 
\]
\item[(b)] Backward distribution. The distribution of
  $X_{t_j}|X_{t_{j+1}},y_{0:j}$ is
  $N(\tilde{a}_{t_j},\tilde{C}_{t_j})$, where
\begin{align*}
\tilde{a}_{t_j} &= a_{t_j}+C_{t_j}P_{t_{j+1}}^TV_{t_{j+1}}^{-1}(x_{t_{j+1}}-\eta_{t_{j+1}}), \\
\tilde{C}_{t_j} &= C_{t_j}-C_{t_j}P_{t_{j+1}}^TV_{t_{j+1}}^{-1}P_{t_{j+1}}C_{t_{j}}.
\end{align*}
Draw $x_{t_j}$ from $X_{t_j}|X_{t_{j+1}},y_{0:j}\sim
N(\tilde{a}_{t_j},\tilde{C}_{t_j})$.
\end{itemize} 
\end{enumerate}

\section{Additional graphics from the orange tree growth example}

\begin{figure}
\begin{center}
\begin{minipage}[b]{0.28\linewidth}
        \centering
        \includegraphics[scale=0.22]{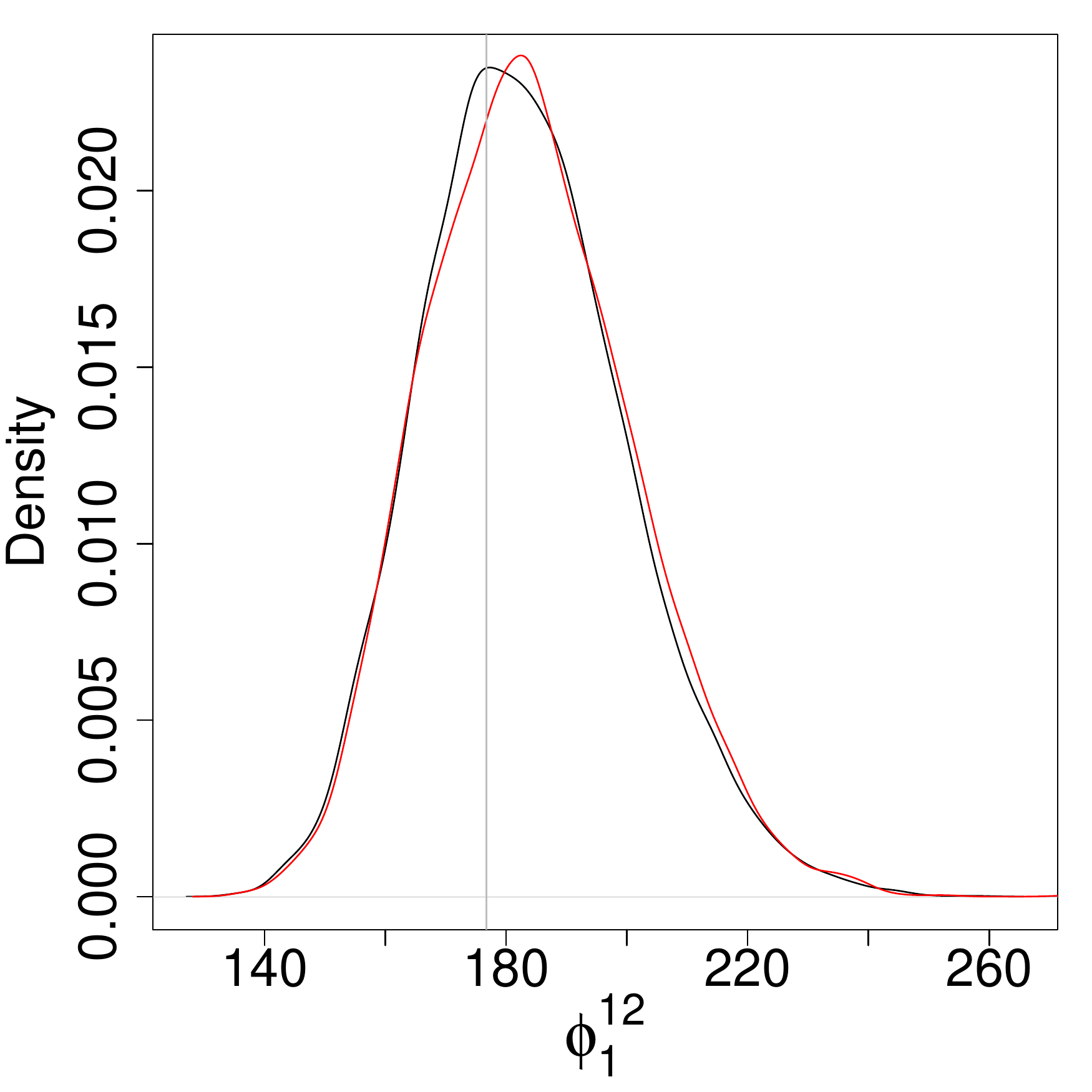}
\end{minipage} 
\hspace{-0.25cm}
\begin{minipage}[b]{0.28\linewidth}
        \centering
        \includegraphics[scale=0.22]{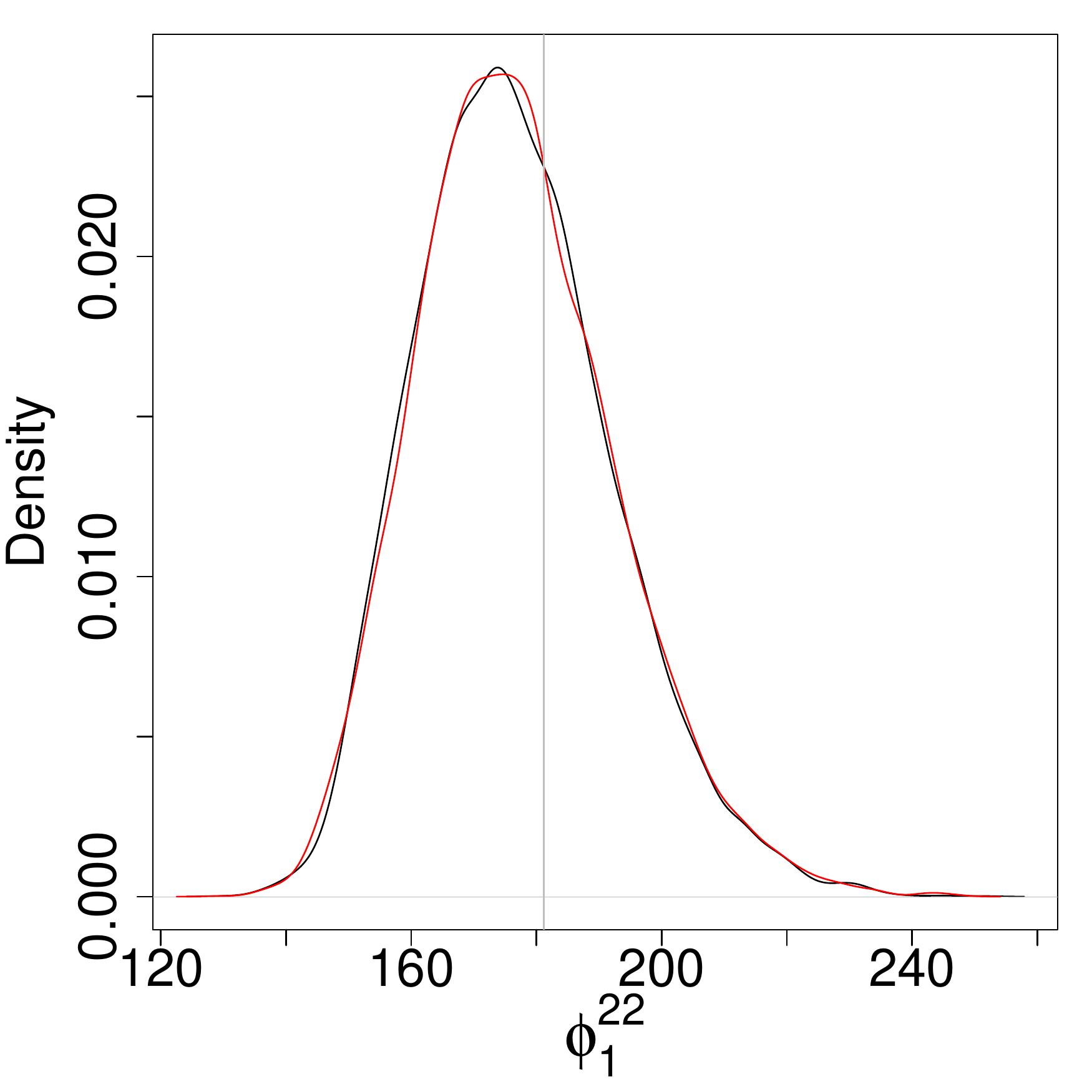}
\end{minipage} 
\hspace{-0.25cm}
\begin{minipage}[b]{0.28\linewidth}
	\centering
        \includegraphics[scale=0.22]{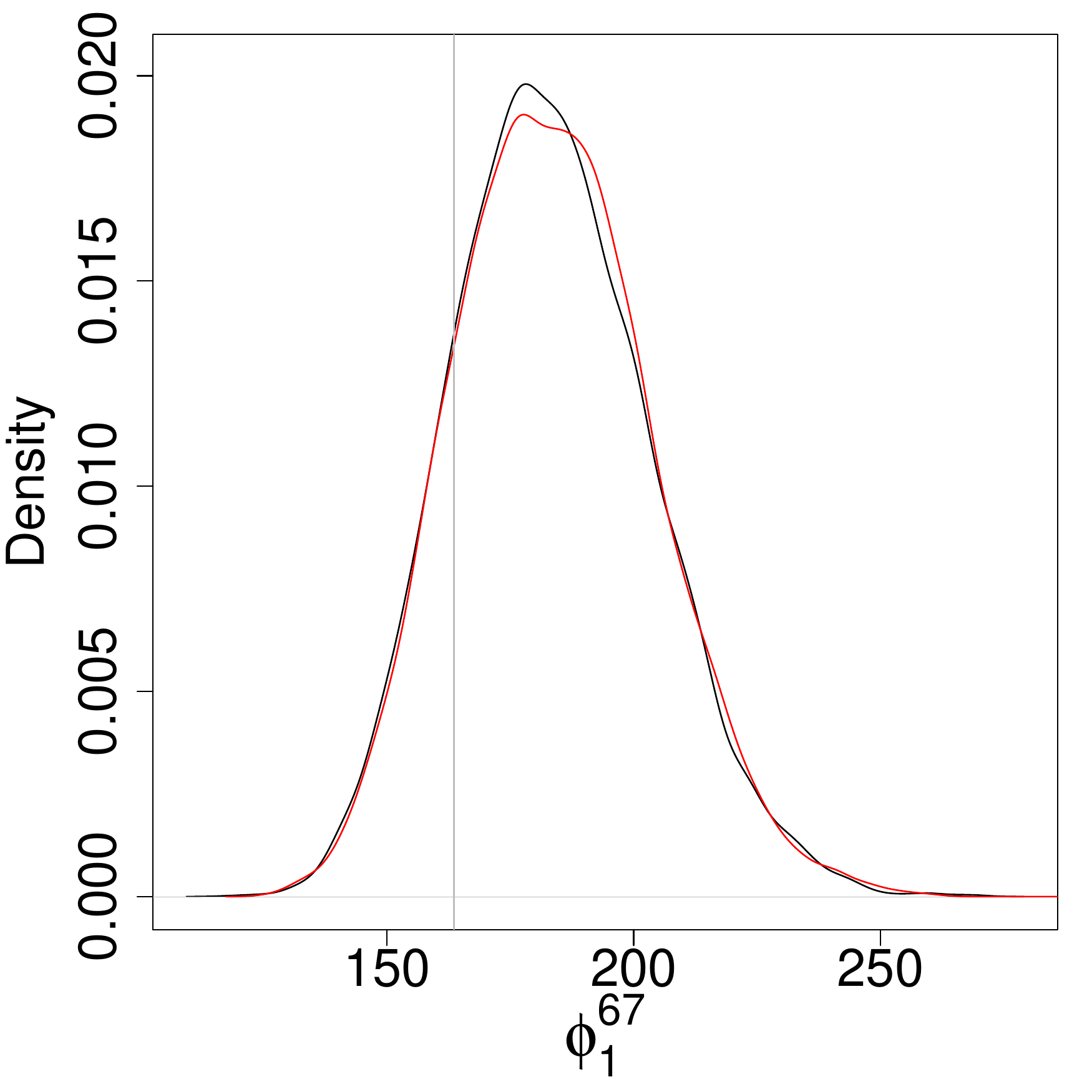}
\end{minipage}\\
\vspace{0.2cm}
\begin{minipage}[b]{0.28\linewidth}
	\centering
        \includegraphics[scale=0.22]{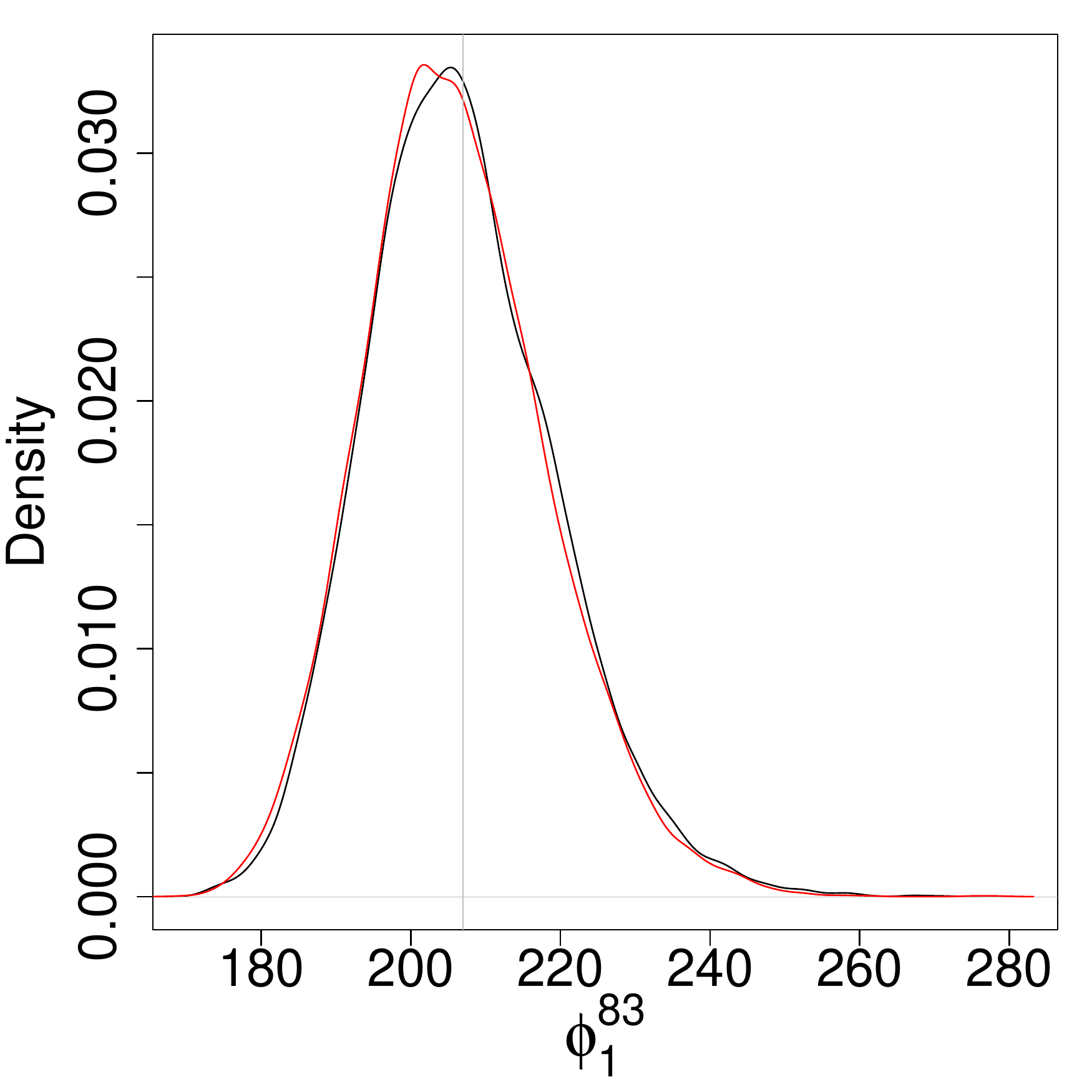}
\end{minipage}
\hspace{-0.25cm}
\begin{minipage}[b]{0.28\linewidth}
        \centering
        \includegraphics[scale=0.22]{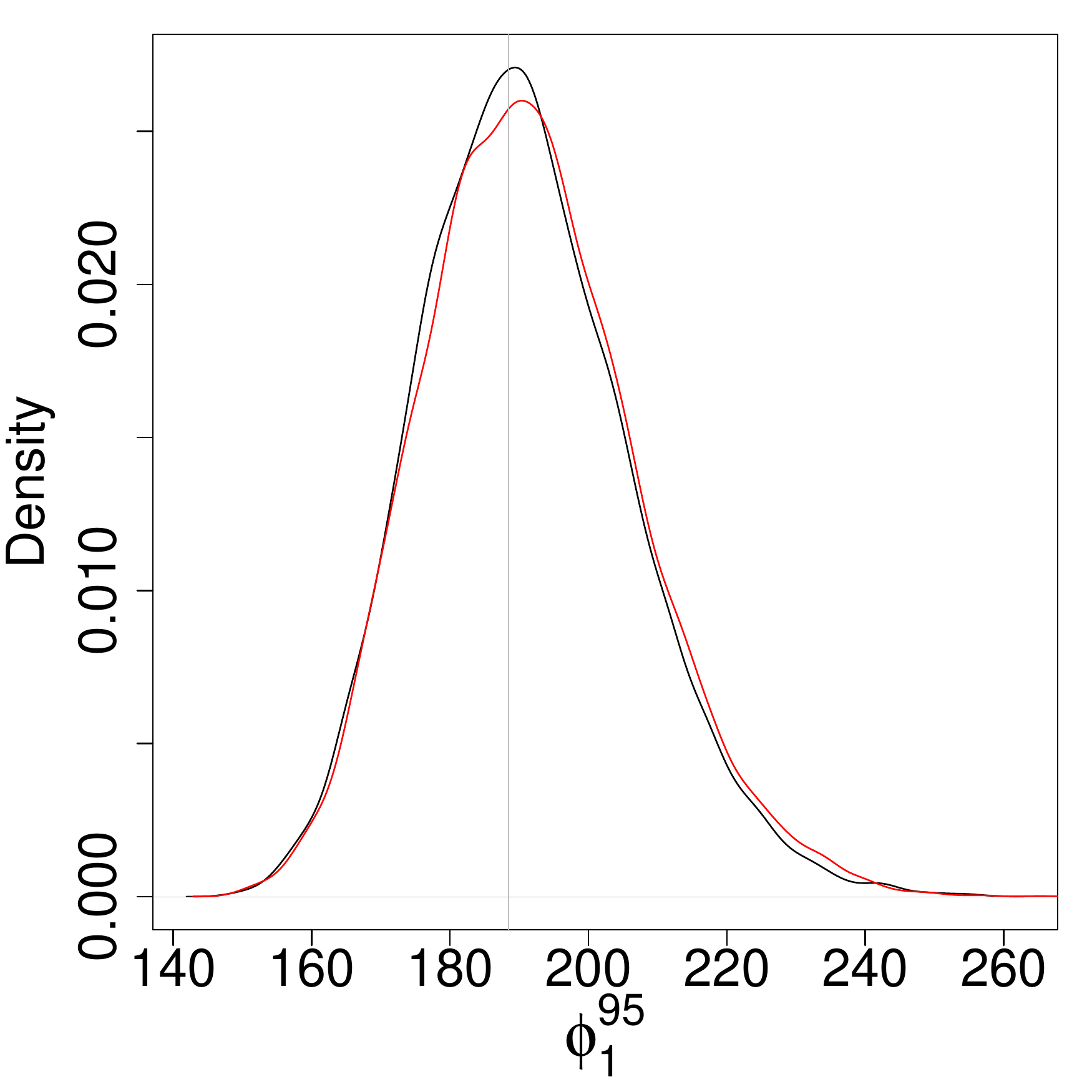}
\end{minipage}
\hspace{-0.25cm}
\begin{minipage}[b]{0.28\linewidth}
        \centering
        \includegraphics[scale=0.22]{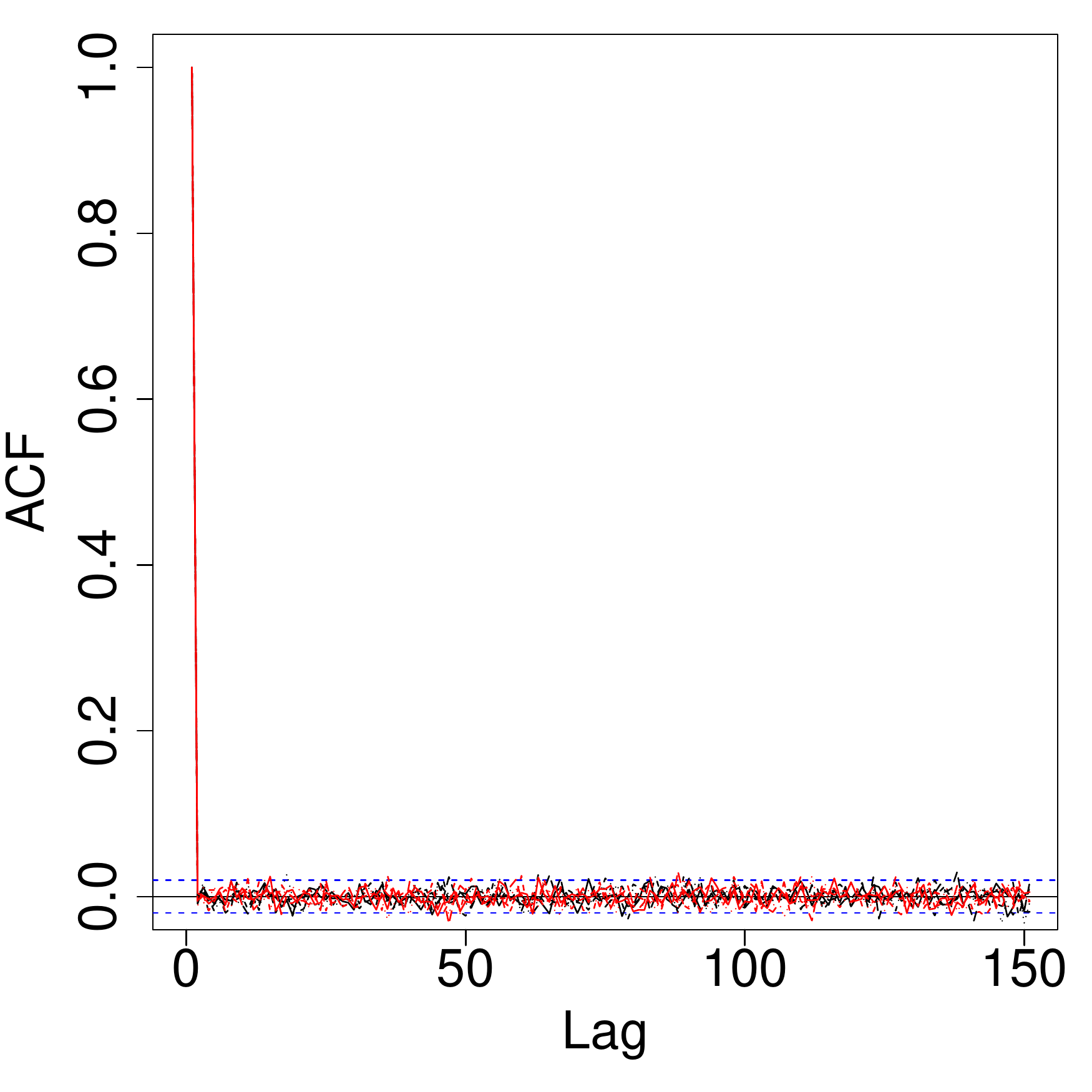}
\end{minipage}\\
\vspace{0.8cm}
\begin{minipage}[b]{0.28\linewidth}
        \centering
        \includegraphics[scale=0.22]{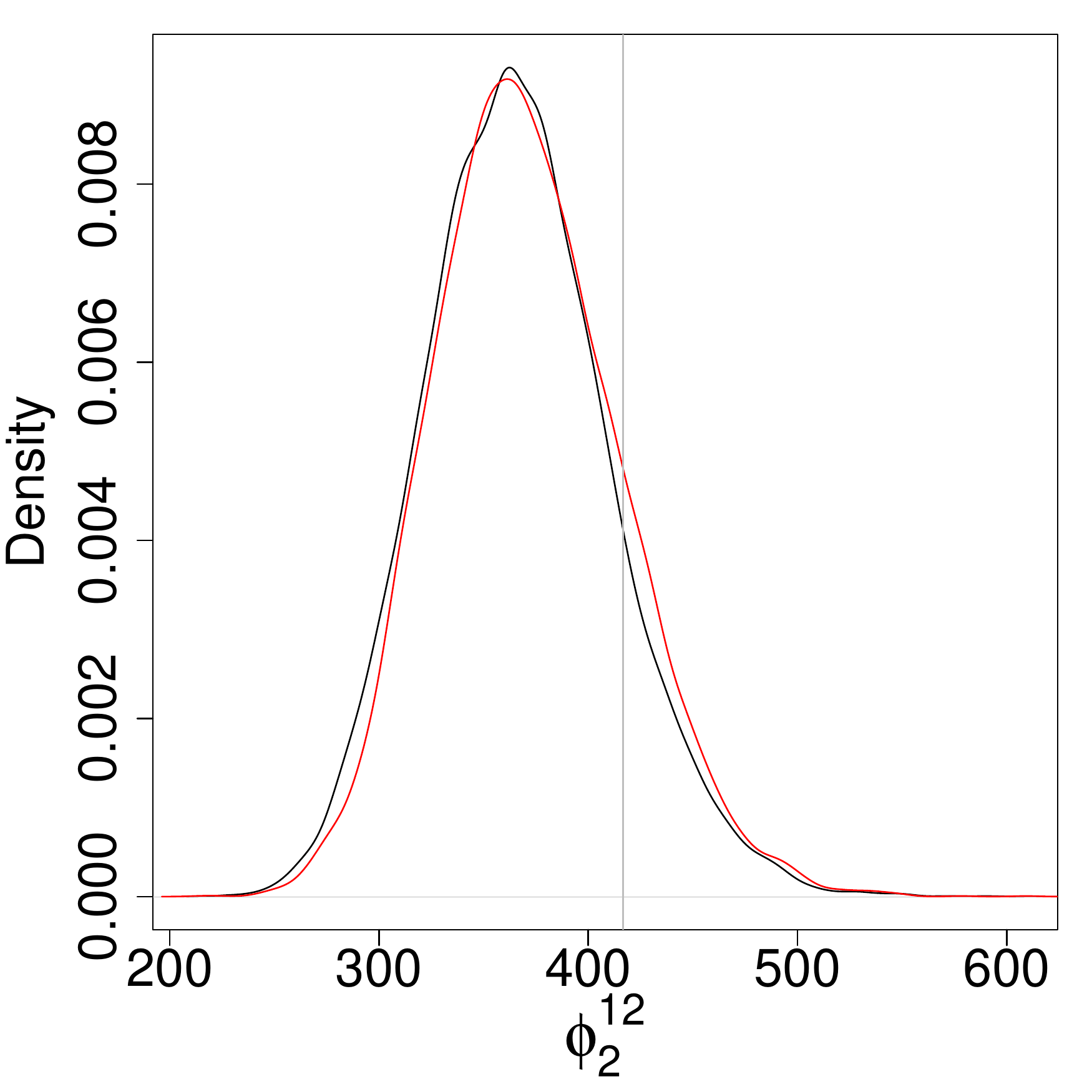}
\end{minipage} 
\hspace{-0.25cm}
\begin{minipage}[b]{0.28\linewidth}
        \centering
        \includegraphics[scale=0.22]{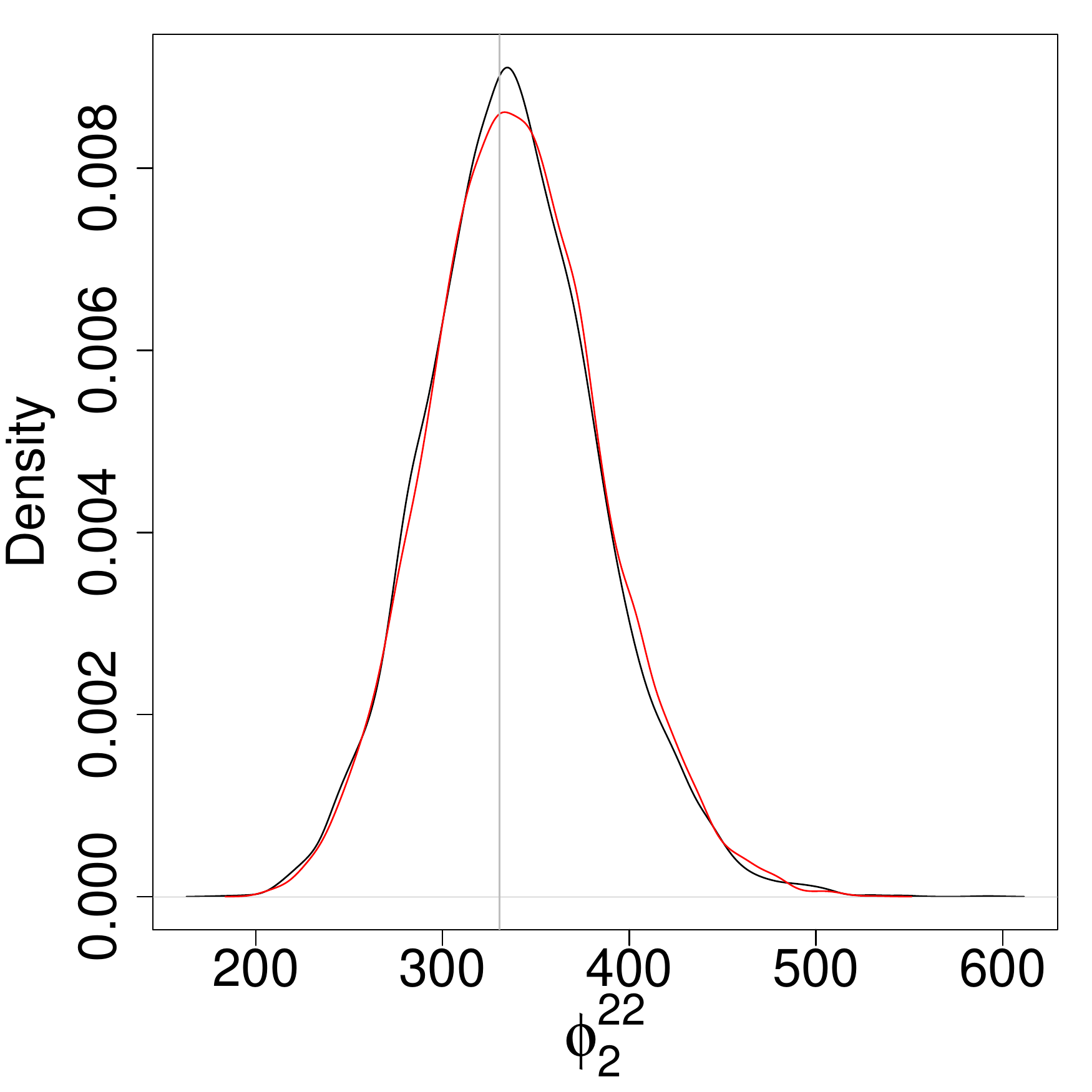}
\end{minipage} 
\hspace{-0.25cm}
\begin{minipage}[b]{0.28\linewidth}
	\centering
        \includegraphics[scale=0.22]{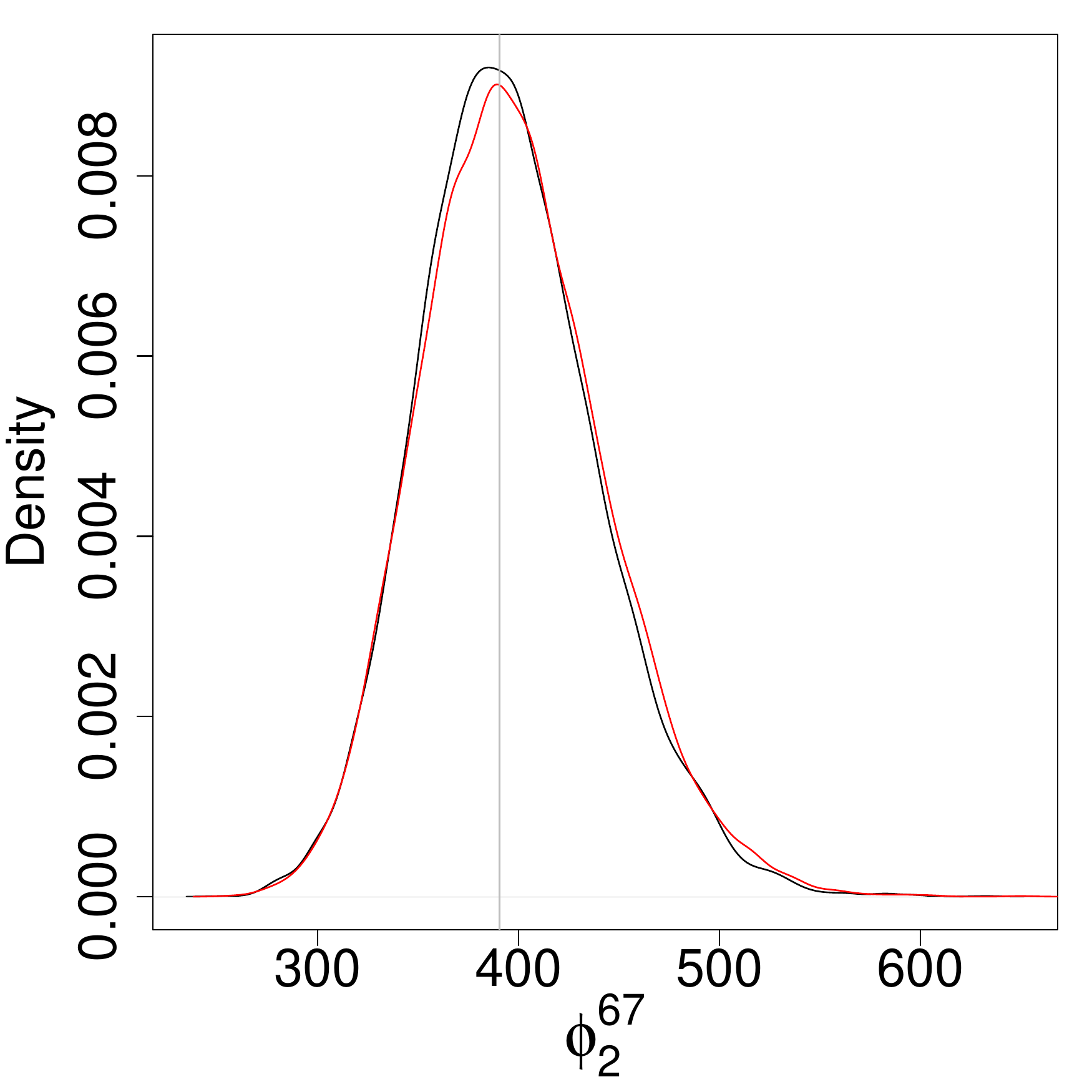}
\end{minipage}\\
\vspace{0.2cm}
\begin{minipage}[b]{0.28\linewidth}
	\centering
        \includegraphics[scale=0.22]{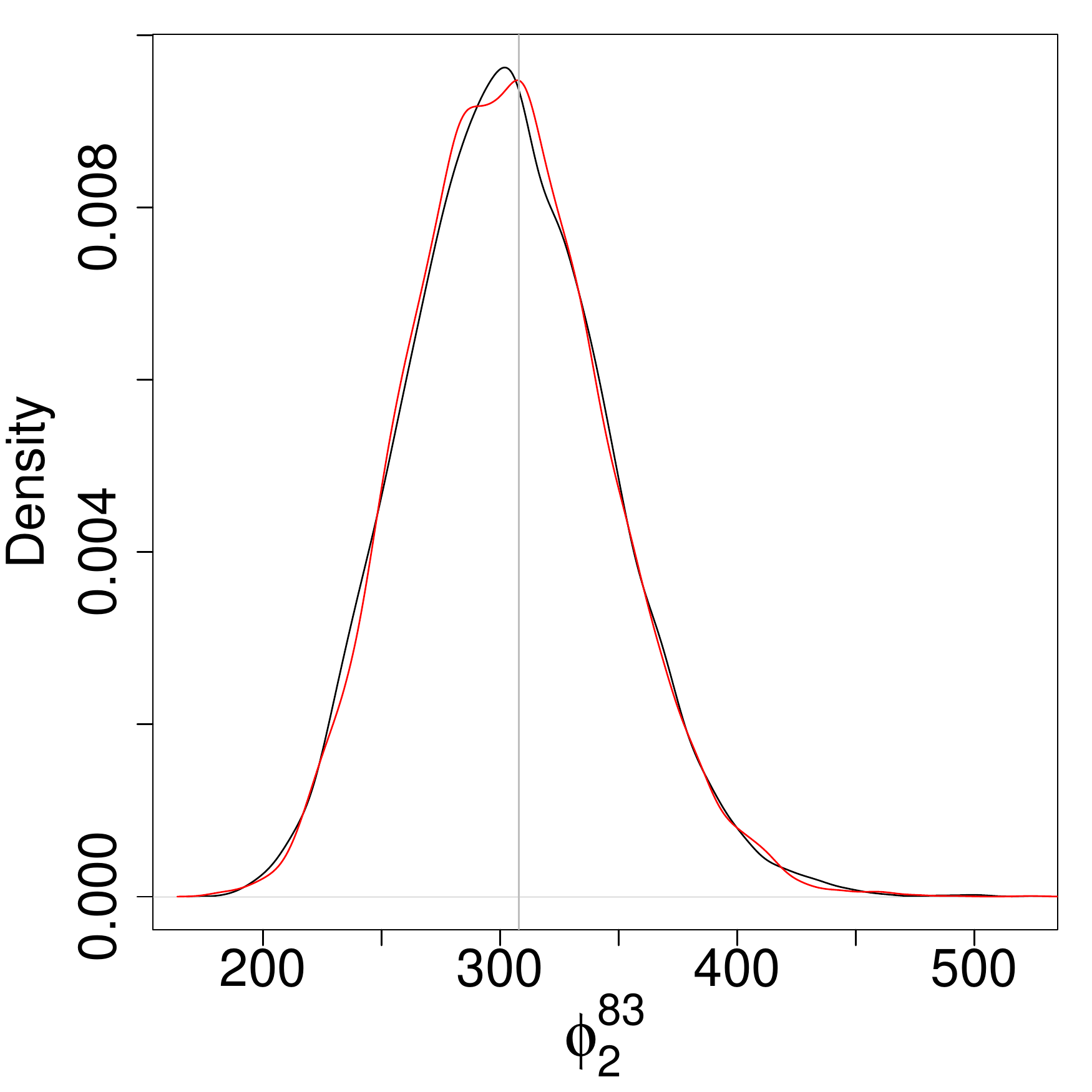}
\end{minipage}
\hspace{-0.25cm}
\begin{minipage}[b]{0.28\linewidth}
        \centering
        \includegraphics[scale=0.22]{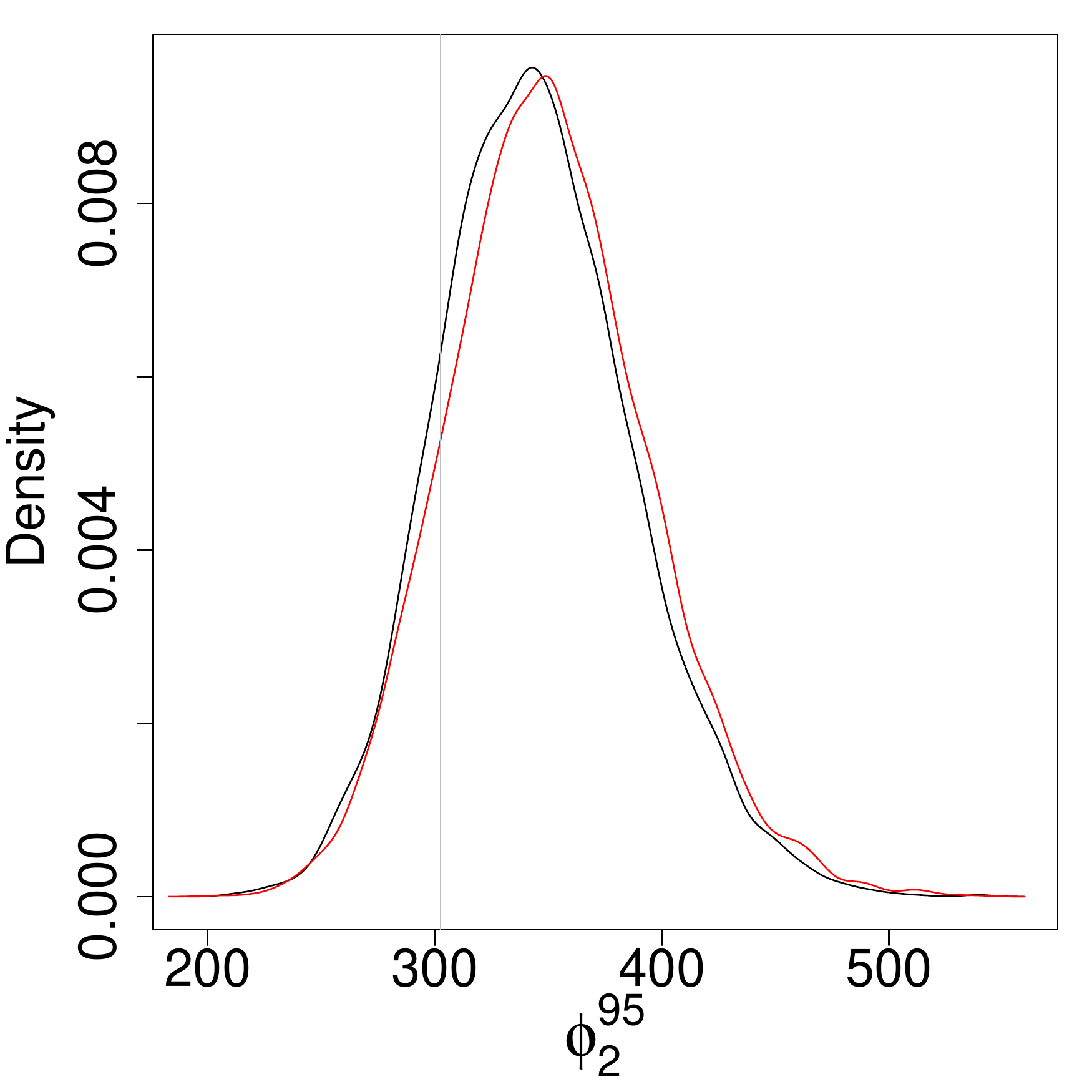}
\end{minipage}
\hspace{-0.25cm}
\begin{minipage}[b]{0.28\linewidth}
        \centering
        \includegraphics[scale=0.22]{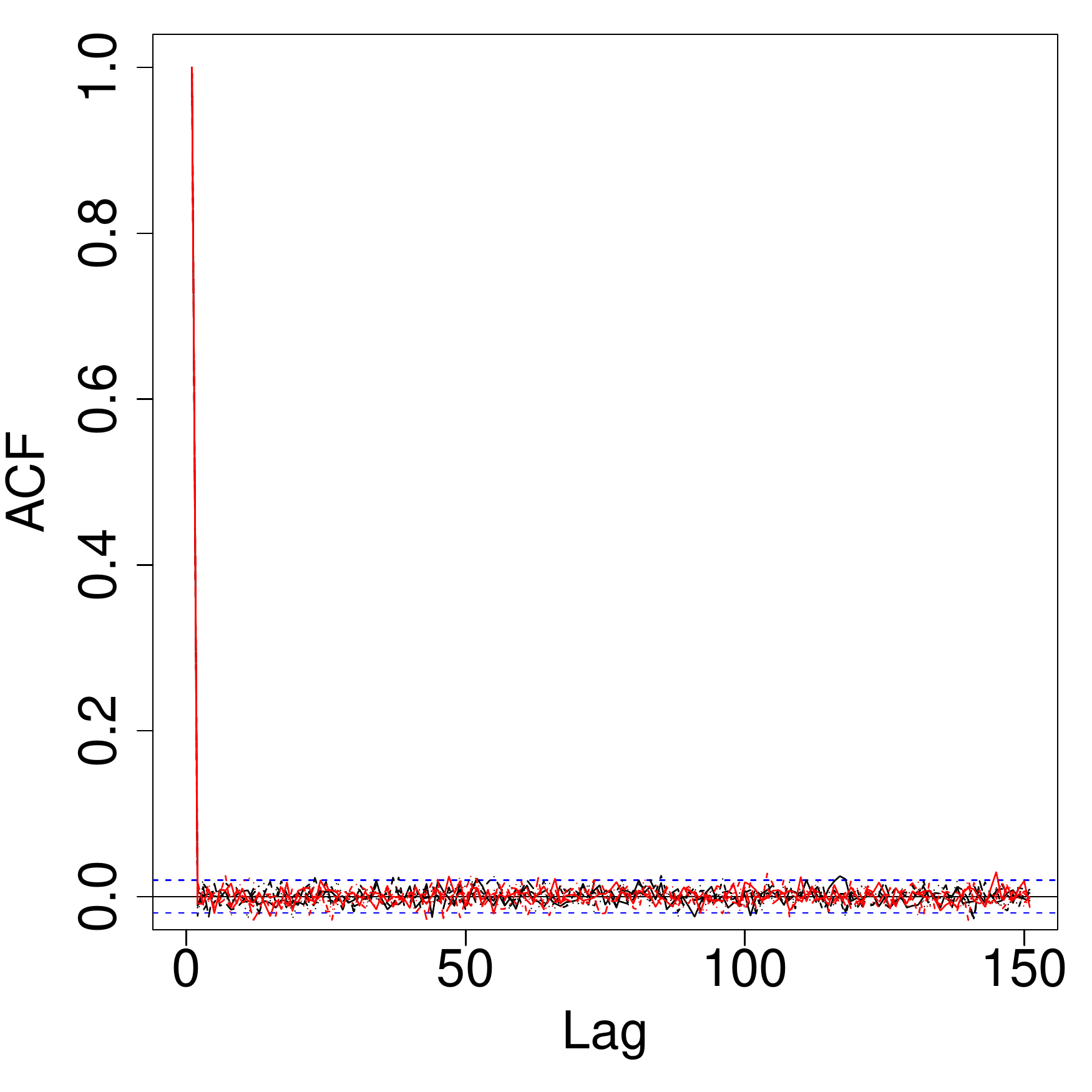}
\end{minipage}
\caption{Marginal posterior densities for a random selection of
  $\phi_1^i$ (top 2 rows) and $\phi_2^i$ (bottom 2 rows) in the orange
  tree growth SDMEM, together with their (overlayed) autocorrelation
  functions. Black: Bayesian imputation. Red: LNA. The vertical grey lines
  indicate the ground truth. } \label{orange fig_phi_i}
\end{center}
\end{figure}

Figure~\ref{orange fig_phi_i} shows the marginal posterior densities
of five randomly chosen random effects, based on synthetic data generated from 
the SDMEM of orange tree growth given in Section 5.1 of the main article.

\section{Simulation study}

In this section, we investigate further differences between the Bayesian 
imputation approach and an inference scheme based on the LNA using
synthetic data generated from (24) in the main paper. For simplicity, we consider 
a fixed treatment (low water, blanket nitrogen) and three blocks. We therefore 
write $X_t^{11k}=(N_t^{11k},C_t^{11k})^T$ and consider the \mbox{SDMEM}
\[
dX_t^{11k}=\alpha(X_t^{11k},b^{11k})\,dt+\sqrt{\beta(X_t^{11k},b^{11k})}\,dW_t^{11k}, \quad k\in\{1,2,3\},
\]
where
\begin{align*}
\alpha(X_t^{11k},b^{11k})&=\begin{pmatrix}
\lambda^{11k} N_t^{11k}-\mu^{11k} N_t^{11k}C_t^{11k} \\
\lambda^{11k} N_t^{11k} \\ \end{pmatrix},\\
\beta(X_t^{11k},b^{11k})&=\begin{pmatrix}
\lambda^{11k} N_t^{11k}+\mu^{11k} N_t^{11k}C_t^{11k} & \lambda^{11k} N_t^{11k}  \\
\lambda^{11k} N_t^{11k} & \lambda^{11k} N_t^{11k} \\ \end{pmatrix}.
\end{align*}
The fixed effects $b^{11k}=(\lambda^{11k},\mu^{11k})^T$ have a standard 
structure to incorporate block effects, with
\[
\lambda^{11k}=\lambda+\lambda_{B_k} \quad\textrm{and}\quad \mu^{11k}=\mu+\mu_{B_k},
\]
where we again impose the corner constraints $\lambda_{B_1} = \mu_{B_1} = 0$ 
to allow for identifiability. 

To mimic the real dataset, we took $\lambda=1.75$, $\mu=0.00095$, $\lambda_{B2}=-0.1154$, $\lambda_{B3}=-0.0225$, 
$\mu_{B2}=-0.0004$ and $\mu_{B3}=0.0002$. For each block, we generated five observations 
(on a regular grid) by using the Euler-Maruyama approximation with a small time-step 
($\Delta t=0.001$) and an initial condition of $x_{0}=(5,5)^T$. To assess the impact 
of measurement error on the quality of inferences that can be made about each parameter, 
we corrupted our data via the observation model
\[
Y_t^{11k}|N_t^{11k},\sigma \indep N(N_t^{11k},\sigma^{2}N_t^{11k}),\quad t=0, 1, 2, 3, 4,
\] 
and took $\sigma\in\{0,0.5,1,5\}$ to give four synthetic datasets. We adopt 
the same prior specification for the unknown parameters as used in the 
real data application.  

\begin{figure}
\begin{center}
\begin{minipage}[b]{0.28\linewidth}
        \centering
        \includegraphics[scale=0.22]{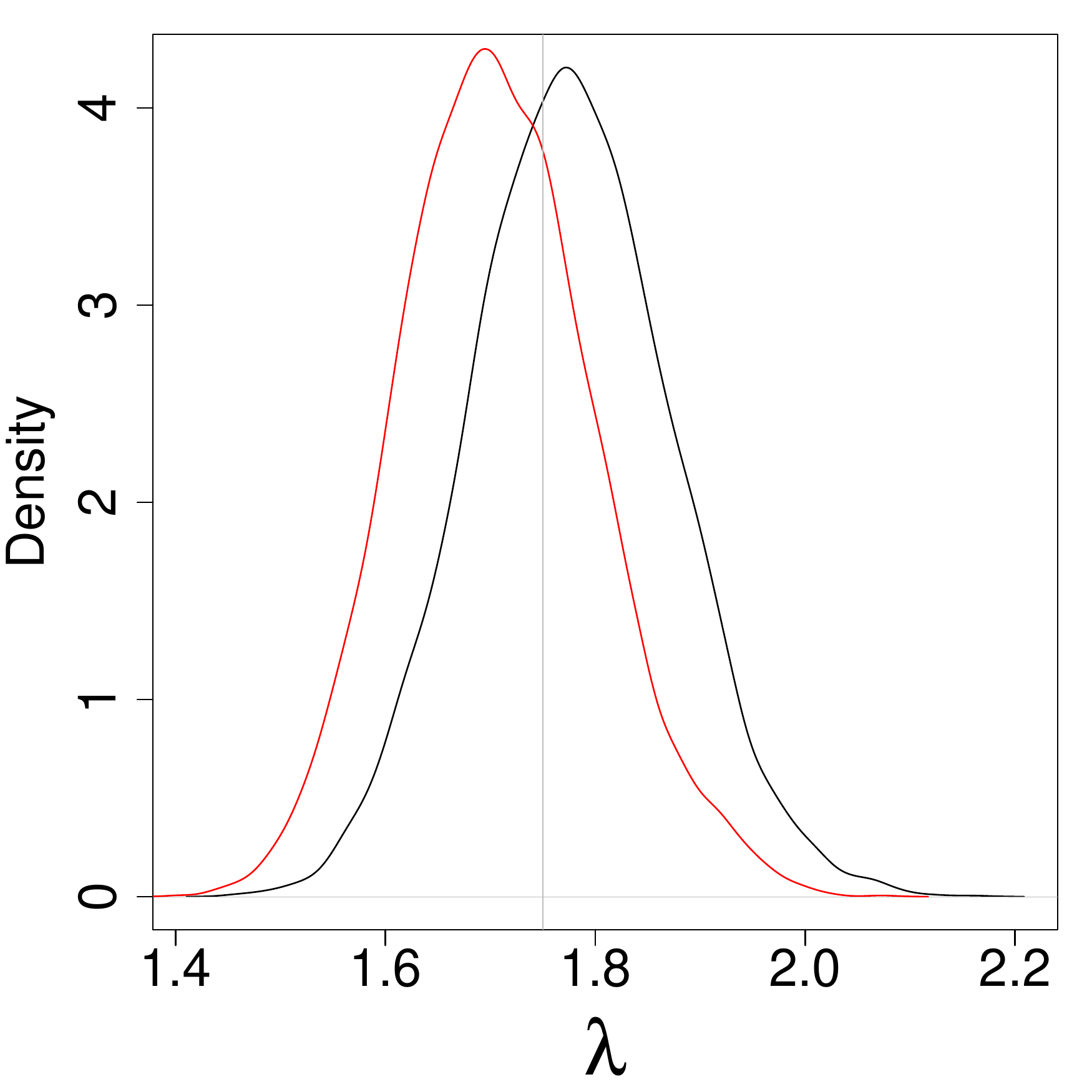}
\end{minipage} 
\hspace{-0.25cm}
\begin{minipage}[b]{0.28\linewidth}
        \centering
        \includegraphics[scale=0.22]{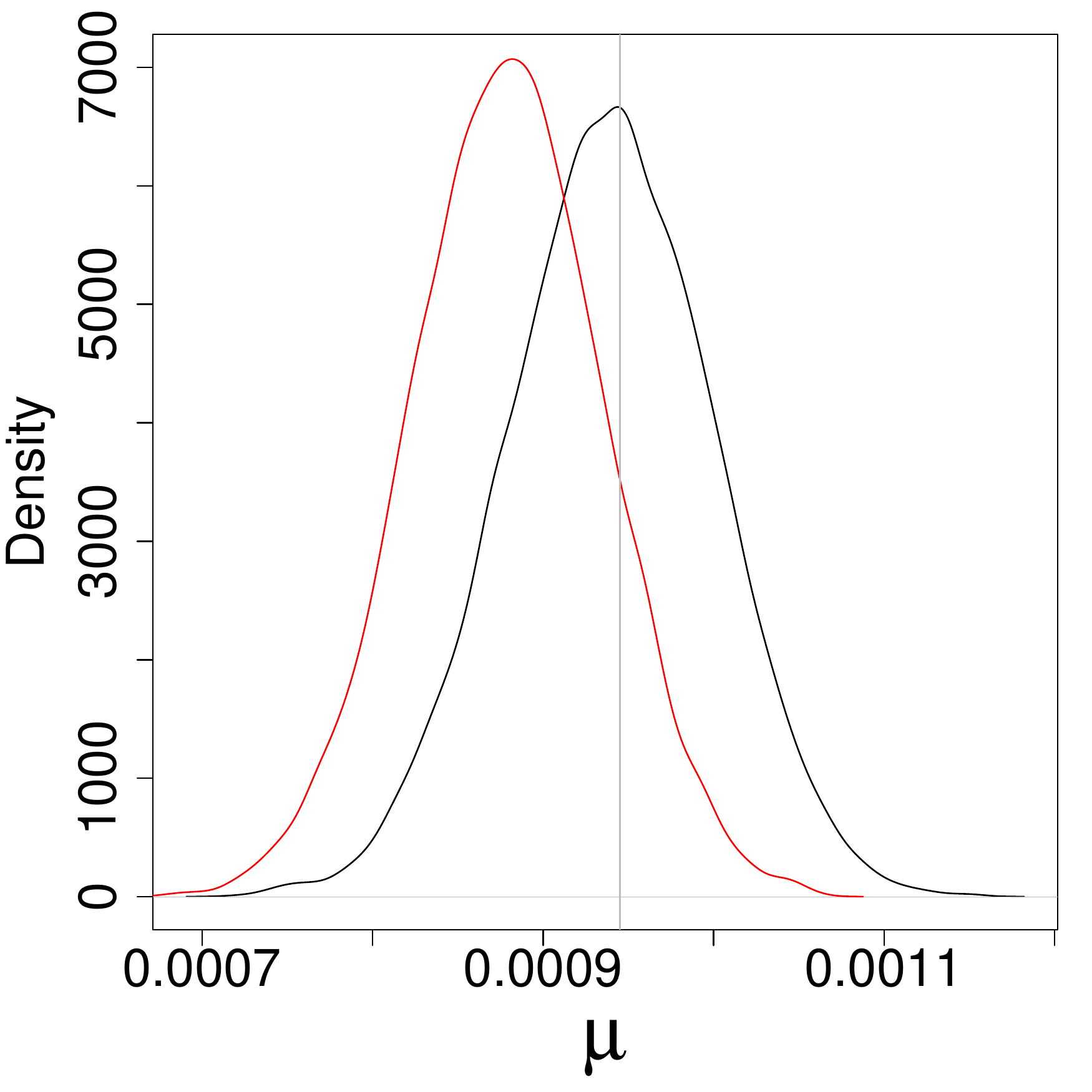}
\end{minipage} 
\hspace{-0.25cm}
\begin{minipage}[b]{0.28\linewidth}
				\hfill
\end{minipage} \\
\vspace{0.2cm}
\begin{minipage}[b]{0.28\linewidth}
        \centering
        \includegraphics[scale=0.22]{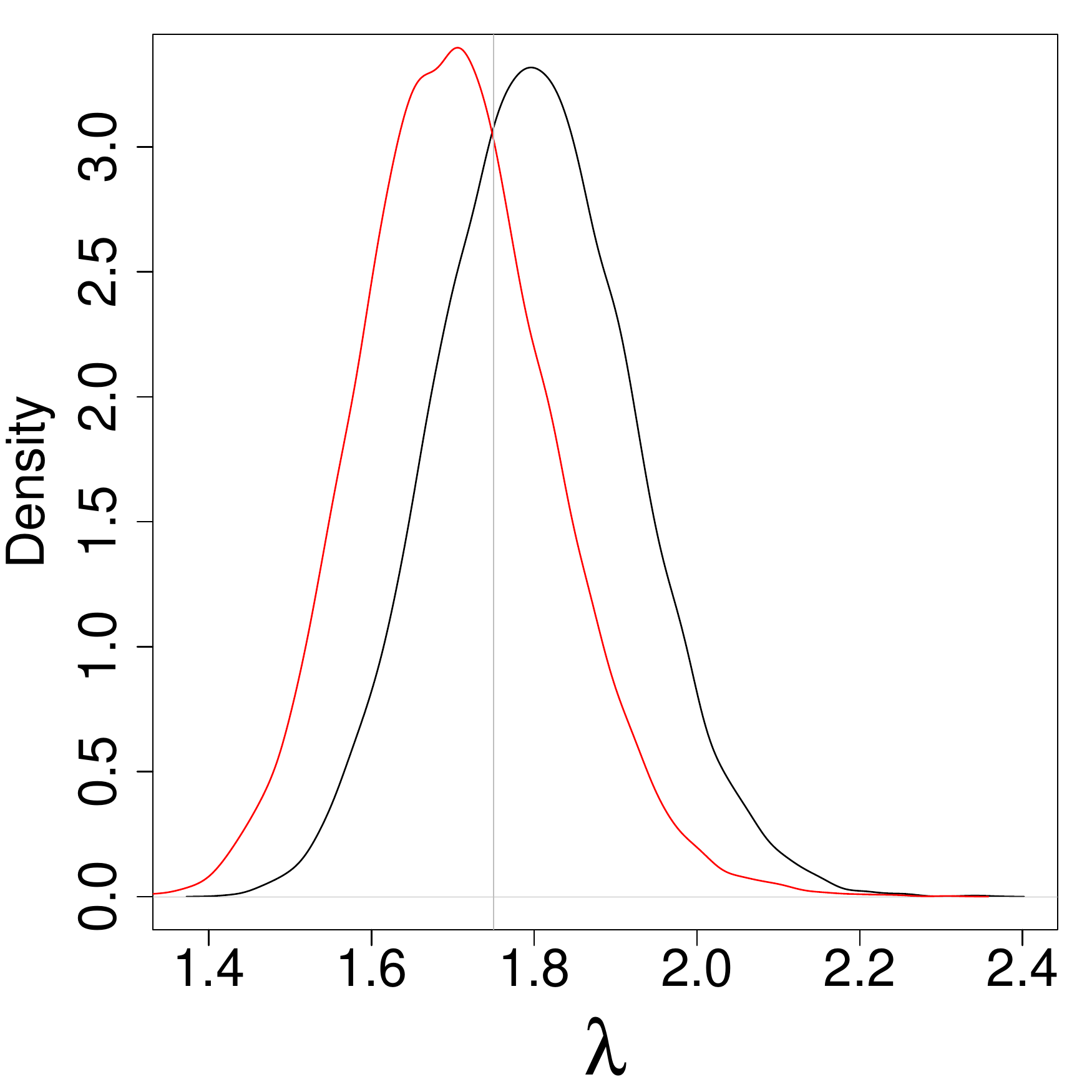}
\end{minipage} 
\hspace{-0.25cm}
\begin{minipage}[b]{0.28\linewidth}
        \centering
        \includegraphics[scale=0.22]{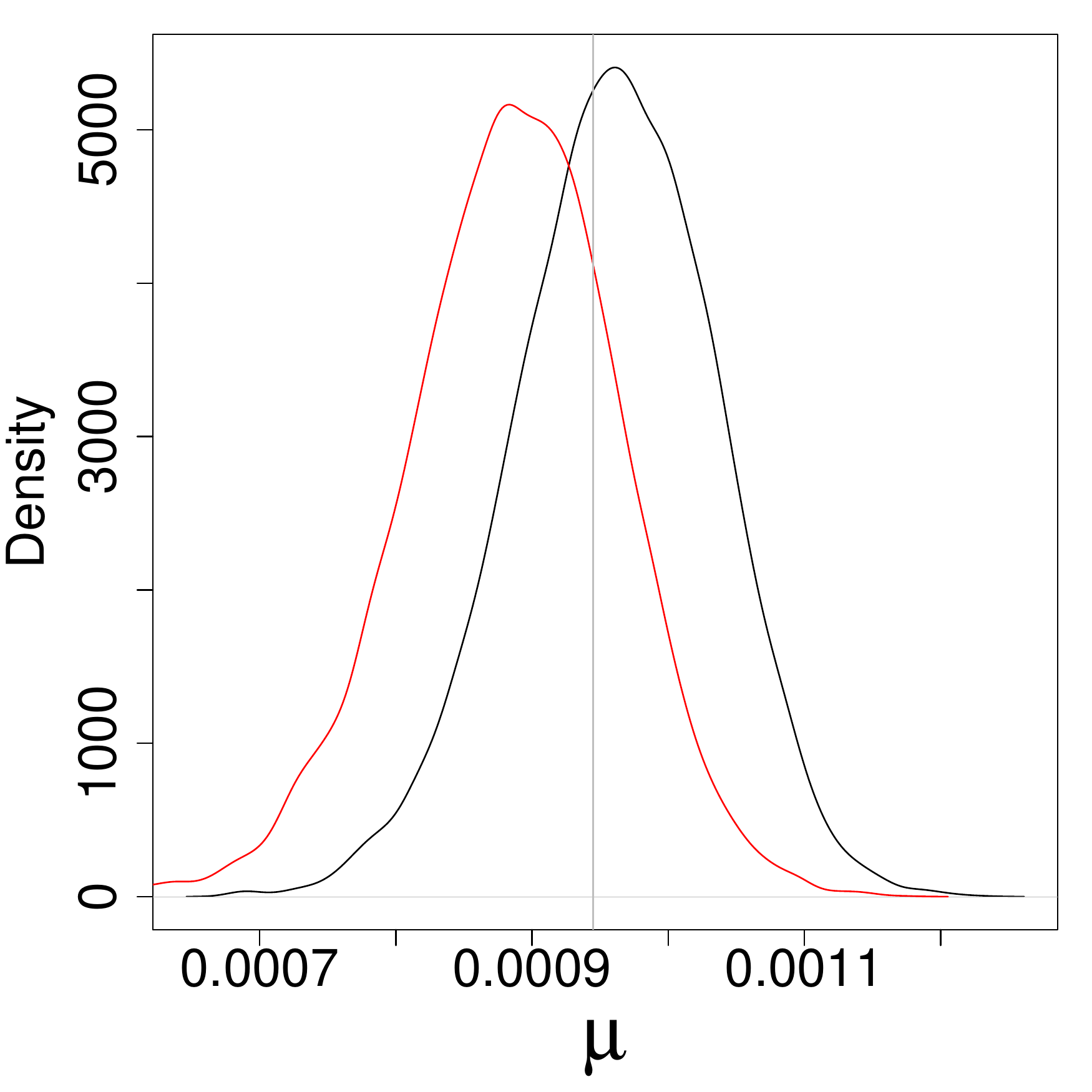}
\end{minipage} 
\hspace{-0.25cm}
\begin{minipage}[b]{0.28\linewidth}
				\centering
        \includegraphics[scale=0.22]{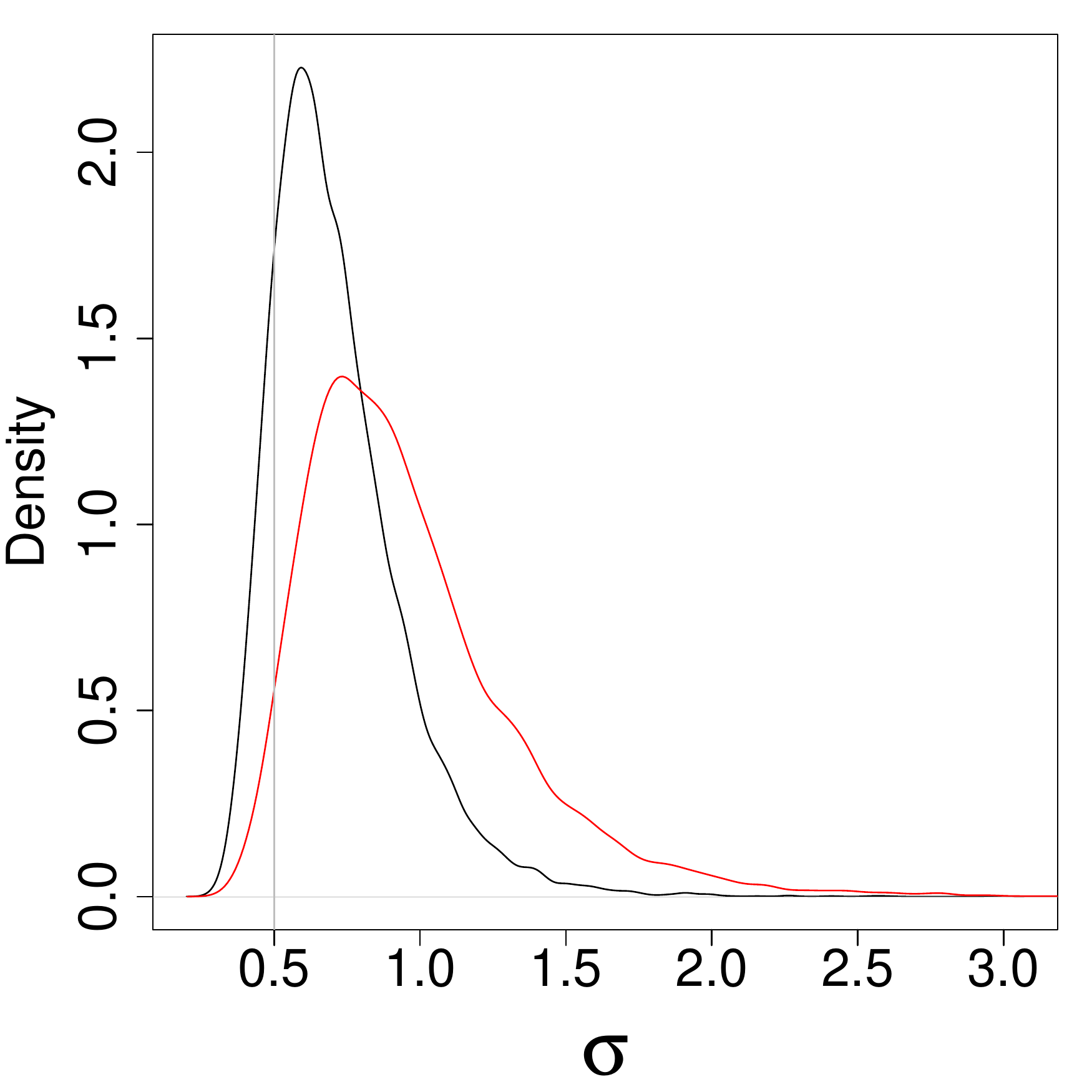}
\end{minipage}\\
\vspace{0.2cm}
\begin{minipage}[b]{0.28\linewidth}
        \centering
        \includegraphics[scale=0.22]{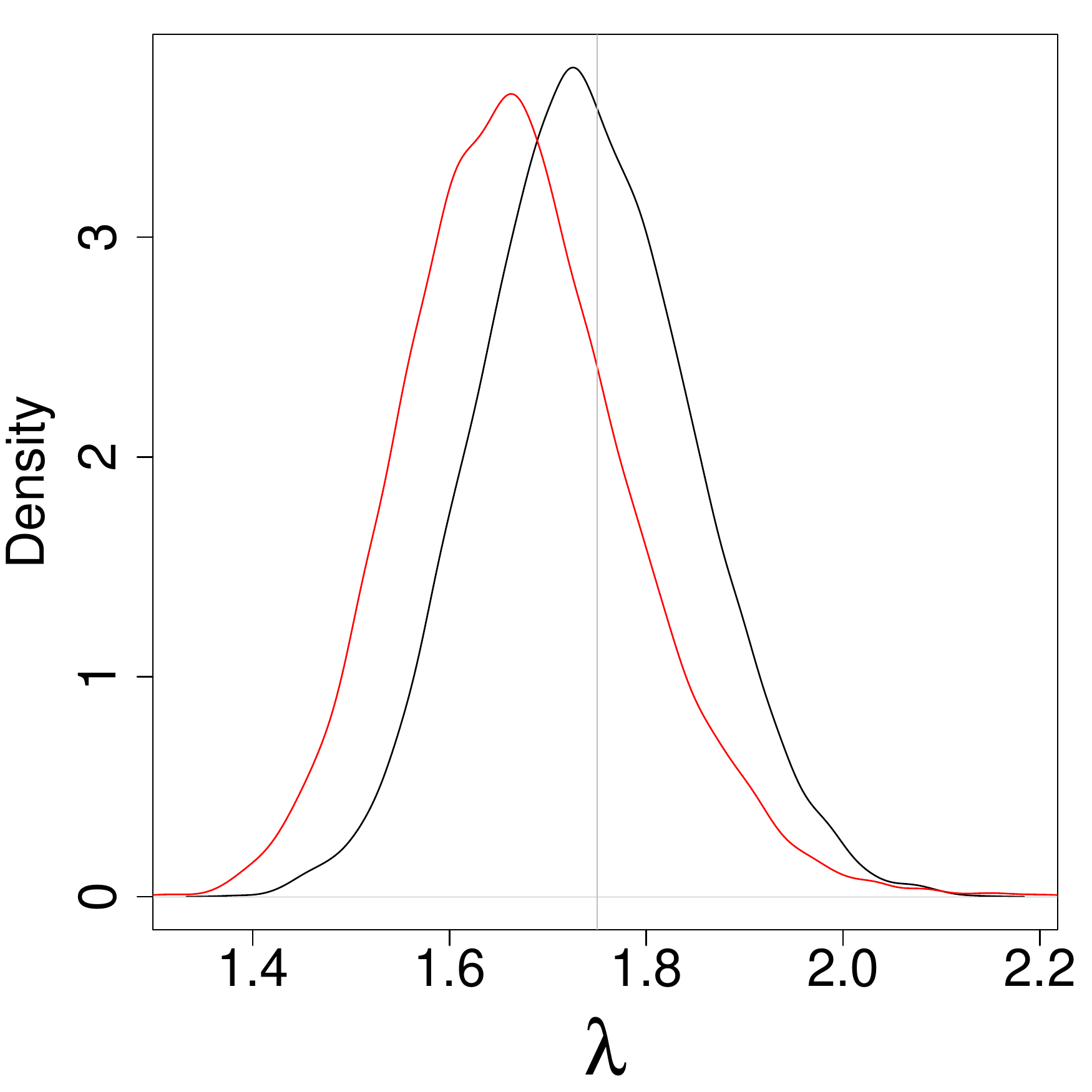}
\end{minipage} 
\hspace{-0.25cm}
\begin{minipage}[b]{0.28\linewidth}
        \centering
        \includegraphics[scale=0.22]{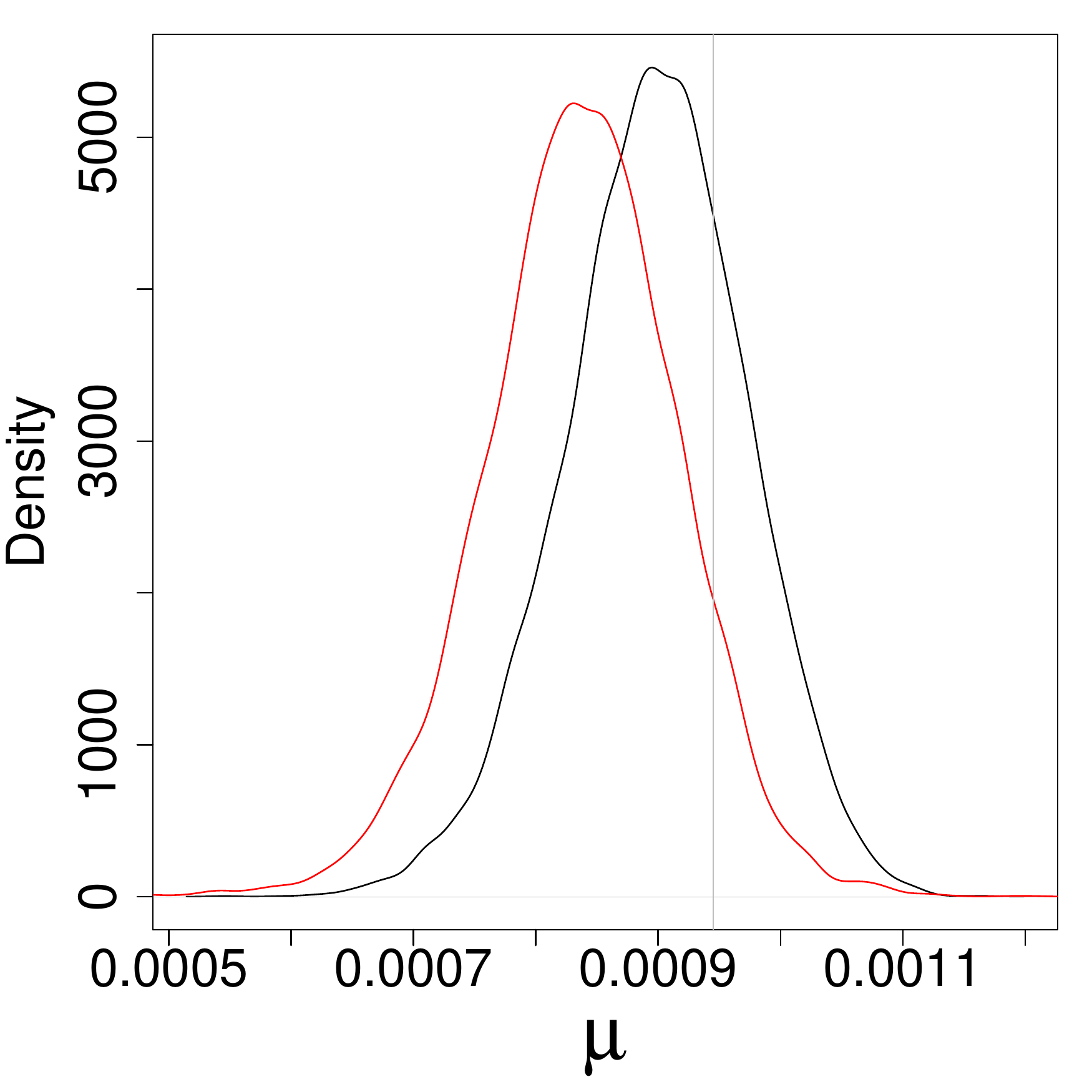}
\end{minipage} 
\hspace{-0.25cm}
\begin{minipage}[b]{0.28\linewidth}
				\centering
        \includegraphics[scale=0.22]{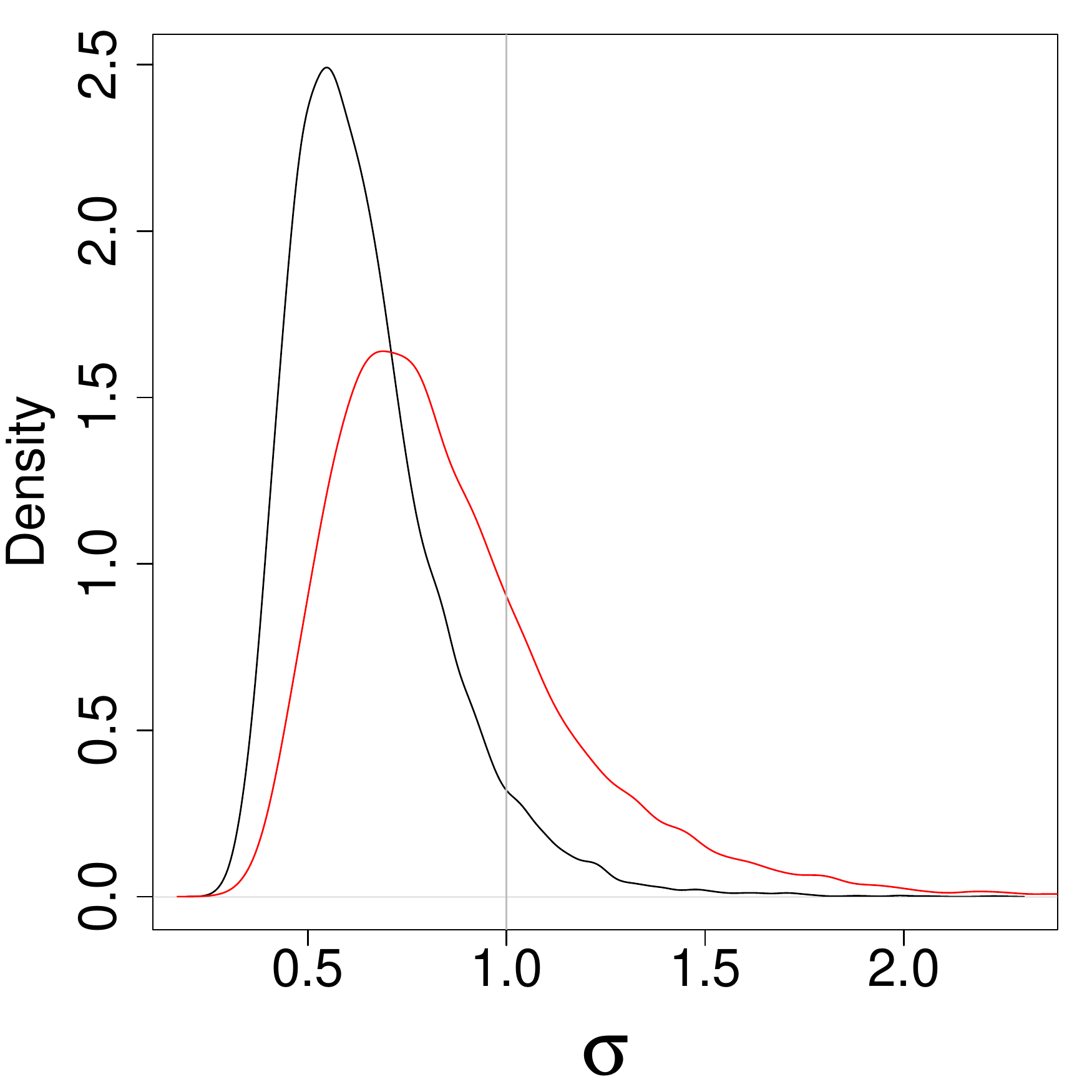}
\end{minipage}\\
\vspace{0.2cm}
\begin{minipage}[b]{0.28\linewidth}
        \centering
        \includegraphics[scale=0.22]{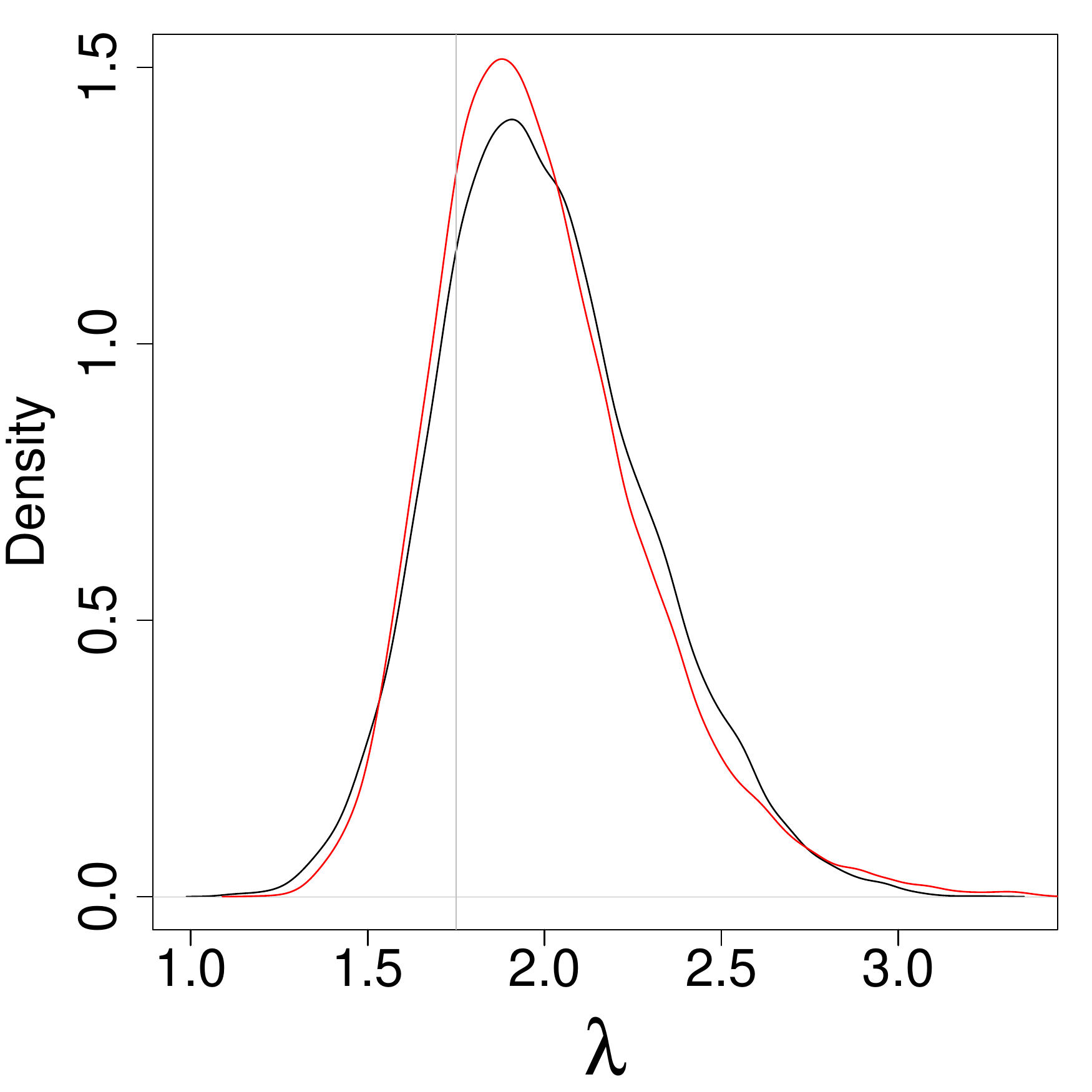}
\end{minipage} 
\hspace{-0.25cm}
\begin{minipage}[b]{0.28\linewidth}
        \centering
        \includegraphics[scale=0.22]{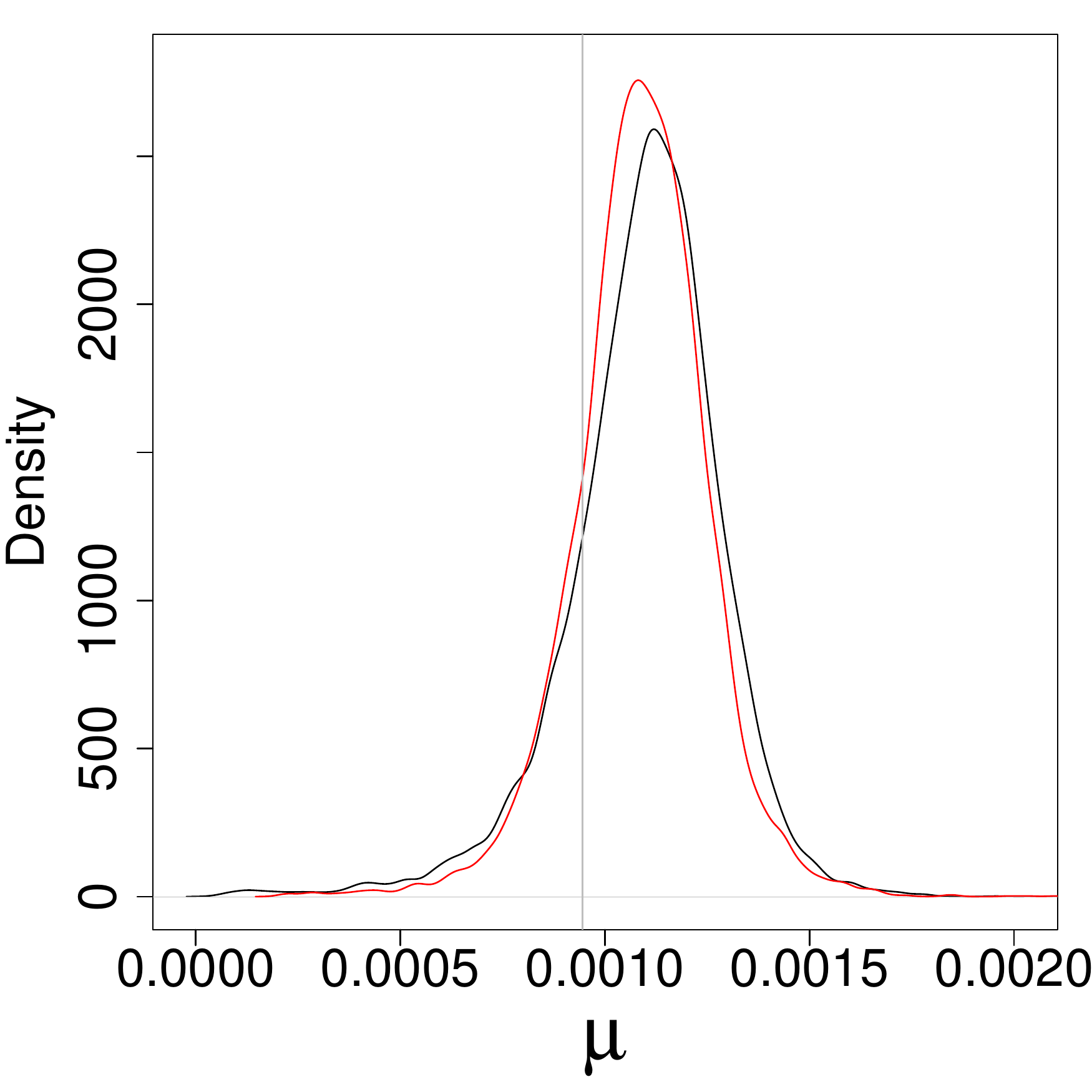}
\end{minipage} 
\hspace{-0.25cm}
\begin{minipage}[b]{0.28\linewidth}
				\centering
        \includegraphics[scale=0.22]{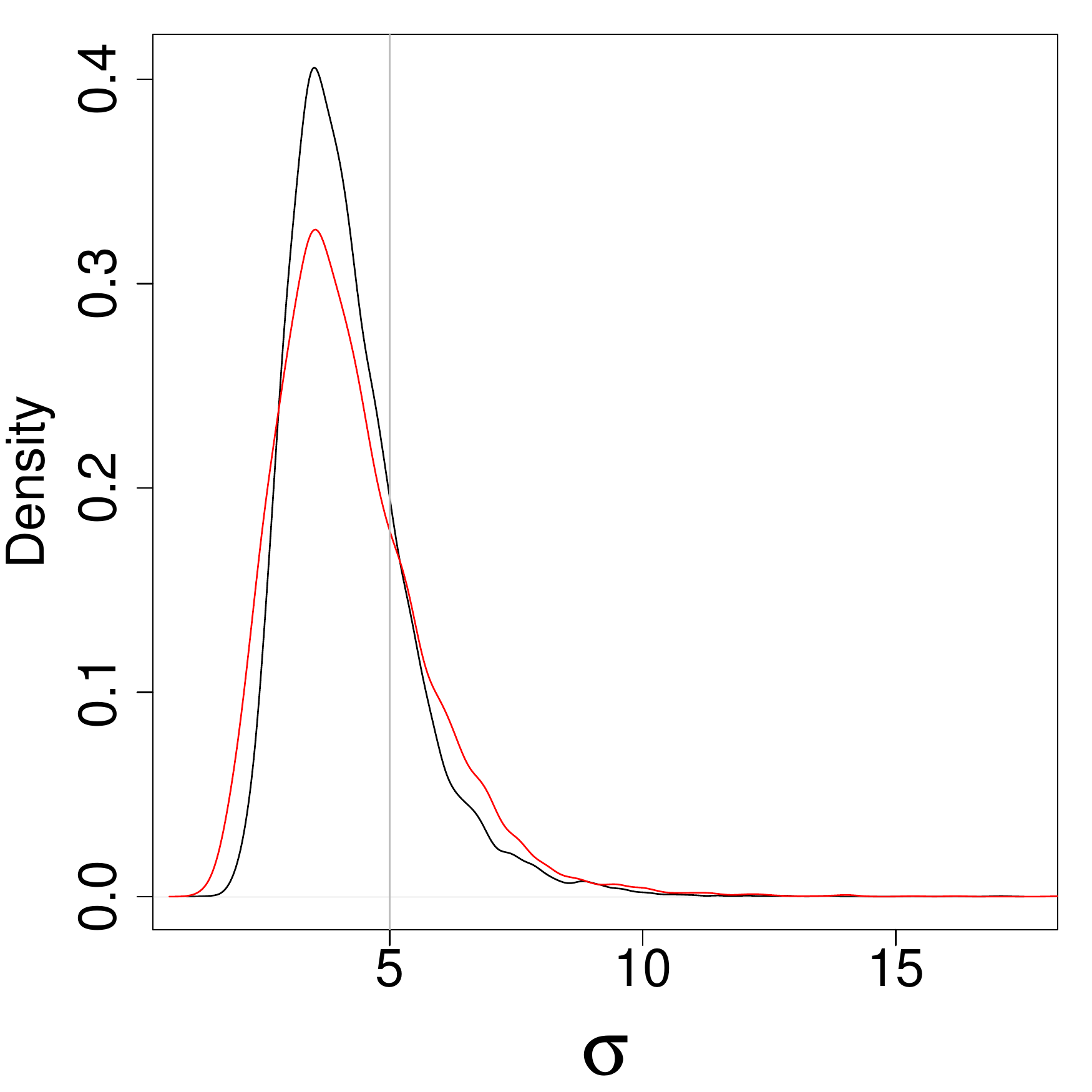}
\end{minipage}
\caption{Marginal posterior densities for the baseline parameters and 
the parameter $\sigma$
controlling the observation error variance
in the aphid simulation study. $\sigma=0$ (1$^{\textrm{st}}$ row), 
$\sigma=0.5$ (2$^{\textrm{nd}}$ row), 
$\sigma=1$ (3$^{\textrm{rd}}$ row), $\sigma=5$ (4$^{\textrm{th}}$ row).
Black: Bayesian imputation. Red: LNA. The vertical grey lines indicate the ground truth. } \label{aphid fig_sim_baselines}
\end{center}
\end{figure}

\begin{figure}[t!]
\begin{center}
\begin{minipage}[b]{0.49\linewidth}
        \centering
				\caption*{\quad $\sigma=0$}\vspace{-0.25cm}
        \includegraphics[scale=0.37]{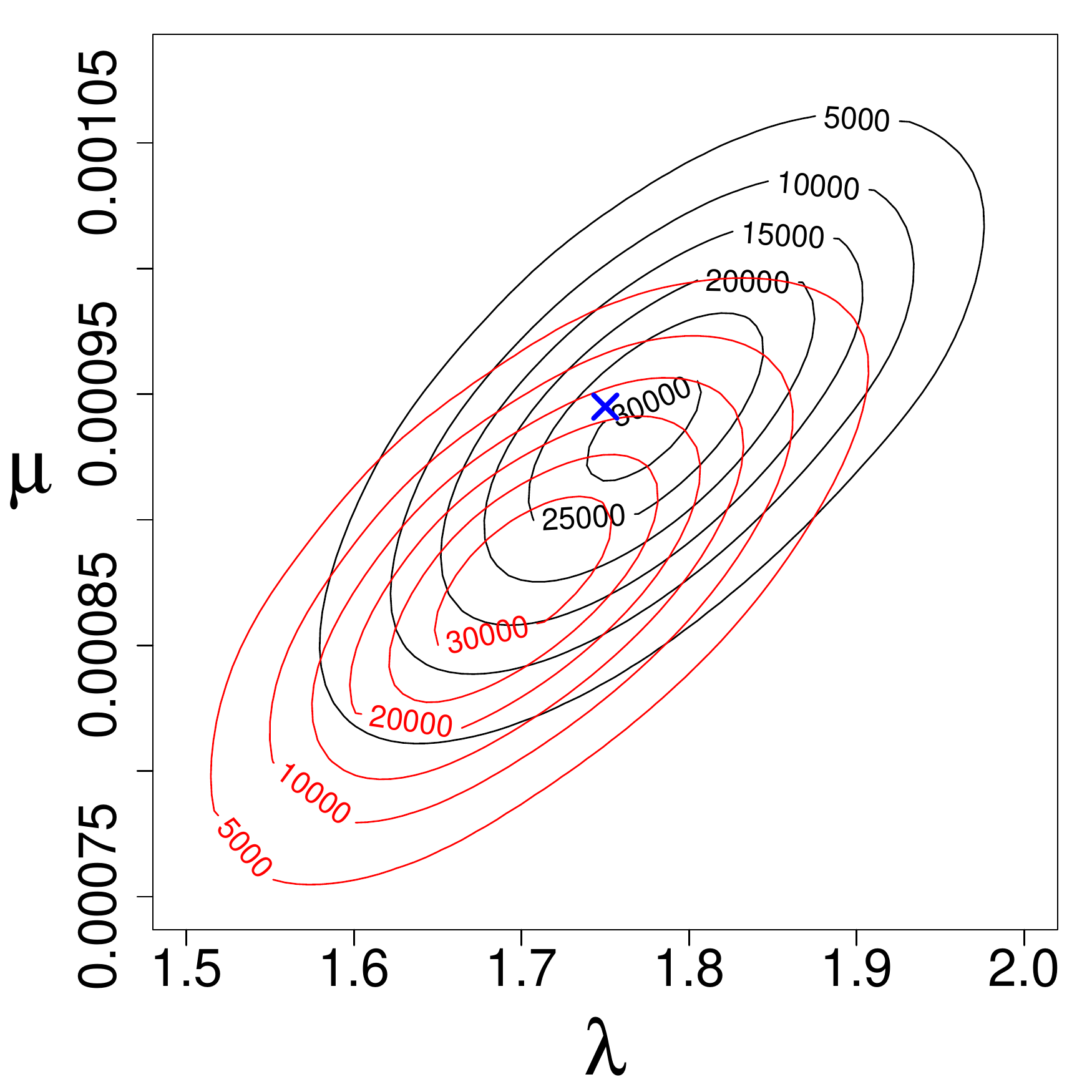}
\end{minipage} 
\hspace{-1cm}
\begin{minipage}[b]{0.49\linewidth}
        \centering
				\caption*{\qquad $\sigma=0.5$}\vspace{-0.25cm}
        \includegraphics[scale=0.37]{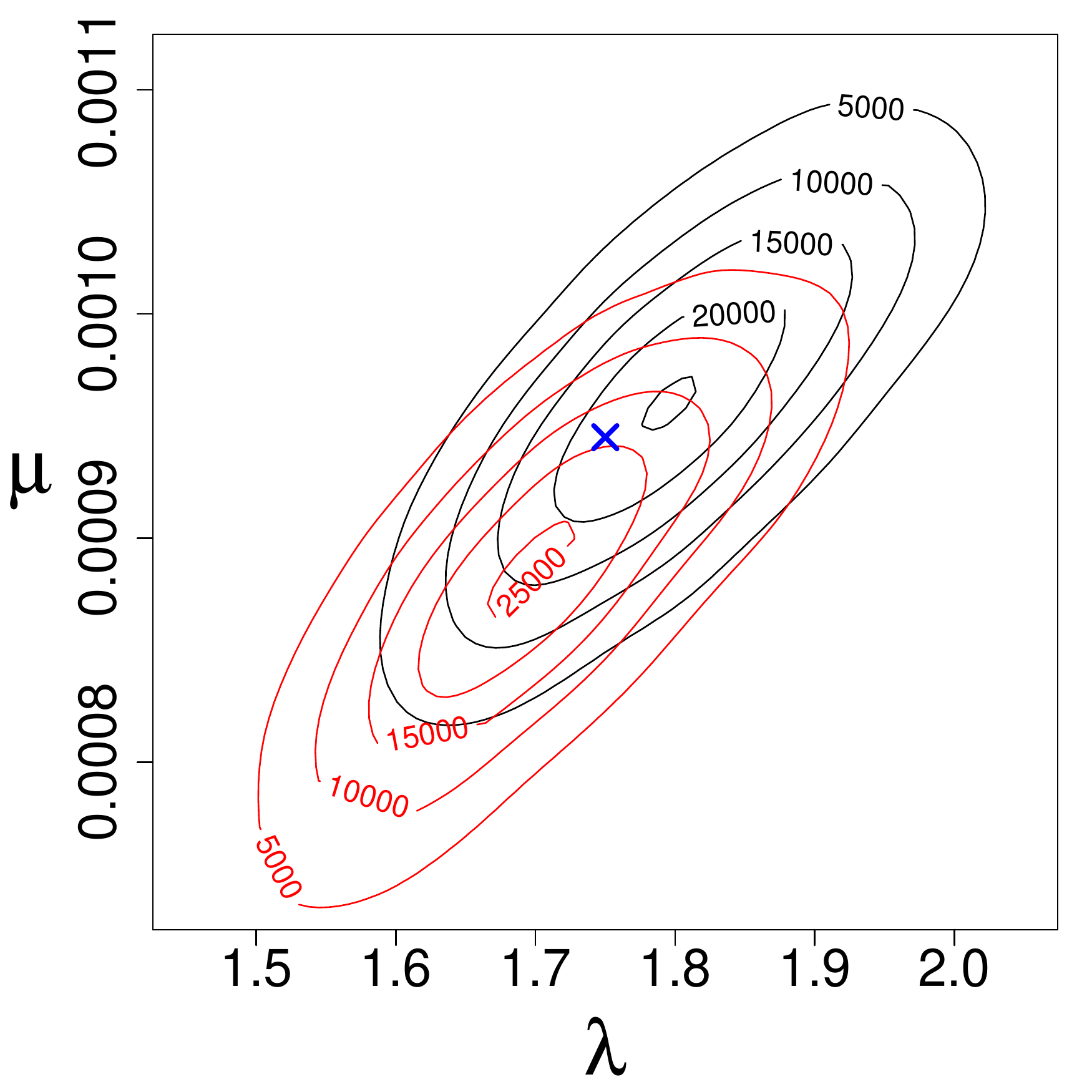}
\end{minipage} \\
\vspace{0.5cm}
\begin{minipage}[b]{0.49\linewidth}
        \centering
				\caption*{\qquad $\sigma=1$}\vspace{-0.25cm}
        \includegraphics[scale=0.37]{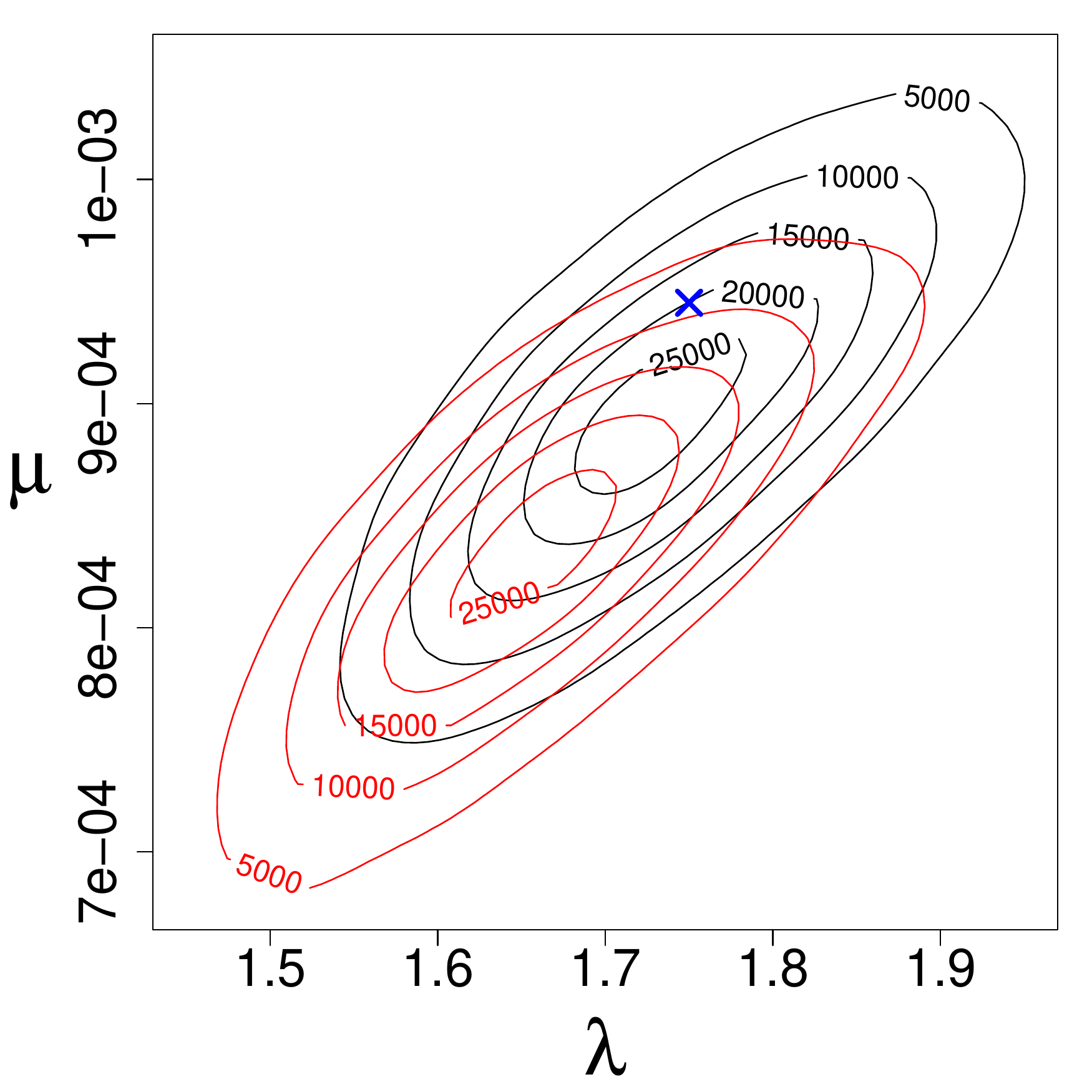}
\end{minipage} 
\hspace{-1cm}
\begin{minipage}[b]{0.49\linewidth}
				\centering
				\caption*{\qquad $\sigma=5$}\vspace{-0.25cm}
        \includegraphics[scale=0.37]{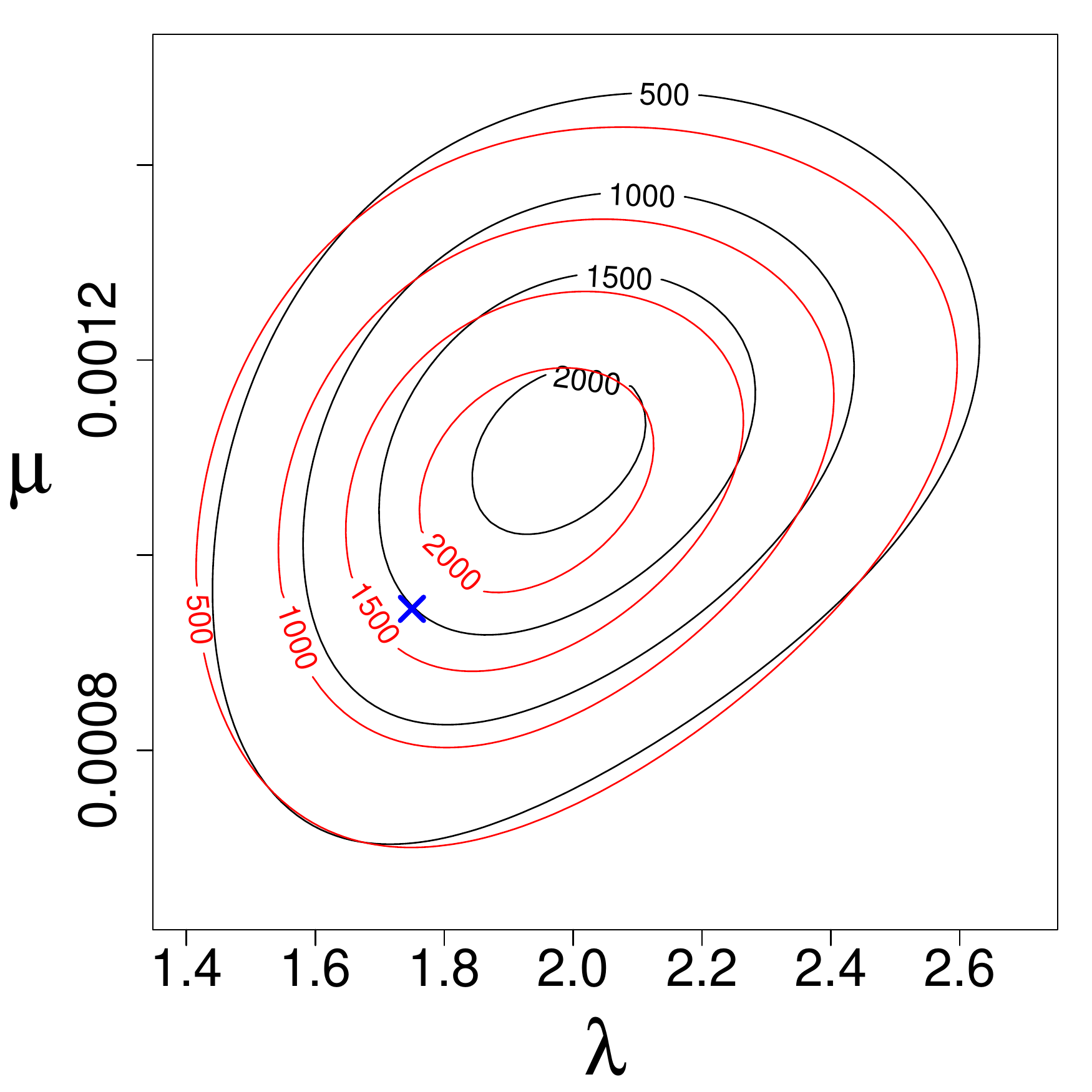}
\end{minipage}
\caption{Bivariate marginal posterior densities for the baseline parameters 
in the aphid simulation study
Black: Bayesian imputation. Red: LNA. The blue cross indicates the ground truth.} \label{aphid fig_sim_joint}
\end{center}
\end{figure}

Both the modified innovation scheme 
(again incorporating the improved bridge construct) 
and the LNA-based inference scheme were 
run long enough to yield a sample of approximately 10K independent posterior draws. 
For the former, we fixed the discretisation level by taking $m=20$ and note 
that $m>20$ gave little difference in posterior output. Figure~\ref{aphid fig_sim_baselines} shows 
the marginal posterior densities of the baseline parameters ($\lambda$ 
and $\mu$) and the measurement error variance ($\sigma$). The joint 
posterior densities of $(\mu,\lambda)^T$ are shown in Figure~\ref{aphid fig_sim_joint}. 
It is clear that when fitting the SDMEM using the Bayesian imputation approach, 
the posterior samples obtained are consistent with the ground truth. This is true to 
a lesser extent when using the LNA, with the ground truth found in the tail of 
the posterior distribution in three out of the four scenarios. In fact, 
when using synthetic data with $\sigma<5$, we see substantive differences 
in posterior output. As was observed when using real data, the LNA 
underestimates parameter values compared to those obtained 
under the Bayesian imputation scheme. In this case, the LNA provides 
a relatively poor approximation to the true posterior distribution.

Increasing $\sigma$ to 5 (and beyond) gives output from 
both schemes which is largely in agreement. This is intuitively reasonable, 
since, as the variance of the measurement process is increased, the ability 
of both inference schemes to accurately infer the underlying dynamics 
is diminished. Essentially, the relative difference between the LNA and 
SDE is reduced.    


\end{document}